\documentclass[fleqn,usenatbib]{mnras}

\usepackage{newtxtext,newtxmath}

\usepackage[T1]{fontenc}

\DeclareRobustCommand{\VAN}[3]{#2}
\let\VANthebibliography\thebibliography
\def\thebibliography{\DeclareRobustCommand{\VAN}[3]{##3}\VANthebibliography}

\usepackage{graphicx}	
\usepackage{amsmath}	
\usepackage{lscape}
\usepackage{color}

\newcommand{\tbol}{\mbox{$T_{\rm bol}$}} 
\newcommand{\lsmmbol}{\mbox{$L_{smm}/L_{bol}$}} 

\newcommand{\spitzer}{\emph{Spitzer}}

\newcommand{\degree}{\mbox{$^{\circ}$}}

\newcommand{\kms}{\mbox{km s$^{-1}$}}

\newcommand{\um}{$\mu$m}

\newcommand{\co}{$^{12}$CO}
\newcommand{\coo}{$^{13}$CO}

\newcommand{\cojone}{$^{12}$CO (1--0)}
\newcommand{\cojtwo}{$^{12}$CO~(2--1)}
\newcommand{\cojthree}{$^{12}$CO (3--2)}

\newcommand{\cooojtwo}{C$^{18}$O~(2--1)}

\title[MASSES Outflow Opening Angles]{The Evolution of Protostellar Outflow Opening Angles and the Implications for the Growth of Protostars}

\author[M. M. Dunham et al.]{
Michael M.~Dunham,$^{1,2}$\thanks{E-mail: mdunham@middlebury.edu}
Ian W.~Stephens,$^{3,4}$
Philip C.~Myers,$^{4}$
Tyler L.~Bourke,$^{5,4}$
H\'ector G.~Arce,$^{6}$
\newauthor
Riwaj Pokhrel,$^{7}$
Jaime~E.~Pineda,$^{8}$ \&
Joseph Vargas$^{2}$
\\
$^{1}$ Department of Physics, Middlebury College, Middlebury, VT 05753, USA \\
$^{2}$ Department of Physics, State University of New York at Fredonia, 280 Central Avenue, Fredonia, NY, 14063, USA \\
$^{3}$ Department of Earth, Environment, and Physics, Worcester State University, Worcester, MA 01602, USA \\
$^{4}$ Center for Astrophysics | Harvard \& Smithsonian, 60 Garden Street, Cambridge, MA 02138, USA \\
$^{5}$ SKA Observatory, Jodrell Bank, Macclesfield SK11 9FT, UK \\
$^{6}$ Department of Astronomy, Yale University, P.O. Box 208101, New Haven, CT 06520, USA \\
$^{7}$ Ritter Astrophysical Research Center, Dept of Physics and Astronomy, University of Toledo, OH, USA \\
$^{8}$ Max-Planck-Institut f\"ur extraterrestrische Physik, Giessenbachstrasse 1, D-85748 Garching, Germany
}

\date{Accepted XXX. Received YYY; in original form ZZZ}

\pubyear{2024}

\begin{document}
\label{firstpage}
\pagerange{\pageref{firstpage}--\pageref{lastpage}}
\maketitle

\begin{abstract}
We use \color{black}$1-4$" ($300-1200$~au) resolution \cojtwo\ \color{black}data from the MASSES ({\underline M}ass {\underline A}ssembly of {\underline S}tellar {\underline S}ystems and their {\underline E}volution with the {\underline S}MA) project to measure the \color{black}projected \color{black}opening angles of 46 protostellar outflows in the Perseus Molecular Cloud, 37 of which are measured with sufficiently high confidence to use in further analysis.  We find that there is a statistically significant difference in the distributions of outflow opening angles for Class 0 and Class I outflows, with a distinct lack of both wide-angle Class 0 outflows and highly collimated Class I outflows.  Synthesizing our results with several previous studies, we find that outflows widen with age through the Class 0 stage but do not continue to widen in the Class I stage.  The maximum \color{black}projected \color{black}opening angle reached is approximately 90\degree~$\pm$~20\degree, with the transition between widening and remaining constant occurring near the boundary between the Class 0 and Class I phases of evolution.  While the volume fractions occupied by these outflows are no more than a few tens of percent of the total core volume, at most, \color{black}recent theoretical work suggests outflows may still be capable of playing a central role in setting the low star formation efficiencies of 25\%~$-$~50\% observed on core scales.\color{black}
\end{abstract}




\section{Introduction}\label{sec_intro}

Stars are assembled in molecular clouds through the gravitational collapse of dense cores of gas and dust \citep[e.g.,][]{shu1987:review,difrancesco2007:ppv,wardthompson2007:ppv,dunham2014:ppvi,pineda2022:pp7}.  The final masses of stars are set during the protostellar stage of star formation, when the forming protostar is embedded within, and accreting material from, its parent dense core.  \color{black}\citet{andre1993:class0} divided this protostellar stage into Class 0 and Class I protostars, where Class 0 protostars have accreted less than half of their final mass and Class I protostars have accreted more than half of their final mass.  In practice the observational class of a protostar is typically measured and assigned based on its observed spectral energy distribution \citep[SED; e.g., ][]{andre1993:class0,myers1993:tbol,evans2009:c2d,frimann2016:simulations}, with the observational classes serving as proxies for the physical stages.  In this paper we follow the convention first introduced by \citet{robitaille2006:models} of using ``Class'' when referring to the observational categories and ``Stage'' when referring to physical categories, although we note that Robitaille et al.~used this terminology to distinguish between protostars and more evolved young stellar objects, and did not separately distinguish between Stage 0 and Stage I protostars.

One of the specific methods of classifying protostars into their observational categories \color{black}is based on the bolometric temperature, \tbol, defined to be the temperature of a blackbody with the same flux-weighted mean frequency as the SED of the object \citep{myers1993:tbol}.  It is in essence a protostellar equivalent of stellar effective temperature; it starts at very low values ($\sim 10$ K) for cold, starless cores and eventually increases to the effective temperature of the central (proto)star once the core and disk have fully dissipated.  Observational studies have placed the boundary between Class 0 and Class I protostars at \tbol~$=70$~K \citep{chen1995:tbol}.

A major unsolved problem in star formation is to understand the complex interactions of physical processes that set the final masses of stars, including mass infall, disk formation and growth, the formation of multiple systems through fragmentation on both core and disk scales, and mass loss through jets and outflows.  While numerous studies have examined these processes, a complete understanding of the interplay between all of them remains lacking.  The {\underline M}ass {\underline A}ssembly of {\underline S}tellar {\underline S}ystems and their {\underline E}volution with the {\underline S}ubmillimeter Array (MASSES) project represents an endeavor to study these physical processes using an unbiased sample of protostars with a homoegenous set of observations.  MASSES used the Submillimeter Array \citep[SMA;][]{ho2004:sma} to map all of the protostars in the Perseus Molecular Cloud \citep[$d=300$~pc;][]{zucker2018:distances,zucker2019:distances} in the continuum and a variety of molecular line tracers at both 230 and 345~GHz.  At 230~GHz, MASSES used the combination of the subcompact and extended SMA configurations to achieve an angular resolution of approximately 3$\arcsec$ (900~au at the distance to Perseus) while maintaining sensitivity to emission on scales up to approximately 20$\arcsec$ (6000~au at the distance to Perseus).  A full description of the MASSES observations, data reduction, and public data releases are presented by \citet{stephens2018:masses} and \citet{stephens2019:masses}.  Other MASSES papers to date have focused on the kinematics of protostellar cores \citep{lee2015:masses,heimsworth2022:masses}, the origins of fragmentation \citep{lee2015:masses,lee2016:masses,pokhrel2018:masses}, chemical evidence of an episodic accretion process \citep{frimann2017:masses}, the formation and growth of circumstellar disks \citep{andersen2019:masses}, and the connections between dense cores and the larger filamentary environments in which they form \citep{stephens2017:masses}.

In this paper we use the MASSES data to investigate the growth of protostellar outflows.  Outflows from protostars may play a crucial role in regulating accretion of material onto stars by injecting momentum and energy into the surrounding medium and possibly feeding turbulence in the surrounding gas \citep[e.g.,][]{nakamura2007:outflows,nakamura2011:serpsouth,hansen2012:outflows,plunkett2013:ngc1333,frank2014:ppvi,offner2014:outflows,offner2017:outflows}.  Indeed, numerous studies have compared the dense core mass function (CMF) to the stellar initial mass function (IMF) and concluded that the IMF may be inherited from the CMF, with a star formation efficiency on dense core scales of approximately 25\%~--~50\% \citep[e.g.,][]{motte1998:oph,motte2001:mm,alves2007:cmf,andre2010:gb,konyves2015:herschel,motte2022:cmf}.  In this scenario, outflows may be partly or mostly responsible for clearing the approximately 50\%~--~75\% of the dense core mass that does not become part of the final stellar mass, ultimately terminating the infall and setting the star formation efficiency on dense core scales \citep[e.g.,][]{velumsay1998:outflows,matzner2000:outflows,arce2006:outflows,myers2008:theory,machida2013:outflows}.

If outflows are to remove a significant fraction of the initial core mass, they must be sufficiently wide to remove mass from, and prohibit accretion from, a substantial fraction of the solid angle of the core.  In a seminal investigation of the opening angles of protostellar outflows, \citet{arce2006:outflows} used \co\ observations to show that outflows grow wider with time, reaching opening angles of approximately 180\degree\ by the end of the protostellar stage and thus halting further infall.   Their results were reproduced in simulations \citep{offner2011:outflows}, as well as in other studies measuring outflow cavity opening angles \citep{seale2008:outflows,velusamy2014:outflows,hsieh2017:outflows,hsieh2023:outflows}.  On the other hand, other studies have claimed that outflows are not capable of dispersing a large fraction of core material \citep[e.g.,][]{hatchell2007:outflows,curtis2010:outflows,habel2021:outflows}.  While the former two studies may be affected by underestimates of the total mass and energetics of the outflows in their sample \citep[see, e.g., ][]{dunham2014:outflows}, the latter work by \citet{habel2021:outflows} used infrared scattered light measurements of outflow cavity opening angles to argue against a widening trend with age.

In this present work we revisit the evolution of protostellar outflows using \co\ data from the MASSES project, leveraging the statistical power provided by the fact that MASSES targeted the complete population of Class 0 and Class I protostars in the Perseus Molecular Cloud.  The organization of this paper is as follows: in \S \ref{sec_observations} we briefly summarize the MASSES sample and observational strategy, with references to previous MASSES papers where these are described in more detail.  In \S \ref{sec_method} we describe the fitting method we use to measure outflow opening angles.  We present our results on the growth of protostellar outflow opening angles in \S \ref{sec_results}.  We then discuss our results in \S \ref{sec_discussion}, including a detailed comparison to previous results (\S \ref{sec_discussion_previous}), a synthesis of what the combination of our results and previous results imply for the accretion of mass onto protostars (\S \ref{sec_discussion_implications}), and a discussion of the limitations of the current work (\S \ref{sec_discussion_limitations}).  Finally, we summarize our main findings in \S \ref{sec_summary}.

\section{MASSES Sample and Observational Strategy}\label{sec_observations}

The MASSES survey targeted all 74 known Class 0 and Class I protostellar systems and candidate first hydrostatic cores within the Perseus Molecular Cloud as of 2014, when the survey was designed \citep{stephens2018:masses,stephens2019:masses}.  Perseus, located at a distance of $\sim$300~pc \citep{zucker2018:distances,zucker2019:distances}, was chosen for the MASSES survey due to its relative proximity to Earth, its large number of Class 0 and I protostars, and its declination of approximately $+30\degree$ making it well-suited for optimal SMA observations.  MASSES obtained continuum images at both 230 and 345~GHz, as well as image (spectral) cubes centered on a number of different molecular line transitions, in two different SMA array configurations, with spatial resolutions ranging from approximately 1$\arcsec$--4$\arcsec$ (300--1200~au at the distance to Perseus), depending on the image weighting used.  Full details on the observational, calibration, and imaging strategies are presented by \citet{stephens2018:masses} and \citet{stephens2019:masses}.  All of the MASSES data are now publicly available\footnote{See https://dataverse.harvard.edu/dataverse/full\_MASSES/} \citep[see][for details]{stephens2019:masses}.  In this paper we focus exclusively on the \cojtwo\ data, first presented in \citet{stephens2018:masses} and  \citet{stephens2019:masses}.

Galleries of all \cojtwo\ maps in the MASSES dataset are presented by \citet{stephens2018:masses} and \citet{stephens2019:masses}.  Since many of the protostars in Perseus are located in clustered regions, and individual maps centered on each pointing do not show the larger environment in which each protostar is located, we present in Appendix \ref{sec_app_clusters} large-scale maps showing the MASSES \cojtwo\ images in the IC348, B1, NGC~1333, L1455, and L1448 regions of Perseus.  These are not mosaics, rather they plot the individual pointings on one large map, offset from each other to reflect their positions relative to each other.

In total, we are able to measure outflow opening angles for 46  different outflows in the MASSES sample, using a method we describe below in \S \ref{sec_method}.  Table \ref{tab_sample} reports the name of each driving source (along with common names frequently used for each source), \color{black}the clustered region of Perseus in which this protostar is located (if any), \color{black}the position of the protostar as traced by our MASSES 1.3~mm continuum detections \color{black}when available \citep{pokhrel2018:masses,stephens2018:masses}, or by the VLA Nascent Disk And Multiplicity Survey of Perseus Protostars \citep[VANDAM;][]{tobin2016:vandam,tobin2018:vandam}, \color{black}the bolometric temperatures of the protostars as compiled by the eHOPS survey \citep{pokhrel2022:protostars}\footnote{These updated values of \tbol\ were calculated by \citet{pokhrel2022:protostars} using photometric and spectroscopic measurements in the $1-850$~\um\ wavelength range from the Two Micron All-Sky Survey (2MASS), the {\it Spitzer Space Telescope}, the {\it Herschel Space Observatory}, and the James Clerk Maxwell Telescope (JCMT).}, the image weighting we used to make our measurements (see the following paragraph below), the major and minor axes of the MASSES \cojtwo\ synthesized beam for each target using the image weighting listed in the previous column, \color{black}the systemic velocity of each protostar when available from MASSES \cooojtwo\ fits by \citet{stephens2018:masses} and \citet{stephens2019:masses}, \color{black}and the velocity intervals used for the fitting procedure described below in \S \ref{sec_method}.  

The MASSES survey produced and publicly delivered images with two different weightings, one with a robust parameter of $+$1 and one with a robust parameter of $-$1.  While we refer to \citet{stephens2019:masses} for details, we note here that, compared to the robust~=~$+$1 images, the robust~=~$-$1 images have higher spatial resolution at the cost of reduced sensitivity and reduced recovery of large-scale emission.  For the majority of the outflows identified here, we use the robust~=~$+$1 images to measure opening angles due to their improved sensitivity and improved recovery of large-scale emission.  Eight poorly resolved outflows in the robust~=~$+$1 images are both better resolved and robustly detected in the robust~=~$-$1 images, thus for these eight outflows we use the robust~=~$-$1 images instead and indicate as such in the sixth column of Table \ref{tab_sample}.  For outflows that are well-detected and well-resolved in both image weightings, our measured opening angles agree to within the uncertainties.

\color{black}Notes on the detected outflows (and their driving sources) relevant for measuring their opening angles are given in Appendix \ref{sec_app_fit}.  Table \ref{tab_sample_nondetect} lists the source (and any associated common names), region (if applicable), phase center of the MASSES observations, and \tbol\ of all MASSES targets for which we are unable to obtain outflow opening angle measurements.  Relevant details on these targets are given in Appendix \ref{sec_app_nofit}, focusing on what is seen in the MASSES \cojtwo\ observations and the reason(s) why we are unable to obtain opening angle measurements. \color{black}

\begin{landscape}
\begin{table}
\begin{center}
\caption{MASSES Protostars with Measured Outflow Opening Angles}
\label{tab_sample}
\begin{tabular}{lcccccccccc}
\hline \hline
            & Common            &              & Protostar R.A.$^{\rm b}$   & Protostar Decl.$^{\rm b}$     & \tbol$^{\rm c}$   &        & Beam Major \& Minor & v$_{\rm systemic}$ & v$_{\rm blue}$      & v$_{\rm red}$ \\
Source      & Name              & Region$^{\rm a}$ & (J2000)         & (J2000)            & (K)     & Robust & Axes (arcsec)$^{\rm d}$       & (\kms)$^{\rm e}$              & (\kms)              & (\kms)  \\
\hline
Per-emb~1   & HH211-MMS         & IC348    & 03 43 56.770    & $+$32 00 49.865    & 24.1    & $-$1   & 1.6 $\times$ 1.2 &  9.4 &    2.0  --  5.0  & 13.0  -- 16.0 \\
Per-emb~2   & IRAS~03292$+$3039 &          & 03 32 17.915    & $+$30 49 48.033    & 30.9    & $+$1   & 2.8 $\times$ 2.6 &  7.0 &    2.0  --  5.0  &  9.0  -- 12.0 \\
Per-emb~3   &                   & NGC 1333 & 03 29 00.554    & $+$31 11 59.849    & 33.8    & $+$1   & 3.4 $\times$ 2.0 &  7.3 & $-$2.0  --  2.0  & 10.0  -- 14.0 \\
Per-emb~5   & IRAS~03282$+$3035 &          & 03 31 20.931    & $+$30 45 30.334    & 35.9    & $+$1   & 2.8 $\times$ 2.6 &  7.3 & $-$1.0  --  4.0  & 10.0  -- 15.0 \\
Per-emb~6   &                   & B1       & 03 33 14.404    & $+$31 07 10.715    & 75.8    & $+$1   & 1.9 $\times$ 1.8 &  6.2 &    2.0  --  4.5  &  7.5  -- 10.0 \\
Per-emb~7   &                   &          & 03 30 32.681    & $+$30 26 26.480    & 35.3    & $+$1   & 2.9 $\times$ 2.2 &  6.5 &    2.0  --  5.0  &  7.0  -- 10.0 \\
Per-emb~9   & IRAS~03267$+$3128 & NGC 1333 & 03 29 51.876    & $+$31 39 05.516    & 37.9    & $-$1   & 1.7 $\times$ 1.3 &  8.2 &    1.5  --  5.5  & 11.5  -- 15.5 \\
Per-emb~10  &                   & B1       & 03 33 16.412    & $+$31 06 52.384    & 34.6    & $-$1   & 1.3 $\times$ 1.0 &  6.4 &    2.0  --  5.5  &  8.5  -- 12.0 \\
Per-emb~11  & IC348MMS          & IC348    & 03 43 57.055    & $+$32 03 04.669    & 34.4    & $+$1   & 2.0 $\times$ 1.9 &  9.0 &    2.5  --  5.0  & 12.0  -- 14.5 \\
Per-emb~13  & NGC~1333~IRAS4B   & NGC 1333 & 03 29 11.990    & $+$31 13 08.140    & 25.0    & $-$1   & 1.5 $\times$ 1.3 &  7.1 & $-$1.0  --  2.0  & 12.0  -- 15.0 \\
Per-emb~15  & RNO15-FIR         & NGC 1333 & 03 29 04.190    & $+$31 14 48.430    & 51.5    & $+$1   & 4.7 $\times$ 3.3 &  6.8 &    4.0  --  5.0  & 11.0  -- 12.0 \\
Per-emb~16  &                   & IC348    & 03 43 50.999    & $+$32 03 23.858    & 37.4    & $-$1   & 1.1 $\times$ 0.9 &  8.8 &    5.5  --  7.0  & 10.0  -- 11.5 \\
Per-emb~17  & IRAS 03245+3002   & L1455    & 03 27 39.120    & $+$30 13 02.526    & 54.8    & $+$1   & 2.2 $\times$ 1.8 &  6.0 &    0.0  --  3.0  &  9.0  -- 12.0 \\
Per-emb~19  &                   & NGC 1333 & 03 29 23.498    & $+$31 33 29.173    & 59.5    & $+$1   & 2.9 $\times$ 2.7 &  7.8 &    4.5  --  7.0  & 10.0  -- 12.5 \\
Per-emb~20  & L1455-IRS4        & L1455    & 03 27 43.199    & $+$30 12 28.962    & 62.1    & $+$1   & 2.4 $\times$ 1.9 &  5.3 &    0.0  --  3.0  &  7.0  -- 10.0 \\
Per-emb~21  &                   & NGC 1333 & 03 29 10.690    & $+$31 18 20.150    & 56.5    & $+$1   & 1.9 $\times$ 1.5 &  9.0 & $-$2.0  --  5.0  & 11.0  -- 18.0 \\
Per-emb~22  & L1448-IRS2        &          & 03 25 22.353    & $+$30 45 13.213    & 43.9    & $+$1   & 3.2 $\times$ 2.7 &  4.3 & $-$4.0  --  2.0  &  6.0  -- 12.0 \\
Per-emb~23  & ASR~30            & NGC 1333 & 03 29 17.250    & $+$31 27 46.340    & 41.7    & $-$1   & 1.4 $\times$ 1.1 &  7.8 &    3.5  --  6.5  & 10.5  -- 13.5 \\
Per-emb~24  &                   & NGC 1333 & 03 28 45.297    & $+$31 05 41.693    & 67.0    & $+$1   & 2.5 $\times$ 2.0 &  7.8 &    0.5  --  3.0  & 12.0  -- 14.5 \\
Per-emb~25  & IRAS 03235$+$3004 &          & 03 26 37.490    & $+$30 15 27.900    & 63.4    & $+$1   & 2.0 $\times$ 1.3 &  5.8 &    3.0  --  4.0  &  6.0  --  7.0 \\
Per-emb~26  & L1448C, L1448-mm  &          & 03 25 38.870    & $+$30 44 05.300    & 41.9    & $+$1   & 1.9 $\times$ 1.8 &  5.4 & $-$5.0  --  0.0  & 10.0  -- 15.0 \\
Per-emb~27  & NGC~1333~IRAS2A   & NGC 1333 & 03 28 55.560    & $+$31 14 37.170    & 50.2    & $+$1   & 2.1 $\times$ 2.1 &  8.1 &    1.5  --  4.5  & 10.5  -- 13.5 \\
Per-emb~28  &                   & IC348    & 03 43 50.987    & $+$32 03 07.967    & 59.2    & $+$1   & 2.1 $\times$ 2.0 &  8.6 &    5.5  --  7.5  &  9.5  -- 11.5 \\
Per-emb~29  & B1-c              & B1       & 03 33 17.860    & $+$31 09 32.307    & 47.4    & $+$1   & 3.4 $\times$ 2.8 &  6.4 &    0.5  --  4.5  &  8.5  -- 12.5 \\
Per-emb~30  & HH789           & B1       & 03 33 27.330    & $+$31 07 10.290    & 96.9    & $+$1   & 3.7 $\times$ 3.1 &  6.9 &    1.0  --  4.0  & 10.0  -- 13.0 \\
Per-emb~31  & [JJK2007] 9       & NGC 1333 & 03 28 32.547    & $+$31 11 05.151    & 76.3    & $+$1   & 2.7 $\times$ 2.3 &  8.0 &    2.5  --  6.5  & 10.5  -- 14.5 \\
Per-emb~34  & IRAS~03271$+$3013 &          & 03 30 15.190    & $+$30 23 49.110    & 97.6    & $+$1   & 3.6 $\times$ 2.9 &  5.7 &    1.0  --  4.0  &  8.0  -- 11.0 \\
Per-emb~36  & NGC~1333~IRAS2B   & NGC 1333 & 03 28 57.360    & $+$31 14 15.610    & 95.1    & $+$1   & 3.4 $\times$ 3.0 &  6.9 & $-$3.0  --  3.0  & 11.0  -- 17.0 \\
Per-emb~40  & B1-a, IRAS 03301$+$3057 & B1       & 03 33 16.650    & $+$31 07 54.810    & 128.9   & $+$1   & 3.3 $\times$ 2.8 &  7.4 &    1.5  --  4.0  &  9.0  -- 11.5 \\
Per-emb~41  & B1-b              & B1       & 03 33 20.341    & $+$31 07 21.355    & 212.6   & $+$1   & 3.4 $\times$ 2.8 &  6.5 &    1.0  --  4.0  &  8.0  -- 11.0 \\
Per-emb~42  & L1448C-S          &          & 03 25 39.135    & $+$30 43 57.909    & 48.8    & $+$1   & 1.9 $\times$ 1.8 &  5.8 &    2.0  --  3.0  &  7.0  --  8.0 \\
Per-emb~44  & SVS~13A           & NGC 1333 & 03 29 03.760    & $+$31 16 03.430    & 85.8    & $+$1   & 4.0 $\times$ 3.6 &  8.7 & $-$7.0  --  4.0  & 12.0  -- 23.0 \\
Per-emb~46  &                   & L1455    & 03 28 00.420    & $+$30 08 01.010    & 228.6   & $+$1   & 3.2 $\times$ 2.8 &  5.2 &    1.5  --  4.0  &  7.0  --  9.5 \\
Per-emb~50  &                   & NGC 1333 & 03 29 07.764    & $+$31 21 57.162    & 101.4   & $+$1   & 1.8 $\times$ 1.5 &  7.6 &    1.5  --  4.0  & 11.0  -- 13.5 \\
Per-emb~52  &                   & NGC 1333 & 03 28 39.699    & $+$31 17 31.882    & 335.9   & $+$1   & 1.9 $\times$ 1.5 &  8.3 &    6.0  --  7.5  &  8.5  -- 10.0 \\
Per-emb~53  & B5-IRS1           &          & 03 47 41.577    & $+$32 51 43.745    & 275.4   & $+$1   & 2.0 $\times$ 1.7 & 11.0 &    3.0  --  8.0  & 12.0  -- 17.0 \\
Per-emb~56  & IRAS~03439$+$3233 &          & 03 47 05.422    & $+$32 43 08.330    & 310.4   & $+$1   & 2.2 $\times$ 2.1 & 11.0 &    6.0  -- 10.0  & 12.0  -- 16.0 \\
Per-emb~57  &                   & NGC 1333 & 03 29 03.322    & $+$31 23 14.338    & 202.7   & $+$1   & 3.5 $\times$ 3.0 &      &    1.5  --  5.5  &  9.5  -- 13.5 \\
\hline 
\end{tabular}
\end{center}
\color{black}$^{\rm a}$The clustered region of Perseus in which this protostar is located, if any.  Protostars with listed regions can be found in the figures presented in Appendix \ref{sec_app_clusters}.\\
\color{black}$^{\rm b}$Protostellar positions traced by the MASSES 1.3~mm continuum detections \color{black}when available \citep{pokhrel2018:masses,stephens2018:masses}, or by the VLA Nascent Disk And Multiplicity Survey of Perseus Protostars \citep[VANDAM;][]{tobin2016:vandam,tobin2018:vandam}.\\
\color{black}$^{\rm c}$\tbol\ values tabulated by \citet{pokhrel2022:protostars}.\\
$^{\rm d}$As compiled by \citet{stephens2019:masses}.\\
\color{black}$^{\rm e}$As measured by \citet{stephens2018:masses,stephens2019:masses} using MASSES \cooojtwo\ observations; missing values indicate MASSES \cooojtwo\ nondetections. \color{black}
\end{table}
\end{landscape}

\begin{landscape}
\begin{table}
\begin{center}
\contcaption{MASSES Protostars with Measured Outflow Opening Angles}
\label{tab_sample:continued}
\begin{tabular}{lcccccccccc}
\hline \hline
            & Common            &              & Protostar R.A.$^{\rm b}$   & Protostar Decl.$^{\rm b}$     & \tbol$^{\rm c}$   &        & Beam Major \& Minor & v$_{\rm systemic}$ & v$_{\rm blue}$      & v$_{\rm red}$ \\
Source      & Name              & Region$^{\rm a}$ & (J2000)         & (J2000)            & (K)     & Robust & Axes (arcsec)$^{\rm d}$       & (\kms)$^{\rm e}$              & (\kms)              & (\kms)  \\
\hline
Per-emb~61  &                   & IC348    & 03 44 21.301    & $+$31 59 32.526    & 271.3   & $+$1   & 2.0 $\times$ 1.7 &  9.3 &    4.0  --  7.5  & 10.5  -- 14.0 \\
Per-emb~62  &                   & IC348    & 03 44 12.973    & $+$32 01 35.289    & 320.6   & $+$1   & 2.9 $\times$ 2.5 &  8.6 &    1.0  --  6.5  & 11.5  -- 17.0 \\
B1-bS       &                   & B1       & 03 33 21.340    & $+$31 07 26.440    & 14.7    & $+$1   & 3.4 $\times$ 2.8 &      &    1.0  --  4.0  &  8.0  -- 11.0 \\
L1448N-NW   & L1448N-IRS3C      & L1448    & 03 25 35.673    & $+$30 45 34.357    & 31.7    & $-$1   & 1.4 $\times$ 1.2 &  5.3 &    0.0  --  3.0  &  7.0  -- 10.0 \\
L1448N-B    & L1448N-IRS3B      & L1448    & 03 25 35.330    & $+$30 45 14.810    & 50.3    & $-$1   & 1.4 $\times$ 1.2 &  5.3 &    0.5  --  3.0  &  7.0  --  9.5 \\
Per-Bolo~45 &                   & NGC 1333 & 03 29 06.770    & $+$31 17 29.960    & 15.0    & $+$1   & 3.1 $\times$ 2.2 &      &    7.0  --  7.5  &  9.5  -- 10.0 \\
Per-Bolo~58 &                   & NGC 1333 & 03 29 25.417    & $+$31 28 15.000    & 20.1    & $+$1   & 3.1 $\times$ 2.2 &  8.2 &    3.0  --  6.5  &  9.5  -- 13.0 \\
SVS~13C     &                   & NGC 1333 & 03 29 02.030    & $+$31 15 37.750    & 42.6    & $+$1   & 3.9 $\times$ 3.5 &  8.9 &    4.0  --  7.0  &  9.0  -- 12.0 \\
\hline 
\end{tabular}
\end{center}
\color{black}$^{\rm a}$The clustered region of Perseus in which this protostar is located, if any.  Protostars with listed regions can be found in the figures presented in Appendix \ref{sec_app_clusters}.\\
\color{black}$^{\rm b}$Protostellar positions traced by the MASSES 1.3~mm continuum detections \color{black}when available \citep{pokhrel2018:masses,stephens2018:masses}, or by the VLA Nascent Disk And Multiplicity Survey of Perseus Protostars \citep[VANDAM;][]{tobin2016:vandam,tobin2018:vandam}.\\
\color{black}$^{\rm c}$\tbol\ values tabulated by \citet{pokhrel2022:protostars}.\\
$^{\rm d}$As compiled by \citet{stephens2019:masses}.\\
\color{black}$^{\rm e}$As measured by \citet{stephens2018:masses,stephens2019:masses} using MASSES \cooojtwo\ observations; missing values indicate MASSES \cooojtwo\ nondetections. \color{black}
\end{table}
\end{landscape}

\begin{table*}
\color{black}
\begin{center}
\caption{MASSES Targets without Measured Outflow Opening Angles}
\label{tab_sample_nondetect}
\begin{tabular}{lccccc}
\hline \hline
            & Common            &                  & Phase Center R.A.$^{\rm b}$ & Phase Center Decl.$^{\rm b}$     & \tbol$^{\rm c}$ \\
Source      & Name              & Region$^{\rm a}$ & (J2000)                     & (J2000)                          & (K)             \\
\hline
Per-emb~4   &                   & NGC 1333         & 03 28 39.10                 & $+$31 06 01.80                   & 37.4 \\
Per-emb~8   &                   & IC348            & 03 44 43.62                 & $+$32 01 33.70                   & 47.9 \\
Per-emb~12  & NGC 1333 IRAS 4A  & NGC 1333         & 03 29 10.50                 & $+$31 13 31.00                   & 32.5 \\
Per-emb~14  & NGC 1333 IRAS 4C  & NGC 1333         & 03 29 13.52                 & $+$31 13 58.00                   & 31.6 \\
Per-emb~18  & NGC 1333 IRAS 7   & NGC 1333         & 03 29 10.99                 & $+$31 18 25.50                   & 43.8 \\
Per-emb~32  &                   & IC348            & 03 44 02.40                 & $+$32 02 04.90                   & 80.5 \\
Per-emb~35  & NGC 1333 IRAS 1, IRAS 03255$+$3103   & NGC 1333         & 03 28 37.09                 & $+$31 13 30.70                   & 59.8 \\
Per-emb~37  &                   & NGC 1333         & 03 29 18.27                 & $+$31 23 20.00                   & 32.1 \\
Per-emb~38  &                   &                  & 03 32 29.18                 & $+$31 02 40.90                   & 149.4 \\
Per-emb~39  &                   &                  & 03 33 13.78                 & $+$31 20 05.20                   & 60.9 \\
Per-emb~43  &                   &                  & 03 42 02.16                 & $+$31 48 02.10                   & 192.4 \\
Per-emb~45  &                   & B1               & 03 33 09.57                 & $+$31 05 31.20                   & 223.6 \\
Per-emb~47  & IRAS 03254$+$3050 & NGC 1333         & 03 28 34.50                 & $+$31 00 51.10                   & 234.3 \\
Per-emb~48  & L1455-FIR2        & L1455            & 03 27 38.23                 & $+$30 13 58.80                   & 245.6 \\
Per-emb~49  &                   & NGC 1333         & 03 29 12.94                 & $+$31 18 14.40                   & 325.0 \\
Per-emb~51  &                   & NGC 1333         & 03 28 34.53                 & $+$31 07 05.50                   & 127.0 \\
Per-emb~54  & NGC 1333 IRAS 6   & NGC 1333         & 03 29 01.57                 & $+$31 20 20.70                   & 20.9 \\
Per-emb~55  & IRAS 03415$+$3152 & IC348            & 03 44 43.62                 & $+$32 01 33.70                   & 209.1 \\
Per-emb~58  &                   & NGC 1333         & 03 28 58.44                 & $+$31 22 17.40                   & 303.2 \\
Per-emb~59  &                   &                  & 03 28 35.04                 & $+$30 20 09.90                   & 341.0 \\
Per-emb~60  &                   & NGC 1333         & 03 29 20.07                 & $+$31 24 07.50                   & 363.0 \\
Per-emb~63  &                   & NGC 1333         & 03 28 43.28                 & $+$31 17 33.00                   & 482.9 \\
Per-emb~64  &                   &                  & 03 33 12.85                 & $+$31 21 24.10                   & 440.3 \\
Per-emb~65  &                   & NGC 1333         & 03 28 56.31                 & $+$31 22 27.80                   & 536.0 \\
Per-emb~66  &                   & IC348            & 03 43 45.15                 & $+$32 03 58.60                   & 480.0 \\
B1-bN       &                   & B1               & 03 33 21.19                 & $+$31 07 40.60                   & 10.2  \\
L1448 IRS2E &                   & L1448            & 03 25 25.66                 & $+$30 44 56.70                   & 15.0 \\
L1451-mm    & L1451-MMS         &                  & 03 25 10.21                 & $+$30 23 55.30                   & 15.0 \\
SVS 13B     &                   & NGC 1333         & 03 29 03.42                 & $+$31 15 57.72                   & 20.0 \\
\hline 
\end{tabular}
\end{center}
$^{\rm a}$The clustered region of Perseus in which this target is located, if any.  Protostars with listed regions can be found in the figures presented in Appendix \ref{sec_app_clusters}.\\
$^{\rm b}$The phase center of the SMA MASSES observation of this target.\\
$^{\rm c}$\tbol\ values tabulated by \citet{pokhrel2022:protostars}.\\
\color{black}
\end{table*}

\section{Method}\label{sec_method}

\subsection{Measuring Outflow Opening and Position Angles}\label{sec_method_fitting}

Given the large number of outflows in our sample and the great diversity in observed outflow morphologies, we desire to measure outflow opening angles using a systematic, repeatable method rather than ``by eye" using a protractor.  We also seek a method that can capture with a single angle the relative width of different outflows even if they do not exhibit simple, conical morphologies.  To accomplish this, we adopt a method that was first developed by \citet{offner2011:outflows} using synthetic (simulated) observations of protostellar outflows.  \color{black}This method accurately reproduces the opening angle that would be determined ``by eye'' for a conical outflow, and returns the opening angle of the approximately conical base for curved outflow morphologies \cite[see][especially their Figure~2]{offner2011:outflows}.\color{black}

We begin by using our MASSES \cojtwo\ observations to construct integrated intensity (moment 0) maps of each outflow.  We manually select (by eye) the velocity intervals to use for each source, considering separately the redshifted and blueshifted velocities.  We cut out the low-velocity channels close to the systemic cloud velocity, as these channels typically feature significant artifacts from the ambient cloud emission that is partially or fully resolved out by our interferometric observations.  We include the lowest possible velocity channels that are not affected by ambient cloud emission since many outflows are widest at the lowest velocities.  The exact velocity intervals over which we integrate are listed in the last two columns of Table \ref{tab_sample}, and we note here that the upper bound in velocity has no discernible effect on our results since the integrated intensity is typically dominated by the lowest velocity channels, where the outflows are both the widest and the brightest.

For each pixel in these integrated intensity maps detected above 3$\sigma$ (where $\sigma$ is the rms noise in the integrated intensity maps, calculated using the IDL procedure {\sc sky}), we calculate the angle between that pixel and an initial guess of the outflow axis.  When available, this initial guess is taken from \citet{stephens2017:masses}; otherwise it is chosen from visual inspection of the data.  We manually mask all pixels with detections above 3$\sigma$ that are not part of the outflow, such as those arising from other, nearby outflows, or from sidelobe emission that is not fully removed in the cleaning process.  We also mask all pixels within 4$\arcsec$ of each protostar in order to avoid the effects of excessive beam smearing, \color{black}but otherwise do not apply radial constraints to the pixels included in the fit. \color{black}We then plot the resulting distribution of angles and fit a Gaussian to the distribution.  Following \citet{offner2011:outflows}, we define the opening angle of the outflow as the full width at quarter-maximum of this best-fit Gaussian.  The angle at which the Gaussian peaks is taken as a refinement to the initial guess for the outflow position angle.  We refer to \citet{offner2011:outflows} for a full discussion of the validity of this fitting method for measuring outflow opening angles.

\begin{figure*}
    \resizebox{5.5in}{!}{\includegraphics{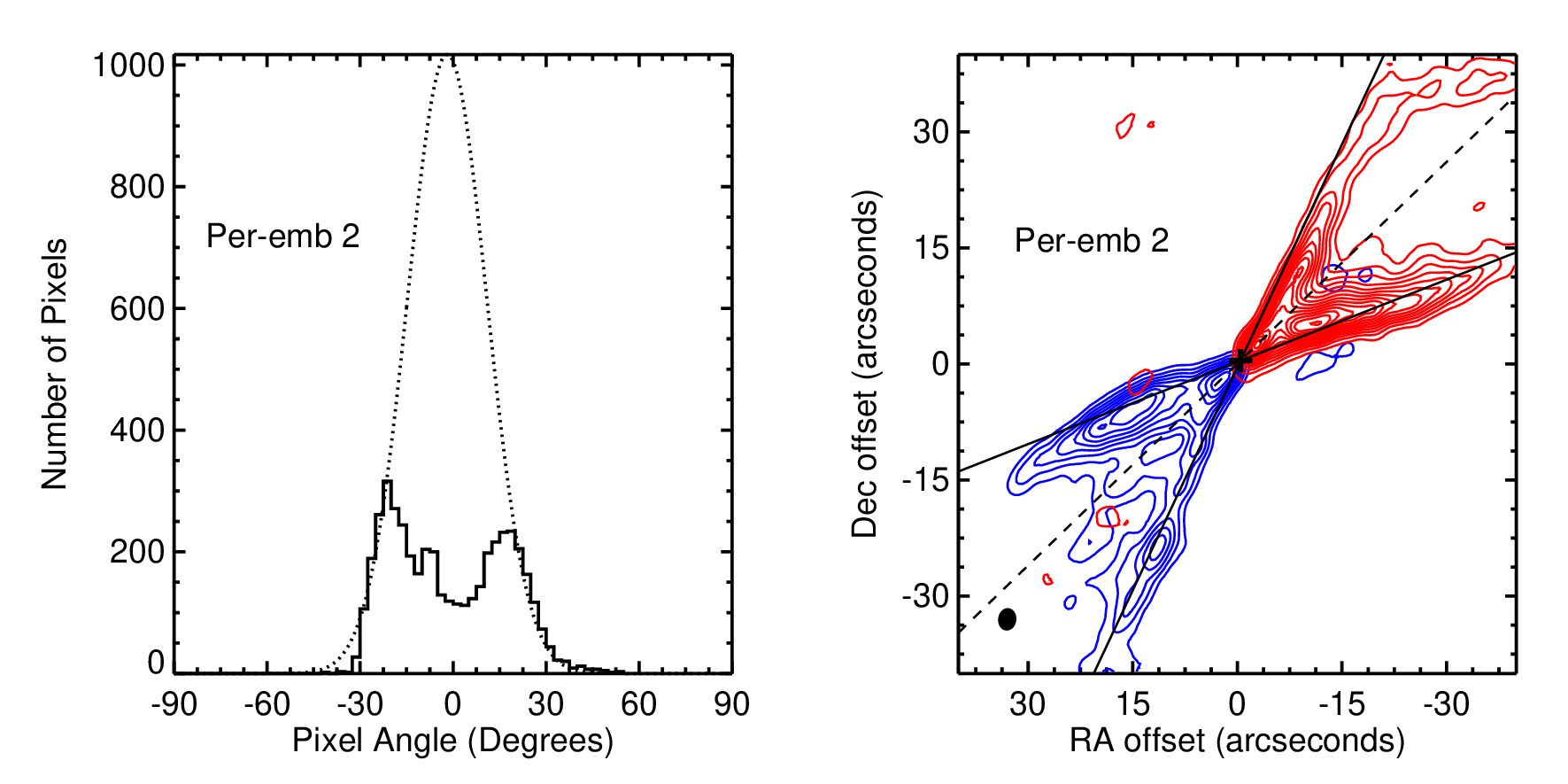}}
    \caption{{\it Left: }The solid histogram shows the distribution of angles to all pixels in the outflow driven by Per-emb~2, as described in the text.  The dotted line shows the best-fit Gaussian distribution, where the fit was restricted to the wings in order to exclude the dip at low angles arising from the resolved, partially empty outflow cavity.  {\it Right: } Integrated blueshifted and redshifted \cojtwo\ emission showing the Per-emb~2 outflow, integrated over the velocity intervals reported in Table \ref{tab_results}. The synthesized beam is shown as the filled ellipse in the lower left, and the cross marks the position of the protostar as traced by the 1.3~mm MASSES continuum detection \citep{pokhrel2018:masses,stephens2018:masses}. The dashed line shows the outflow position angle determined based on the peak angle of the Gaussian fit, and the solid lines show the opening angle determined from the full width at quarter-maximum of the Gaussian fit (see text for details).}
    \label{fig_per2}
\end{figure*}

One complication we must account for is that, with the relatively high spatial resolution of our interferometric observations, the outflows in our sample frequently feature resolved cavity walls and partially or fully empty cavities at small angles close to the outflow axis.  One such example is shown in Figure \ref{fig_per2} for Per-emb~2.  The partially empty cavities result in a double-peaked distribution of angles to detected pixels, with a dip where the Gaussian distribution should peak due to the lack of detections close to the outflow axis.  For these cases, which make up the majority of our sample, we restrict the Gaussian fit to the wings, omitting from the fit the small angles close to the outflow axis where the distribution turns over and decreases.  Exactly what is meant by ``the wings,'' and the implications for the resulting uncertainties in measured opening angles, are discussed in \S \ref{sec_method_uncertainties} below.

{\color{black}
\begin{table}
\begin{center}
\caption{Number of Outflows in Each Group}
\label{tab_confidence}
\begin{tabular}{lc}
\hline \hline
Group & Number \\
\hline
High Confidence (HC)   & 22 \\
Medium Confidence (MC) & 15 \\
Low Confidence (LC)    & 5 \\
Upper Limit            & 4 \\
\hline
\end{tabular}
\end{center}
\end{table}
}

We are able to successfully measure opening angles for 46 outflows 
in the MASSES sample.  We classify these outflows into four different groups based on our confidence in our results.  For outflows like the Per-emb~2 outflow (Figure \ref{fig_per2}), there is no ambiguity in identifying the presence of an outflow, no ambiguity in identifying the pixels contained within the outflow, and the Gaussian fit provides an excellent by-eye match to both the distribution of pixel angles and the morphology of the outflow in our integrated blueshifted and redshifted \cojtwo\ maps.  We classify these objects into the ``High Confidence'' (HC) group.  For other outflows, there may be an asymmetry between the two outflow lobes\footnote{Whenever such asymmetries exist we restrict the fit to the lobe that appears visually wider, in an effort to capture the full opening angles of each outflow \color{black}under the assumption that the asymmetries are due to partial non-detections of intrinsically symmetrical outflows.  If the asymmetries are instead intrinsic the effective opening angle of the two lobes combined would be smaller in these cases than reported here\color{black}.}, or ambiguity in whether or not emission in some part of the map is part of the outflow, that require us to make subjective choices about what emission to include in the fit.  Even though the quality of the fits are generally similar, the subjective choices add extra systematic uncertainties that are difficult to quantify, leading us to classify these outflows into the ``Medium Confidence'' (MC) group. As a third category, some protostars drive outflows that, while clearly present, feature substantial ambiguity in their morphologies. We classify these outflows into the ``Low Confidence'' group and emphasize that our measurements of them may be highly uncertain, with uncertainties in many cases that are difficult to quantify due to the subjective decisions in interpreting the complex \cojtwo\ maps. Finally, some outflows are partially or fully unresolved in one or both of their dimensions, and are thus classified as upper limits.  Table \ref{tab_confidence} reports the numbers of outflows in each of these four categories.

\subsection{Discussion of Uncertainties}\label{sec_method_uncertainties}

\begin{figure*}
    \resizebox{5.5in}{!}{\includegraphics{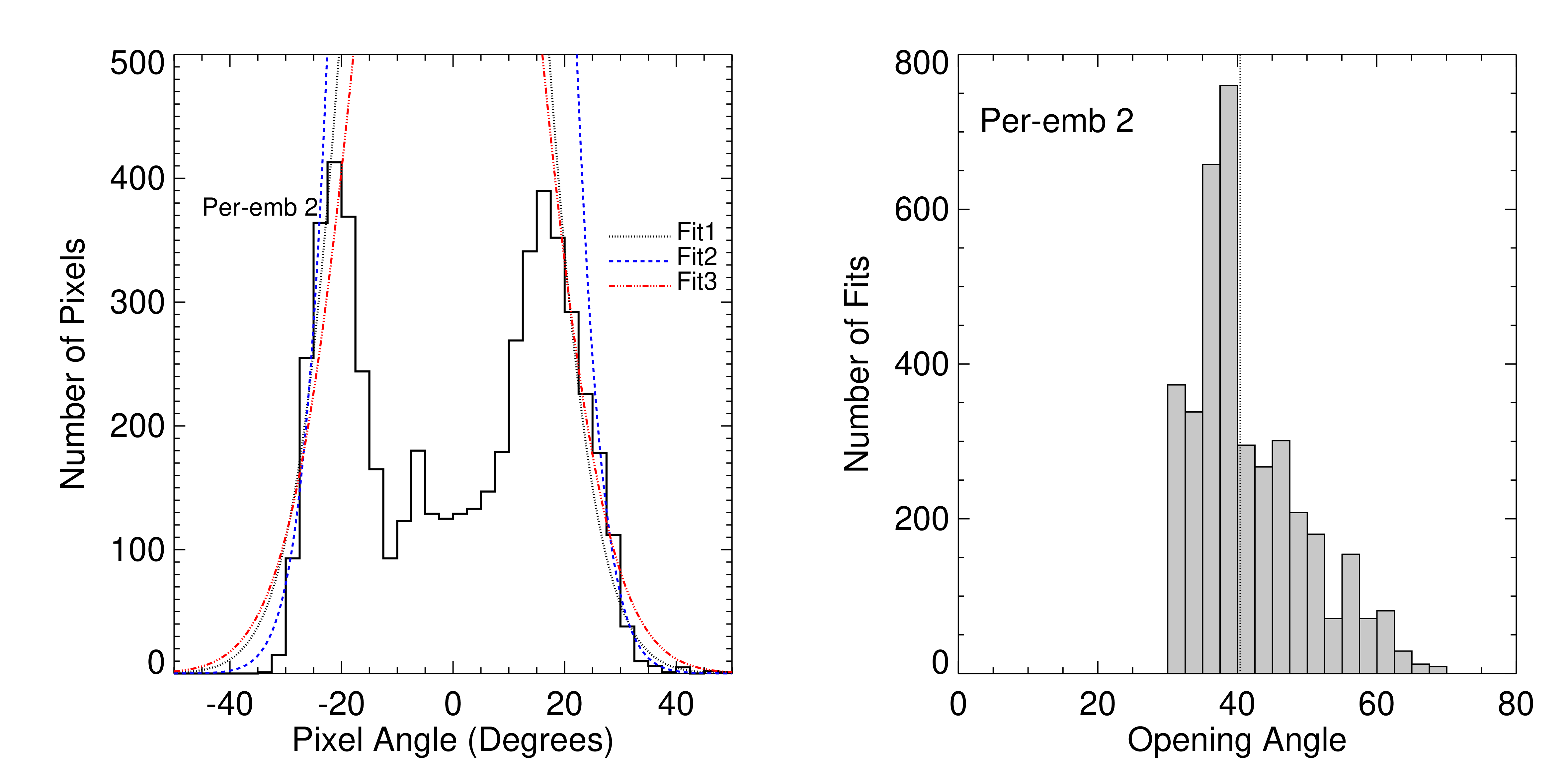}}
    \caption{{\it Left: }The solid histogram shows the distribution of angles to all pixels in the outflow driven by Per-emb~2, as described in the text (this is the same distribution as that shown in Figure \ref{fig_per2}). The three different dashed lines show three different best-fit Gaussian distributions, with each of the three fits performed over slightly different intervals of pixel angles. The intervals used for each fit are listed in Table \ref{tab_uncertainties}.  {\it Right: }The distribution of all possible opening angles for the Per-emb~2 outflow, determined as described in the text.  The dotted vertical line shows the opening angle determined from our fit to the nominal, manually selected interval over which to perform the Gaussian fit.}
    \label{fig_uncertainties}
\end{figure*}

As described above, we restrict the majority of our Gaussian fits to the distributions of outflow pixel angles to the wings of the distributions.  These ``wings'' are selected manually by specifying two different intervals over which to simultaneously fit (one interval at negative pixel angles and the other at positive pixel angles).  The intervals are chosen such that they each start where the distributions first rise above zero, and they end just before the distributions turn over at angles close to the outflow axis.  However, small adjustments to the bounds of these intervals do result in small changes to the best-fit opening angles.  We demonstrate this below in the left panel of Figure \ref{fig_uncertainties}, which shows three different best-fit Gaussians for Per-emb~2 that result from three slightly different fitting intervals (the intervals themselves are specified in Table \ref{tab_uncertainties}).

\begin{table}
\begin{center}
\caption{Intervals for the Fits Shown in Figure \ref{fig_uncertainties}}
\label{tab_uncertainties}
\begin{tabular}{lcccc}
\hline \hline
 & Negative & Negative & Positive & Positive \\
 & Lower & Upper & Lower & Upper \\
 & Bound & Bound & Bound & Bound \\
Fit & (Degrees) & (Degrees) & (Degrees) & (Degrees) \\
\hline
Fit1 & $-$38.75 & $-$23.75 & 18.75 & 38.75 \\
Fit2 & $-$48.75 & $-$23.75 & 25.50 & 50.00 \\
Fit3 & $-$48.75 & $-$17.00 & 25.50 & 50.00 \\
\hline
\end{tabular}
\end{center}
\end{table}

All three of the fits shown in the left panel of Figure \ref{fig_uncertainties} adequately fit the data, but Fit2 gives an opening angle 7\degree\ smaller than Fit1, whereas Fit3 gives an opening angle 5\degree\ larger than Fit1.  Given that the formal fit uncertainties are typically less than 1\degree, the choice of fitting interval is the dominant source of uncertainty in our results.  To properly quantify this, we vary each of the four interval bounds, in steps of the bin size, by up to $\pm$20\degree\ from our nominal, manually chosen bins.  For each possible combination of the four bounds (except those where the negative or positive lower bounds exceed the corresponding upper bounds, which we remove from consideration), we perform the Gaussian fit, manually
rejecting those fits that are obviously bad but keeping all others.  The right panel of Figure \ref{fig_uncertainties} shows the resulting distribution of best-fit opening angles for the outflow driven by Per-emb~2, with the result obtained from the nominal, manually chosen interval indicated with a dotted line.  The standard deviation of this distribution is 8\degree.

We repeat this process for each outflow in our sample.  We then add in quadrature the standard deviation of the distribution of possible outflow opening angles, determined as described above, with the formal fit uncertainty for the fit performed over the nominal, manually chosen interval. We treat this resulting value as the uncertainty for each outflow opening angle.  We do not apply this method for the LC outflows, and thus do not determine or list uncertainties for the LC outflow opening angle measurements, since the true uncertainties for these outflows are dominated by the subjective choices made in identifying the pixels containing outflow emission and are thus not easily quantifiable.

While the full results are tabulated below, we note here that the uncertainties determined via this method range from 5--31\degree, but most typically range between 5--15\degree.

\subsection{Validation of Method}\label{sec_method_validation}

While our method for measuring opening angles was developed and validated by \citet{offner2011:outflows}, we further validate this method in two ways: (1) by comparing to outflow opening angles measured by hand from the same data, and (2) by applying our method to the outflow maps observed by \citet{arce2006:outflows}.

\subsubsection{Comparison to Measurements by Hand}\label{sec_method_validation_byhand}

The results from applying our method to the MASSES data were compared to measurements of the same outflow opening angles made by the more traditional (and also more subjective) “by eye with a protractor” method. A student with no prior experience with protostellar outflows attempted to independently identify outflows by manually inspecting the \cojtwo\ spectral cubes, collapsing the cubes into integrated intensity (moment 0) maps over the relevant velocity channels, and then measuring the opening angles of the outflows manually with a protractor.  We note that the use of a student with no prior experience was intentional in order to apply a ``sanity check'' on our automated method.  Comparing to the results from this automated method, which will be presented and discussed in the following sections of this paper, we find that the mean difference between the automated and ``by-eye'' methods is 13.3\degree~$\pm$~5.6\degree. 
Given that the mean difference is within 3$\sigma$ of 0\degree, we find that our automated method does generally agree with manual ``by-eye'' measurements.

Digging more deeply into the results, we note that four outflows (Per-emb~27, Per-emb~30, Per-emb~44, and Per-emb~62) featured ``by-eye'' measurements that disagree with the results from our automated method by greater than 20\degree.  All four cases can be explained as errors resulting from the use of a student who did not possess background knowledge on these objects, including confusion due to multiple outflows in a binary system (Per-emb~27), unfamiliarity with the results from previously published, lower-resolution maps that recover more extended emission (Per-emb~44), and improper selection of the velocity intervals for the integrated intensity (moment 0) maps (Per-emb~30 and Per-emb~62).  When excluding these four outflows, 
the mean difference between these ``by-eye'' measurements and the results of our automated method is 3.5\degree~$\pm$~2.1\degree. Thus we find excellent agreement between the two methods, indicating that our method does pass this basic ``sanity check.''

\subsubsection{Comparison to \citet{arce2006:outflows}}\label{sec_method_validation_as06}

\citet{arce2006:outflows} obtained interferometric observations of nine protostellar outflows and measured their opening angles manually using a protractor.  While we have already demonstrated that our method provides visually good fits to the opening angles of our outflows, we further quantify this by comparing the results of applying our method to the \citet{arce2006:outflows} data with their manual opening angle measurements.

Out of their nine observed outflows, \citet{arce2006:outflows} were able to measure opening angles for seven of them.  Five out of these seven would be classified into the HC group by our method.  The remaining two (RNO~43 and T~Tau) would be classified into either the MC or LC groups based on the presence of asymmetrical outflow lobes and substantial subjective uncertainties in what emission is and is not interpreted to be part of the outflow.  Since the MC and LC groups include resulting systematic uncertainties that are difficult to fully quantify, we restrict our comparison to the five outflows that we would have placed in the HC group.


For all five of the HC outflows our method results in opening angles that agree with \citet{arce2006:outflows} to within 15\degree, with four out of the five agreeing to within 10\degree.  The mean difference between our measurements and those reported by \citet{arce2006:outflows} is 4.2\degree~$\pm$~3.8\degree. Our measurements thus show strong statistical agreement with those of \citet{arce2006:outflows}, with a mean difference that is within 2$\sigma$ of 0\degree. Furthermore, the dispersion in differences is comparable to many of our opening angle uncertainties that were determined as discussed in \S \ref{sec_method_uncertainties}.

Finally, we note that there is one outflow in both our sample and the \citet{arce2006:outflows} sample: Per-emb~5 (IRAS~03282+3035).  Using our MASSES data and our method, we measured an opening angle of 50\degree.  Using the \citet{arce2006:outflows} data and our method, we measured an opening angle 53\degree.  Using their data and their manual method, \citet{arce2006:outflows} measured an opening angle of 50\degree. Given the typical uncertainties of 5--10\degree\ reported in this paper, combined with the above analysis, we thus conclude that our method is in excellent agreement with methods based on manual, ``by-eye'' measurements using a protractor.

\section{Results}\label{sec_results}

\begin{table}
\begin{center}
\caption{Position and Opening Angle Fitting Results}
\label{tab_results}
\begin{tabular}{lcccc}
\hline \hline
            & \tbol$^{\rm a}$ & Position            & Full Opening                      & Confidence    \\
Source      & (K)             & Angle ($^{\rm{o}}$) & Angle ($^{\rm o}$)$^{\rm b}$ & Group$^{\rm c}$ \\
\hline
Per-emb~1   &  24.1           & 115                 &  $31 \pm 5$                & HC \\
Per-emb~2   &  30.9           & 131                 &  $40 \pm 8$                & HC \\
Per-emb~3   &  33.8           &  96                 &  $49 \pm 11$               & HC \\
Per-emb~5   &  35.9           & 123                 &  $50 \pm 8$                & HC \\
Per-emb~6   &  75.8           &  58                 &  $60 \pm 5$                & MC \\
Per-emb~7   &  35.3           & 171$^{\rm d}$       &  $<25$$^{\rm d}$                     & UL \\
Per-emb~9   &  37.9           &  56                 &  $31 \pm 5$                & MC \\
Per-emb~10  &  34.6           &  49                 &  $41 \pm 11$               & HC \\
Per-emb~11  &  34.4           & 162                 &  $29 \pm 5$                & HC \\
Per-emb~13  &  25.0           & 172$^{\rm d}$       &  $<34$$^{\rm d}$                     & UL \\
Per-emb~15  &  51.5           & 150                 &  $89 \pm 5$                & HC \\
Per-emb~16  &  37.4           & 177                 &  $91 \pm ...$              & LC \\
Per-emb~17  &  54.8           &  64                 &  $69 \pm 8$                & HC \\
Per-emb~19  &  59.5           & 146                 &  $73 \pm 13$               & MC \\
Per-emb~20  &  62.1           & 126                 &  $57 \pm 14$               & HC \\
Per-emb~21  &  56.5           &  57$^{\rm d}$       &  $46 \pm ... $$^{\rm d}$             & LC \\
Per-emb~22  &  43.9           & 131                 &  $56 \pm 7$                & HC \\
Per-emb~23  &  41.7           &  61                 &  $39 \pm 12$               & HC \\
Per-emb~24  &  67.0           &  92                 &  $62 \pm 5$                & HC \\
Per-emb~25  &  63.4           & 107                 &  $51 \pm 11$               & HC \\
Per-emb~26  &  41.9           & 160$^{\rm d}$       &  $40 \pm 6$$^{\rm d}$                & HC \\
Per-emb~27  &  50.2           &  14$^{\rm d}$       &  $65 \pm 11$$^{\rm d}$               & MC \\
Per-emb~28  &  59.2           & 121$^{\rm d}$       &  $16 \pm 5$$^{\rm d}$                & MC \\
Per-emb~29  &  47.4           & 133                 &  $37 \pm 7$                & MC \\
Per-emb~30  &  96.9           & 115                 &  $107 \pm 9$              & HC \\
Per-emb~31  &  76.3           & 149$^{\rm d}$       &  $45 \pm 13$$^{\rm d}$               & MC \\
Per-emb~34  &  97.6           &  54                 &  $84 \pm 9$                & MC \\     
Per-emb~36  &  95.1           &  13$^{\rm d}$       & $162 \pm 31$$^{\rm d}$              & MC \\
Per-emb~40  & 128.9           & 109$^{\rm d}$       &  $43 \pm 8$$^{\rm d}$               & MC \\
Per-emb~41  & 212.6           &  25                 &  $<35$                     & UL \\
Per-emb~42  &  48.8           &  50$^{\rm d}$       &  $49 \pm 14$$^{\rm d}$               & MC \\
Per-emb~44  &  85.8           & 140$^{\rm d}$       & $144 \pm 20$$^{\rm d}$               & MC \\
Per-emb~46  & 228.6           & 133                 &  $79 \pm 11$               & HC \\
Per-emb~50  & 101.4           &  95                 &  $55 \pm 12$               & HC \\
Per-emb~52  & 335.9           &  12                 &  $51 \pm 5$                & MC \\
Per-emb~53  & 275.4           &  60                 &  $90 \pm 11$               & HC \\
Per-emb~56  & 310.4           & 145                 &  $63 \pm 11$               & HC \\
Per-emb~57  & 202.7           & 148$^{\rm d}$       &  $25 \pm ...$$^{\rm d}$              & LC \\
Per-emb~61  & 271.3           &  10                 &  $60 \pm 5$                & MC \\
Per-emb~62  & 320.6           &  18                 &  $73 \pm 6$                & HC \\
B1-bS       &  14.7           & 110$^{\rm d}$       &  $<63$$^{\rm d}$                     & UL \\
L1448N-NW   &  31.7           & 130$^{\rm d}$       &  $42 \pm 8$$^{\rm d}$                & HC \\
L1448N-B    &  50.3           & 129$^{\rm d}$       &  $54 \pm 14$$^{\rm d}$               & MC \\
Per-Bolo~45 &  15.0           & 142$^{\rm d}$       &   $9 \pm ...$$^{\rm d}$              & LC \\
Per-Bolo~58 &  20.1           &  87                 &  $23 \pm 5$                & HC \\
SVS~13C     &   42.6          &   7$^{\rm d}$       &  $67 \pm ...$$^{\rm d}$              & LC \\
\hline 
\end{tabular}
\end{center}
$^{\rm a}$Repeated from Table \ref{tab_sample} for ease of access.\\
$^{\rm b}$No uncertainties are presented for the measured opening angles in the low confidence group.  See \S \ref{sec_method_uncertainties} for details.\\
$^{\rm c}${\sc HC}: high confidence. {\sc MC}: medium confidence. {\sc LC}: low confidence. {\sc UL}: upper limit.\\
$^{\rm d}$Based on one outflow lobe (see Appendix \ref{sec_app_fit} for details).\\
\end{table}


Table \ref{tab_results} reports the positions and opening angles, along with the uncertainties in the opening angles and the confidence group assigned to each outflow, for the 46 outflows in the MASSES sample for which we are able to obtain valid measurements.  The \tbol\ values for each object are also repeated from Table \ref{tab_sample} for convenience.  Notes on each outflow for which we measure an opening angle, including justification for the confidence group assigned for that outflow, are presented in Appendix \ref{sec_app_fit}. Additional notes on MASSES pointings for which we are unable to measure outflow opening angles are given in Appendix \ref{sec_app_nofit}.  \color{black}We emphasize here that all opening angles reported in this paper are not corrected for inclination and are thus projected onto the plane of the sky.  The effects of inclination corrections are discussed in \S \ref{sec_inclination}.\color{black}

\begin{figure*}
    \resizebox{7.3in}{!}{\includegraphics{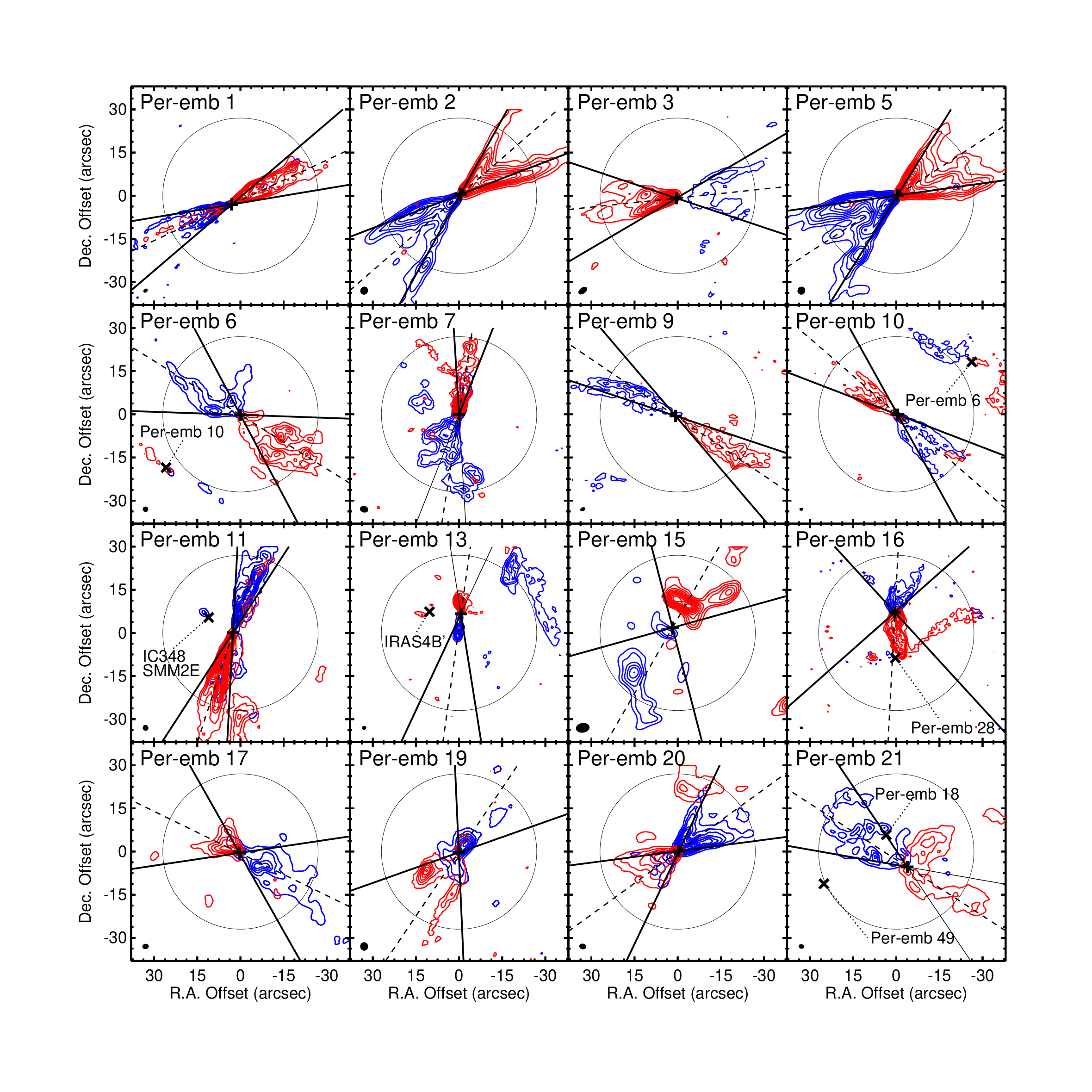}}
    \vspace{-0.6in}
    \caption{Blueshifted (blue contours) and redshifted (red contours) integrated intensity (moment 0) \cojtwo\ maps for Per-emb~1, 2, 3, 5, 6, 7, 9, 10, 11, 13, 15, 16, 17, 19, 20, and 21.  The velocity ranges over which the emission is integrated are given in Table \ref{tab_sample}. Driving source positions are marked with a cross, and the locations of other sources of interest in each field are marked with an x. The axes are plotted using coordinate offsets from the phase centers of the observations.  The large circles show the full-width half-maximum (FWHM) size of the SMA primary beam at 230~GHz, centered on the phase centers.  The cones plotted with thick solid lines show the opening angles of the outflows as measured in this work, with the position angles plotted using dashed lines.  \color{black}If an outflow lobe is not used in the fit, the corresponding cone is plotted with a thin, rather than a thick, solid line.  \color{black}The synthesized beams are shown in the bottom left corner of each panel.  All emission contours and overplotted lines indicating outflow opening and position angles are masked at declination offsets greater than 30$\arcsec$ to improve the clarity of the labeling of each panel.  Contours start at 3$\sigma$ and increase in steps of 4$\sigma$, where $\sigma$ represents the 1$\sigma$ rms in the blueshifted (for blue contours) or redshifted (for red contours) integrated intensity maps.  The only exceptions are for Per-emb~7 and Per-emb~15, where the contours start at 2.5$\sigma$ and increase in steps of 1$\sigma$.}
    \label{fig_result1}
\end{figure*}

\begin{figure*}
    \resizebox{7.3in}{!}{\includegraphics{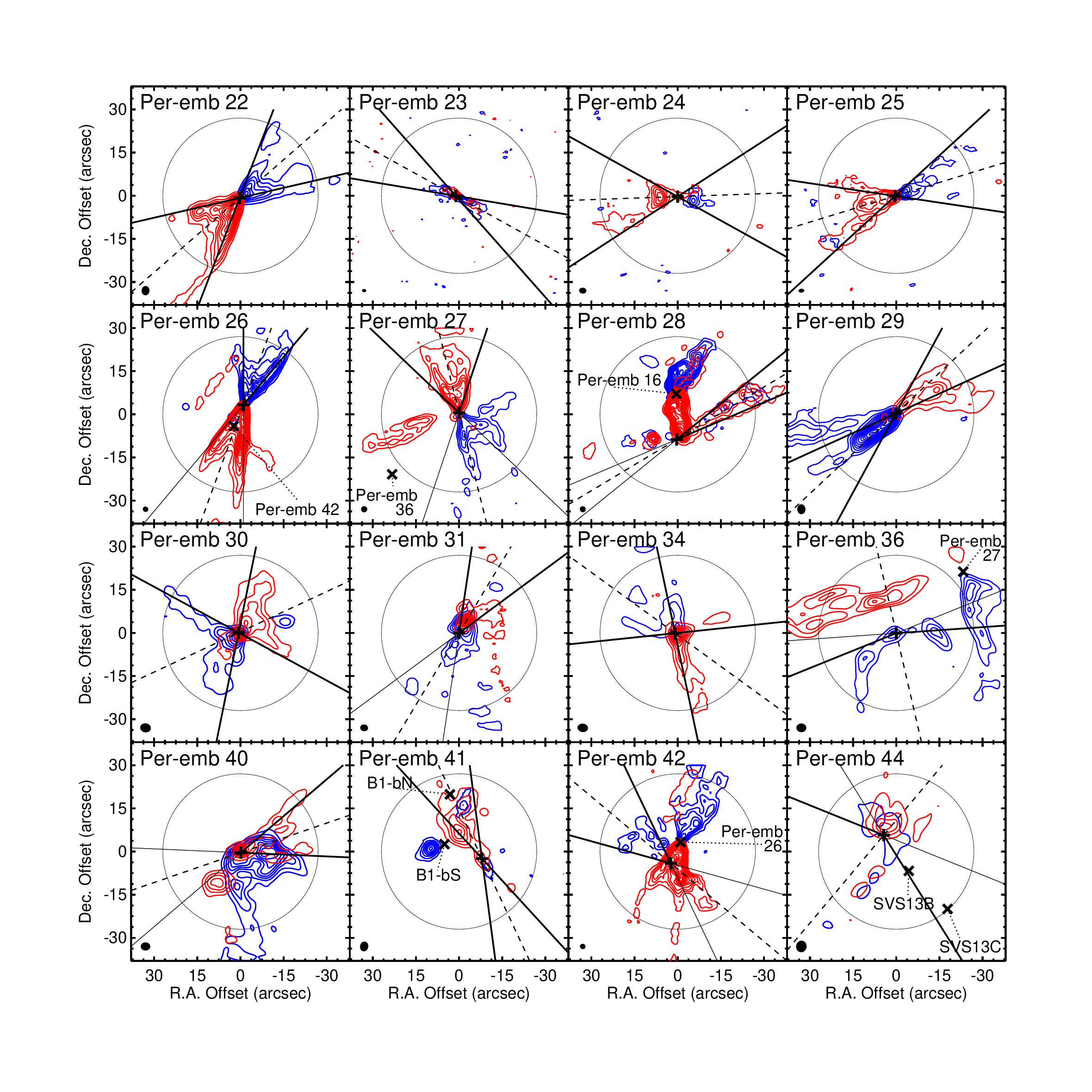}}
    \vspace{-0.6in}
    \caption{Same as Figure \ref{fig_result1}, except for Per-emb~22, 23, 24, 25, 26, 27, 28, 29, 30, 31, 34, 36, 40, 41, 42, and 44.  Contours start at 3$\sigma$ and increase in steps of 4$\sigma$, where $\sigma$ represents the 1$\sigma$ rms in the blueshifted (for blue contours) or redshifted (for red contours) integrated intensity maps.  The only exception is for Per-emb~36, where the contours start at 2.5$\sigma$ and increase in steps of 2.5$\sigma$.}
    \label{fig_result2}
\end{figure*}

\begin{figure*}
    \resizebox{7.3in}{!}{\includegraphics{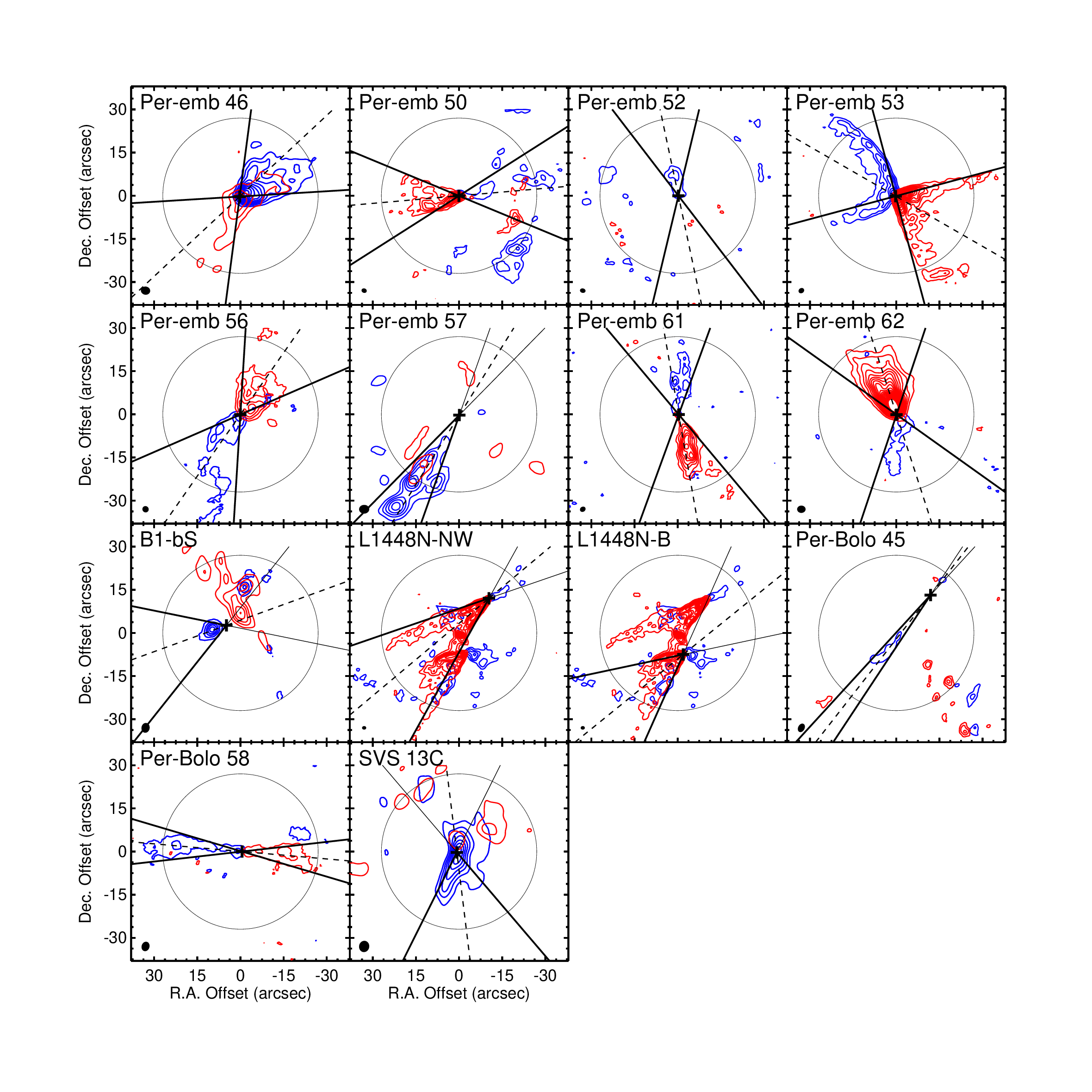}}
    \vspace{-0.6in}
    \caption{Same as Figure \ref{fig_result1}, except for Per-emb~46, 50, 52, 53, 56, 57, 61, 62, B1-bS, L1448-NW, L1448-B, Per-Bolo~45, Per-Bolo~58, and SVS~13C.  Contours start at 3$\sigma$ and increase in steps of 4$\sigma$, where $\sigma$ represents the 1$\sigma$ rms in the blueshifted (for blue contours) or redshifted (for red contours) integrated intensity maps.}
    \label{fig_result3}
\end{figure*}

Figures \ref{fig_result1} -- \ref{fig_result3} plot blueshifted and redshifted integrated intensity (moment 0) contours for each of the 46 outflows, along with lines showing the position and opening angles measured for each outflow. Visual inspection of these figures reveals two key points about our method: (1) Our method agrees very well with the cones that would be drawn by hand for conical outflows (examples include Per-emb~2, Per-emb~5, Per-emb~20, among others), in agreement with the validation results presented in \S \ref{sec_method_validation}, and (2) our method accurately measures the opening angles of the conical (or approximately conical) bases of outflows displaying more parabolic morphologies (examples include Per-emb~3, Per-emb~53, and Per-emb~62, among others).

\begin{figure*}
    \resizebox{6.0in}{!}{\includegraphics{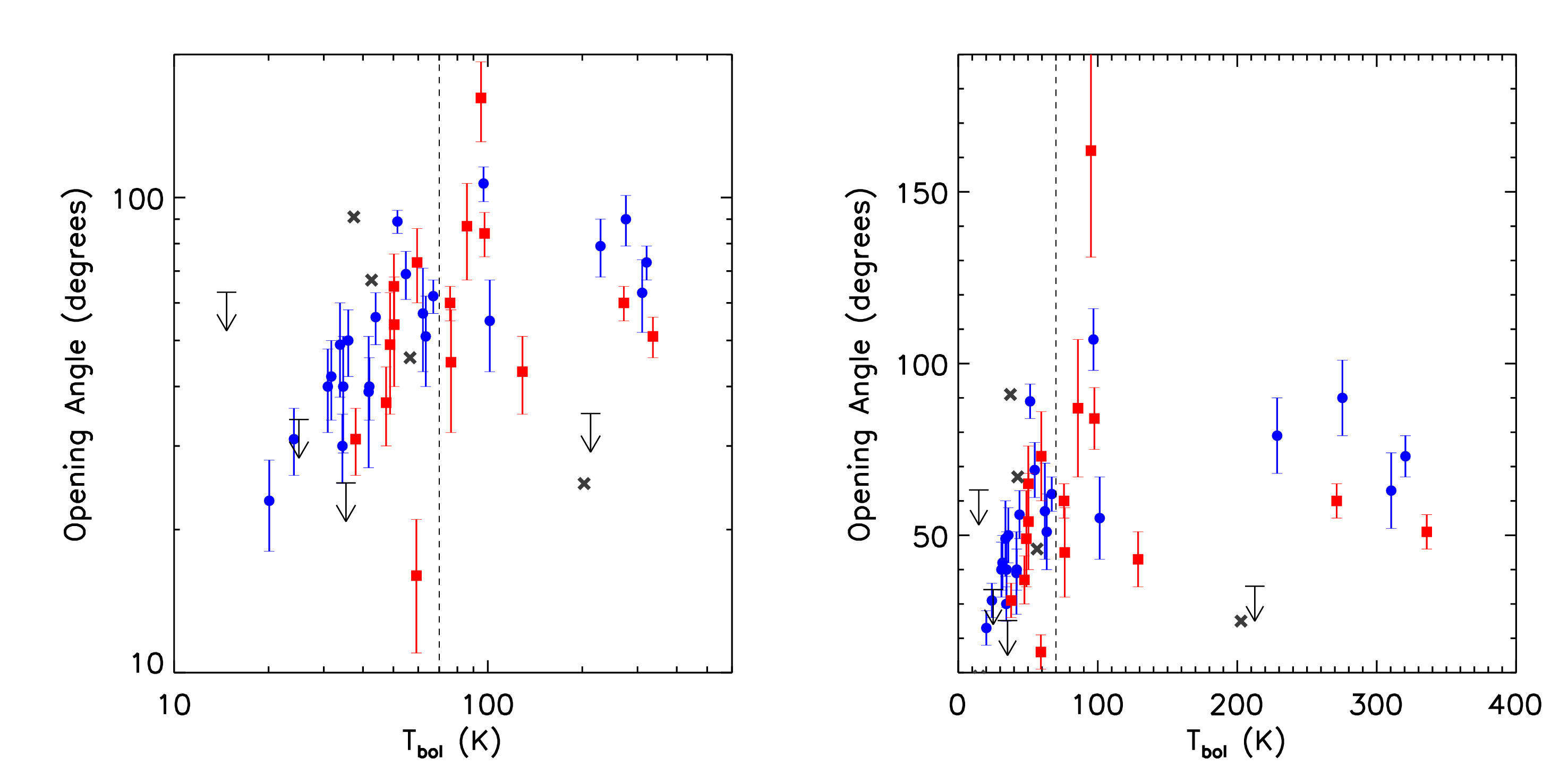}}
    \vspace{0.0in}
    \caption{Measured outflow opening angles plotted versus the bolometric temperatures of the driving sources, on both logarithmic (left) and linear (right) axes.  In both panels the blue circles show high confidence measurements, the red squares show medium confidence measurements, the black x's show low confidence measurements, and the arrows show upper limits.  The dashed vertical line in each panel marks \tbol~$=70$~K, the Class 0/I boundary in bolometric temperature \citep{chen1995:tbol}, with Class 0 protostars to the left (\tbol~$<70$~K) and Class I protostars to the right (\tbol~$\geq 70$~K).}
\label{fig_angletbol}
\end{figure*}

Figure \ref{fig_angletbol} plots, for each of the 46 outflows in our sample, the measured opening angle versus the bolometric temperature (\tbol) of the driving source on both logaritmic and linear axes.  Due to the large uncertainties with the measurements in the low confidence group, for the remainder of this paper we focus only on the results for the 37 outflows in the high and medium confidence groups.

\begin{figure*}
    \resizebox{6.5in}{!}{\includegraphics{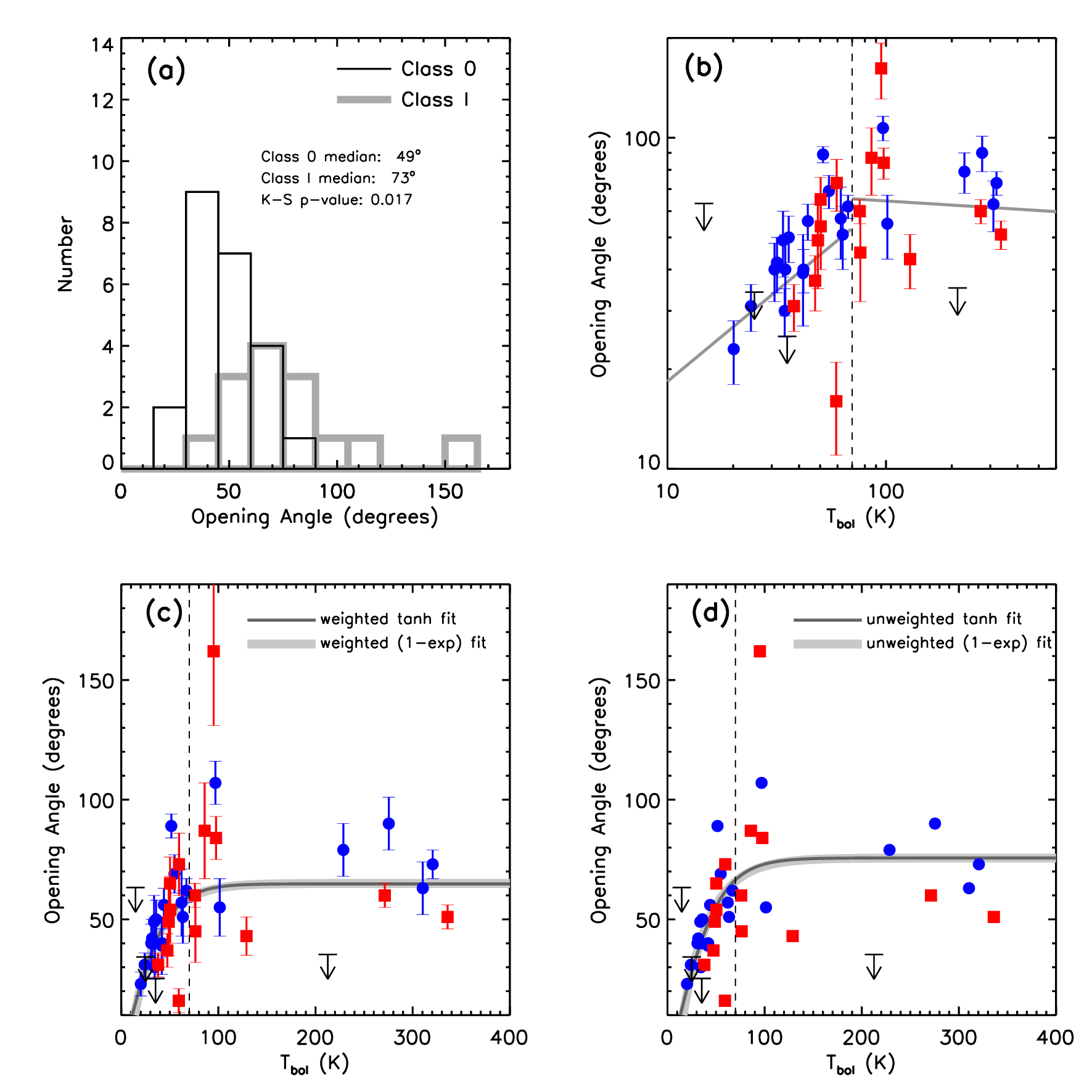}}
    \vspace{0.0in}
    \caption{(a) Histograms showing the distributions of our measured Class 0 (thin black lines) and Class I (thick gray lines) outflow opening angles for the medium and high confidence groups.  (b) Measured outflow opening angles versus \tbol\ plotted on logarithmic axes.  The best fit two-component power-law, with the break point fixed at \tbol~$=~70$~K and the fit weighted by the error bars on each measurement, is overplotted with the thick gray line.   (c) Measured outflow opening angles versus \tbol\ plotted on linear axes.  The best-fit hyperbolic tangent and exponential functions are overplotted with thin dark gray and thick light gray lines, respectively (see text for details).  The fits are weighted by the error bars on each measurement.  (d) Same as panel (c), except the fits are unweighted.  The error bars are not plotted to further emphasize that these are unweighted fits.  \color{black}In panels (b) -- (d), medium confidence measurements (red squares), high confidence measurements (blue circles), and upper limits (downward arrows) are all plotted but only the medium and high confidence opening angles are used in the fits.\color{black}}
\label{fig_angletbol_fits}
\end{figure*}

Initial visual inspection of these figures suggests a significant difference between the opening angles of the outflows driven by the Class 0 and Class I protostars.  There appears to be an increasing trend in the Class 0 opening angles, with a distinct lack of the widest outflows.  There is no clear visual evidence for a continuation of this trend in Class I, although there is a distinct lack of the most collimated outflows.  To further illustrate this difference, panel (a) of Figure \ref{fig_angletbol_fits} plots the distributions of the 23 Class 0 and 14 Class I opening angles (medium and high confidence only).  The median opening angles are 49\degree\ for Class 0 and 73\degree\ for Class I.  A K-S test returns $p=0.017$, indicating that the null hypothesis that the two distributions are drawn from the same parent distribution can be rejected with 98.3\% confidence.  Thus we conclude there is a statistically significant difference in the opening angles of Class 0 and Class I outflows, with the Class I outflows wider than the Class 0 outflows.

We now examine trends between outflow opening angles and bolometric temperatures.  For the Class 0 outflows, we calculate correlation coefficients of approximately 0.5 (0.53 in linear-linear space, 0.49 in log-log space), indicating a moderate positive linear correlation.  For the Class I outflows, we calculate correlation coefficients of $-0.30$ in linear-linear space and $-0.21$ in log-log space, indicating either no linear correlation or a very weak negative linear correlation.  Panel (b) of Figure \ref{fig_angletbol_fits} displays the results of separate power-law fits (linear least squares fits in log-log space) to the Class 0 and Class I results.  Defining $\theta$ to be the outflow opening angles in degrees, the best-fit lines are:

\begin{equation}
\log(\theta) = (0.55 \pm 0.26)\, \log(\tbol) + (0.72 \pm 0.42), \quad {\rm Class\, 0}
\end{equation}

\begin{equation}
\log(\theta) = (-0.05 \pm 0.12)\, \log(\tbol) + (1.92 \pm 0.27), \quad {\rm Class\, I}
\end{equation}
A Class 0 slope that is 2.1$\sigma$ above zero and a Class I slope that is below zero by a statistically insignificant amount (0.4$\sigma$) support our above statements that the Class 0 outflows have opening angles that exhibit a moderate positive linear correlation with \tbol, whereas the Class I outflows have opening angles that exhibit either no, or a very weak negative, linear correlation with \tbol.

To better examine where the transition occurs between an increasing trend of outflow opening angle with \tbol\ and no trend, we repeated the linear least squares fits in log-log space with the break point treated as an additional free parameter rather than holding it fixed at \tbol$~=~70$~K.  Both the parameters of the fits and the  quality of the fits were statistically consistent over a broad range of \tbol\ values for the break point, between approximately $40~{\rm K}<~\tbol~<~100~{\rm K}$, lending support to the notion that there is a transition in the growth of outflow opening angles near the Class 0/I boundary.

Continuing with this analysis, we next tried fitting the following two functions, both of which increase at low values of \tbol\ and asymptote to constant values at high values of \tbol:

\begin{equation}\label{eq_tanh}
    \theta = \theta_0 \, {\rm tanh} \left(\frac{\tbol\ - T_{\rm bol, shift}}{T_{\rm bol,0}} \right)
\end{equation}

and

\begin{equation}\label{eq_exp}
    \theta = \theta_0 \, \left( 1 - e^{-\frac{\left(T_{\rm bol} - T_{\rm bol, shift} \right)}{T_{\rm bol,0}}} \right) \qquad .
\end{equation}
In both fits the parameter $\theta_0$ is the opening angle that the functions asymptote to at large values of \tbol, the parameter $T_{\rm bol, shift}$ sets the horizontal shift of the functions, and the parameter $T_{\rm bol,0}$ controls how quickly the functions transition from increasing to approximately constant.  We emphasize here that our fits are not meant to actually describe or determine the underlying mathematical relationship between outflow opening angles and \tbol; they are simply example functions that have turnover from increasing with \tbol\ to constant in \tbol, allowing us to explore when in the evolution of a protostar this transition occurs \citep[see, e.g.,][for a similar fitting strategy to study the evolution of protostellar properties.]{heimsworth2022:masses}.

\begin{table}
\begin{center}
\caption{Best-Fit Parameters for Hyperbolic Tangent and Exponential Fits}
\label{tab_fits}
\begin{tabular}{lccc}
\hline \hline
                                          & $\theta_0$ & $T_{\rm bol, shift}$ & $T_{\rm bol,0}$ \\
Function                                  & (degrees)  & (K)                  & (K)             \\
\hline
Weighted tanh (Equation \ref{eq_tanh})     & 65         & 5                    & 42              \\
Weighted (1-$e$) (Equation \ref{eq_exp})   & 65         & 10                    & 26              \\
Unweighted tanh (Equation \ref{eq_tanh})   & 78         & 9                   & 44              \\
Unweighted (1-$e$) (Equation \ref{eq_exp}) & 78         & 14                   & 28              \\
\hline 
\end{tabular}
\end{center}
\end{table}

The best-fit parameters are tabulated in the first two rows of Table \ref{tab_fits}, and the best-fit functions are overplotted on our measurements in panel (c) of Figure \ref{fig_angletbol_fits}.  The two fits show remarkable consistency with each other: they both asymptote to opening angles of 65\degree, and they reach 90\% of these maximum \color{black}projected \color{black}opening angles at \tbol\ values of 67 K and 70 K for the hyperbolic tangent and ($1-e$) fits, respectively.  Thus for both functions nearly all of the increase in opening angle occurs during the Class 0 stage.

Since our error bars likely do not represent true statistical uncertainties (see \S \ref{sec_method_uncertainties}), and thus may not be appropriate to use in weighted fits, we repeat the hyperbolic tangent and ($1-e$) fits with uniform weighting of each measurement, rather than weighting by the uncertainties.  The results of these fits are tabulated in the third and fourth rows of Table \ref{tab_fits}, and the best-fit functions are overplotted on our measurements in panel (d) of Figure \ref{fig_angletbol_fits}.  The unweighted functions aymptote to somewhat larger opening angles (78\degree\ for both the hyperbolic tangent and ($1-e$) fits), and they reach 90\% of their maximum \color{black}projected \color{black}opening angles at larger values of \tbol\ (74 K and 79 K for the hyperbolic tangent and ($1-e$) fits, respectively).  However, as these \tbol\ values are still very close to the Class 0/I boundary of 70 K, our unweighted fits agree with the general conclusion from the weighted fits that nearly all of the increase in opening angle occurs in Class 0.

\section{Discussion}\label{sec_discussion}

\subsection{Comparison to Previous Results}\label{sec_discussion_previous}

Several studies have investigated the widths of outflow opening angles since the seminal work by \citet{arce2006:outflows}.  Here we compare our results to those obtained from these studies, focusing on studies that have used large samples for statistical analyses.  \color{black}Corrections for inclination and methodology used to measure opening angles that will be necessary when comparing results from different studies are described in Appendix \ref{sec_app_corrections}, and will be referenced as necessary throughout this section.\color{black}

\subsubsection{Comparison to \citet{seale2008:outflows}}\label{sec_discussion_comparison_sl08}

\citet{seale2008:outflows} measured the opening angles of 27 outflows using mid-infrared images from the {\it Spitzer Space Telescope} \color{black}(approximately 2$"$ resolution) \color{black}that detect scattered light off the outflow cavity walls.  They plotted angular intensity profiles at the largest radius from each protostar where the cavity walls were clearly detected, and then measured the angular distance between the two peaks corresponding to each of the cavity walls.  While this method will give different results for parabolic outflows than our method, which measures the opening angle at the base of the outflow, the majority of their outflows appear conical, and indeed they noted that their results are relatively insensitive to the exact radii at which the measurements are made.  \color{black}\citet{seale2008:outflows} used their measurements to conclude that outflows do widen with age, with a distinct lack of the widest outflows at the youngest ages and a distinct lack of the most collimated outflows at the oldest ages (see in particular their Figure~4).  While they did not find any evidence for a transition between widening and constant opening angles, their sample is dominated by younger objects (Class 0 or Class 0/I transition objects) and thus such a transition would not be apparent in their data.  Thus, at least in terms of general trends, their findings are consistent with our results.

Direct comparisons between their results and ours are hindered by the fact that they do not measure or report \tbol\ values for their driving sources, along with the fact that their classifications are based on two different luminosity ratios ($L_{\rm bol}/L_{\rm 1.3}$ and $L_{\rm smm}/L_{\rm bol}$) that are calculated using incompletely sampled SEDs  \citep[especially in the far-infrared and submillimeter; see][for detailed discussions about the reliability of luminosity calculations using such SEDs]{dunham2008:lowlum,enoch2009:protostars,dunham2013:luminosities}.  Nevertheless, using their classifications, the median outflow opening angle is 46\degree\ for the 9 outflows in the \citet{seale2008:outflows} sample that are classified as Class 0 by both of their luminosity ratios (nearly identical to our Class 0 median opening angle of 49\degree).  Furthermore, the median opening angle for the 4 outflows in the \citet{seale2008:outflows} sample that are classified as either Class 0/I or Class I by both of their luminosity ratios is 54\degree, supporting the conclusion that the median opening angle is narrower for the youngest sources (a quantitative comparison to our median Class I opening angle of 73\degree\ is not possible since their sample is dominated by younger objects).

\begin{table}
\color{black}
\begin{center}
\caption{Common Sources in MASSES and \citet{seale2008:outflows}}
\label{tab_compare_sl08}
\begin{tabular}{lcc}
\hline \hline
               & MASSES                  & \citet{seale2008:outflows} \\
Driving Source & Opening Angle (\degree) &  Opening Angle (\degree)   \\
\hline
Per-emb~2      & 40 $\pm$ 8              & 82 $\pm$ 8                                          \\
Per-emb~17     & 69 $\pm$ 8              & 24 $\pm$ 8                                          \\
Per-emb~25     & 51 $\pm$ 11             & 66 $\pm$ 8                                          \\
Per-emb~26     & 40 $\pm$ 6              & 46 $\pm$ 4                                          \\
Per-emb~30     & 107 $\pm$ 9             & 49 $\pm$ 8                                          \\
Per-emb~31     & 45 $\pm$ 13             & 82 $\pm$ 4                                          \\
Per-emb~34     & 84 $\pm$ 9              & 49 $\pm$ 6                                          \\
Per-emb~40     & 43 $\pm$ 8              & 57 $\pm$ 8                                          \\
\hline 
\end{tabular}
\end{center}
\color{black}
\end{table}

Finally, Table \ref{tab_compare_sl08} compares the MASSES and \citet{seale2008:outflows} opening angles for the 8 outflows common to both samples.  While the agreement appears poor, four outflows (Per-emb~2, Per-emb~17, Per-emb~30, and Per-emb~34) are very weakly detected by \citet{seale2008:outflows} in their mid-infrared scattered light images, rendering the reliability of this comparison too low for further analysis.  We thus conclude that the general trends are in agreement between the two samples, with a more detailed quantitative comparison simply not possible.\color{black}

\subsubsection{Comparison to \citet{velusamy2014:outflows}}\label{sec_discussion_comparison_velusamy}

\begin{figure*}
    \resizebox{\hsize}{!}{\includegraphics{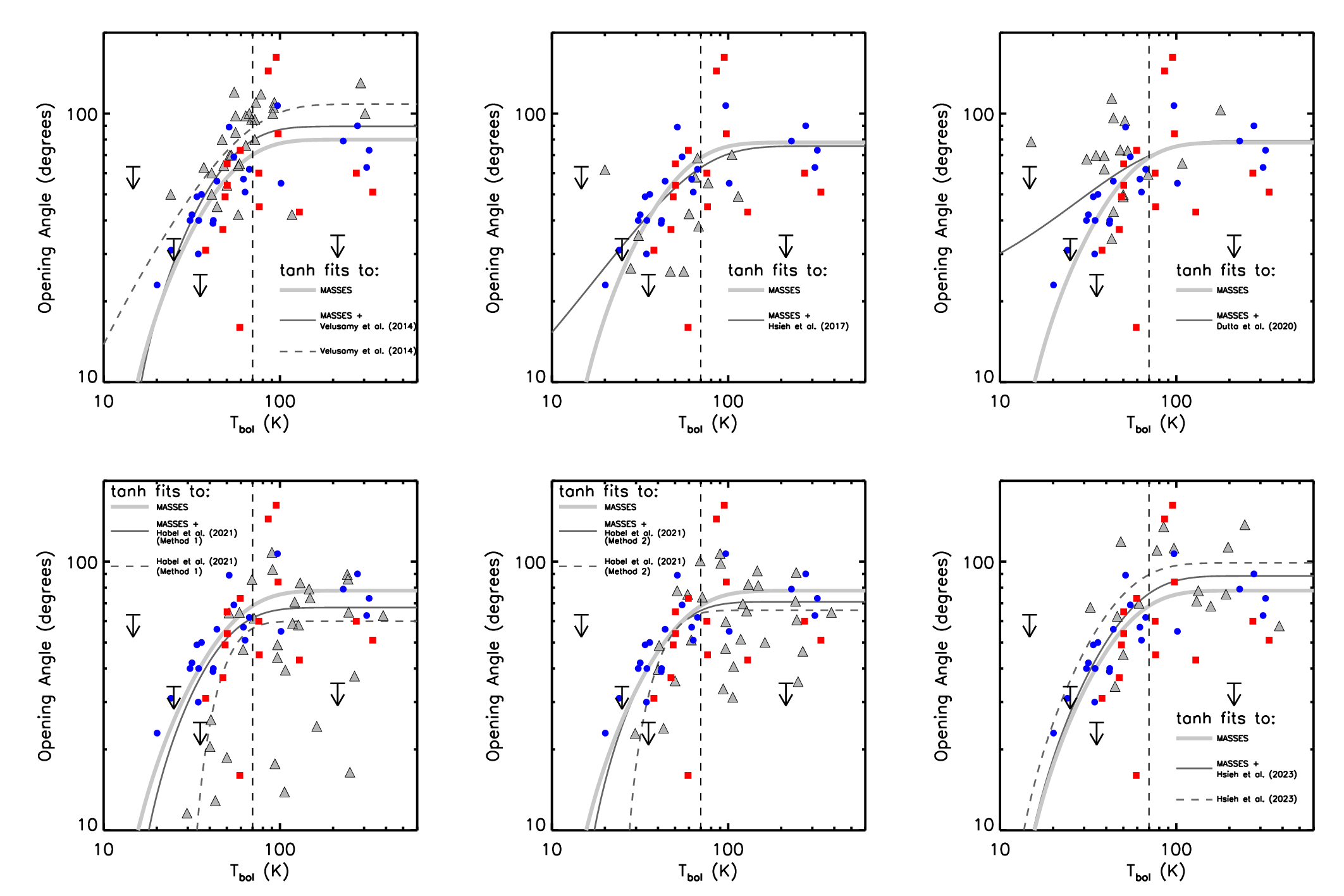}}
    \vspace{0.0in}
     \caption{\color{black}Outflow opening angles (projected onto the plane of the sky) plotted versus the bolometric temperatures of their driving sources, on logarithmic axes.  \color{black}In all panels the blue circles plot the MASSES high confidence measurements from this work, the red squares plot the MASSES medium confidence measurements from this work, the downward arrows plot the MASSES upper limits from this work, and the gray triangles plot the measurements from the previous study noted in the text in each panel.  \color{black}The dashed vertical line in each panel marks \tbol~$=70$~K, the Class 0/I boundary in bolometric temperature \citep{chen1995:tbol}, with Class 0 protostars to the left (\tbol~$<70$~K) and Class I protostars to the right (\tbol~$\geq 70$~K).  \color{black}The thick gray line in each panel shows the same unweighted hyperbolic tangent fit to the MASSES measurements that is shown in panel (d) of Figure \ref{fig_angletbol_fits}.  The other lines in each panel show other unweighted hyperbolic tangent fits as noted in the text in each panel, and as described in the text in \S \ref{sec_discussion_previous}.  \color{black}Upper limits are not used in any of the fits presented here.\color{black}}
\label{fig_angletbol_compare}
\end{figure*}

\begin{table*}
\color{black}
\begin{center}
\caption{Comparison to Previously Published Surveys of Outflow Opening Angles}
\label{tab_compare}
\begin{tabular}{lccccc}
\hline \hline
                                          & \multicolumn{2}{c}{Median Projected Opening Angle$^{\rm a}$} & \multicolumn{3}{c}{Unweighted Hyperbolic Tangent Fits} \\
                                          & Class 0       & Class I                  & $\theta_0$    & $T_{\rm bol, shift}$ & $T_{\rm bol,0}$ \\
Dataset                                   & (degrees)     & (degrees)                & (degrees)     & (K)                  & (K)             \\
\hline
\color{black}MASSES\color{black}                                          & \color{black}49\color{black}   & \color{black}73\color{black}   & \color{black}78\color{black}   & \color{black}5\color{black}    & \color{black}42\color{black}       \\
\citet{velusamy2014:outflows}                                           & 68                           & 105                          & 108                          & 2                            & 59                               \\
MASSES + \citet{velusamy2014:outflows}                                  & ...$^{\rm b}$                & ...$^{\rm b}$                & 90                           & 12                           & 41                               \\
\citet{hsieh2017:outflows}                                              & 38                           & 55                           & ...$^{\rm c}$                & ...$^{\rm c}$                & ...$^{\rm c}$                    \\
MASSES + \citet{hsieh2017:outflows}                                     & ...$^{\rm b}$                & ...$^{\rm b}$                & 76                           & $-$2                         & 61                               \\
\citet{dutta2020:almasop}                                               & 70                           & 84                           & ...$^{\rm c}$                & ...$^{\rm c}$                & ...$^{\rm c}$                    \\
MASSES + \citet{dutta2020:almasop}                                      & ...$^{\rm b}$                & ...$^{\rm b}$                & 79                           & $-$15                        & 62                               \\
\color{black}\citet{habel2021:outflows} (Method 1)\color{black}           & \color{black}26\color{black}   & \color{black}62\color{black}   & \color{black}60\color{black}   & \color{black}30\color{black}   & \color{black}23\color{black}       \\
\color{black}MASSES + \citet{habel2021:outflows} (Method 1)\color{black}  & ...$^{\rm b}$                & ...$^{\rm b}$                & \color{black}67\color{black}   & \color{black}13\color{black}   & \color{black}37\color{black}       \\
\color{black}\citet{habel2021:outflows} (Method 2)\color{black}           & \color{black}49\color{black}   & \color{black}65\color{black}   & \color{black}66\color{black}   & \color{black}24\color{black}   & \color{black}21\color{black}       \\
\color{black}MASSES + \citet{habel2021:outflows} (Method 2)\color{black}  & ...$^{\rm b}$                & ...$^{\rm b}$                & \color{black}71\color{black}   & \color{black}12\color{black}   & \color{black}35\color{black}       \\
\citet{hsieh2023:outflows}                                              & 65                           & 110                          & 99                           & 10                           & 40                               \\
MASSES + \citet{hsieh2023:outflows}                                     & ...$^{\rm b}$                & ...$^{\rm b}$                & 88                           & 10                           & 46                               \\
\color{black}MASSES + literature (Habel method 1)\color{black}            & \color{black}57\color{black}   & \color{black}76\color{black}   & \color{black}79\color{black}   & \color{black}9\color{black}    & \color{black}38\color{black}       \\
\color{black}MASSES + literature (Habel method 2)\color{black}            & \color{black}57\color{black}   & \color{black}76\color{black}   & \color{black}81\color{black}   & \color{black}9\color{black}    & \color{black}39\color{black}       \\
\color{black}MASSES + literature (no Habel)\color{black}                  & \color{black}58\color{black}   & \color{black}90\color{black}   & \color{black}90\color{black}   & \color{black}4\color{black}    & \color{black}51\color{black}       \\
\hline 
\end{tabular}
\end{center}
$^{\rm a}$\color{black}All opening angles are full opening angles, and are projected onto the plane of the sky and thus uncorrected for inclination.\color{black}\\
$^{\rm b}$Median opening angles not calculated.\\
$^{\rm c}$No fit performed due to a lack of Class I objects.\\
\color{black}
\end{table*}

\citet{velusamy2014:outflows} measured the \color{black}projected \color{black}opening angles of 31 outflows using mid-infrared images from the {\it Spitzer Space Telescope} processed using the HiRes deconvolution algorithm \citep{backus2005:hires} \color{black}in order to achieve approximately 1$"$ resolution (native {\it Spitzer} resolution is approximately 2$"$).  \color{black}They measured the opening angles of the scattered light cavities by hand, and they explicitly noted that they measured the opening angle at the base for parabolic outflows.  Since our method agrees with opening angles measured by hand in this manner \color{black}\citep[see][in particular their Figure~2]{offner2011:outflows}, \color{black}and also measures projected opening angles, \color{black}our results can be directly compared \color{black}without any conversions or adjustments.\color{black}

\color{black}The top left panel \color{black}of Figure \ref{fig_angletbol_compare} plots the outflow opening angles versus the \tbol\ of the driving sources for both our measurements and those of \citet{velusamy2014:outflows}.  \color{black}The median opening angles in the \citet{velusamy2014:outflows} sample are 68\degree\ for Class 0 and 105\degree\ for Class I, as reported in Table \ref{tab_compare}, thus like the MASSES sample the \citet{velusamy2014:outflows} sample shows an increasing trend between Class 0 and Class I.  Overplotted in Figure \ref{fig_angletbol_compare} are unweighted hyperbolic tangent fits to just the MASSES opening angles, to the combination of the MASSES and \citet{velusamy2014:outflows} opening angles, and just the \citet{velusamy2014:outflows} opening angles. The parameters of these fits are reported in Table \ref{tab_compare}. 
\color{black}The qualitative agreement between our results is very good, and indeed \citet{velusamy2014:outflows} also concluded that there is a widening trend for younger outflows that flattens considerably for more evolved outflows (although like our sample, theirs also contains about two times more Class 0 outflows than Class I outflows).  While they did not explicitly attempt to constrain the location of this transition, they did present power-law fits (linear fits in logarithmic space) that assumed this transition occurs very close to the Class 0/I boundary.  Thus we find good agreement between their conclusions and the results from our present study.

\color{black}Inspection of Figure \ref{fig_angletbol_compare} suggests there may be a systematic offset to higher opening angles in the \citet{velusamy2014:outflows} measurements compared to our own measurements, most notable when comparing the medians (68\degree\ versus 49\degree\ for Class 0; 105\degree\ versus 73\degree\ for Class I) and in the offset between the unweighted hyperbolic tangent fits solely to the MASSES results and solely to the \citet{velusamy2014:outflows} results.  Although this offset could be due to the different samples, there is some evidence that it is in fact due to the different outflow tracers used (mid-infrared scattered light vs. \co\ emission).  \color{black}Of the four outflows contained in both samples, the opening angles reported by \citet{velusamy2014:outflows} are 30\degree\ wider (Per-emb~5), 4\degree\ wider (Per-emb~16), 19\degree\ wider (Per-emb~1), and 40\degree\ wider (Per-emb~53), giving mean and median differences of 23.25\degree\ and 24.5\degree, respectively.  \citet{velusamy2014:outflows} also noted this offset based on comparing to previously published \co\ maps and suggested that opacity effects in the lowest velocity \co\ channels where outflows are typically the widest may limit the ability of \co\ to probe the full extent of outflow opening angles.  Indeed, \citet{arce2013:hh46} \color{black}and \citet{zhang2016:hh4647} \color{black}present a specific instance where a \color{black}narrower opening angle is seen in \co\ emission compared to both \coo\ emission and mid-infrared scattered light (HH 46/47; see in particular Figure~8 of Arce et al. and Figure~16 of Zhang et al.).  Future work is needed to more explicitly test the true extent to which opening angles measured in \co\ underestimate the intrinsic opening angles due to both opacity and/or chemistry \citep[e.g.,][]{tychoniec2021:tracers} effects, using larger samples of outflows whose opening angles are measured using a variety of different tracers.\color{black}

\subsubsection{Comparison to \citet{hsieh2017:outflows}}

\citet{hsieh2017:outflows} measured the opening angles of 12 outflows driven by low-luminosity protostars using near-infrared and mid-infrared images \color{black}with resolutions of approximately $1'' - 2''$, \color{black}nine of which are Class 0 outflows and three of which are Class I outflows.  To measure the opening angles they used the two-dimensional \citet{whitney2003:models} radiative transfer models that include \color{black}curved outflow cavities (specifically, outflow cavities following power laws with indices of 1.5) \color{black}fit to the scattered light infrared images.  They adopted the ``Class Late 0'' model from \citet{whitney2003:models_sequence} for all of their objects, with the opening angle \color{black}and inclination \color{black}of the outflow treated as free parameters \color{black}and their reported opening angles corrected for inclination.  \color{black}Since their models adopt \color{black}curved \color{black}rather than conical outflow cavities, their opening angles are defined to be the opening angles of cones that reach the same width at a vertical distance of 5000~au above the midplane \color{black}(see Appendix \ref{sec_app_corrections}).\color{black}

\color{black}Despite the significant methodology differences we can still perform a direct comparison between our results because the two methods give opening angles that are consistent within the typical uncertainties of 5\degree\ -- 15\degree, as shown in the appendix of \citet{myers2023:outflows}.  Since our measurements are not corrected for inclination, in order for this comparison to be valid we take the opening angles reported by \citet{hsieh2017:outflows} and project them back onto the plane of the sky, using \color{black}Equation \ref{eq_inclination} with \color{black}their reported source inclinations.  The top middle panel of Figure \ref{fig_angletbol_compare} plots the outflow opening angles versus the \tbol\ of the driving sources for both our measurements and those of \citet{hsieh2017:outflows} (after projection back onto the plane of the sky).  The median projected opening angles in the \citet{hsieh2017:outflows} sample are 38\degree\ for Class 0 and 55\degree\ for Class I, as reported in Table \ref{tab_compare}.
Overplotted in Figure \ref{fig_angletbol_compare} are unweighted hyperbolic tangent fits to just the MASSES opening angles and to the combination of the MASSES and \citet{hsieh2017:outflows} opening angles (we do not fit to just the \citet{hsieh2017:outflows} opening angles since their sample contains so few Class I objects).  The parameters of these fits are reported in Table \ref{tab_compare}.

\color{black}Based on their measurements, \citet{hsieh2017:outflows} concluded that there is a trend of increasing opening angle with \tbol.  They do not find any clear evidence for a flattening of this trend, but their sample only contains 3 Class I outflows, all of which are close to the Class 0/I boundary.  \color{black}When considering their results combined with ours, the two samples given generally consistent results.\color{black}

\subsubsection{Comparison to \citet{dutta2020:almasop}}

\color{black}\citet{dutta2020:almasop} measured the outflow lobe widths and bolometric temperatures of the driving sources for 17 protostars in Orion originally identified as Orion Planck Galactic Cold Clumps.  They measured the outflow lobe widths at distances of both 400 and 800~au from the driving source using subarcsecond resolution \cojtwo\ images obtained with the Atacama Large Millimeter/submillimeter Array (ALMA).  We calculated opening angles at 800~au distances from the driving sources, using their measured widths at this distance.  \citet{dutta2020:almasop} did not correct for inclination, so these calculated opening angles can be directly compared to our opening angles, which are also uncorrected for inclination.

The top right panel of Figure \ref{fig_angletbol_compare} plots the outflow opening angles versus the \tbol\ of the driving sources for both our MASSES measurements and those calculated by us using the \citet{dutta2020:almasop} measurements.  Overplotted are unweighted hyperbolic tangent fits to just the MASSES opening angles and to the combination of the MASSES and \citet{dutta2020:almasop} opening angles (we do not fit to just the \citet{dutta2020:almasop} opening angles since their sample contains so few Class I objects).  The parameters of these fits are reported in Table \ref{tab_compare}.  Also reported in Table \ref{tab_compare} are the median Class 0 and Class I opening angles in the \citet{dutta2020:almasop} sample.

Examining the \citet{dutta2020:almasop} results alone does not reveal strong existence of a trend, consistent with their findings of at best a very weak positive correlation between outflow widths and \tbol.  However, as they noted, the majority of their sample spans a very narrow evolutionary range in the latter part of the Class 0 evolutionary stage, with only two Class~I objects and only one Class 0 object with \tbol~$<$~30~K (this one young Class 0 object appears to be a significant outlier in that it has a large opening angle, and this outlier will be discussed further in \S \ref{sec_discussion_comparison_synthesis} below).  Fitting to the combined MASSES and \citet{dutta2020:almasop} samples yields a hyperbolic tangent fit that asymptotes to 79\degree\ in the Class I stage, nearly identical to our value of 78\degree\ for the MASSES sample alone.  Thus while the \citet{dutta2020:almasop} do not by themselves clearly show an increasing trend in Class 0 that transitions to a constant maximum \color{black}projected \color{black}opening angle in Class I, their results are not inconsistent with this result.
\color{black}

\subsubsection{Comparison to \citet{habel2021:outflows}}\label{sec_discussion_comparison_habel}

\citet{habel2021:outflows} measured the opening angles of 30 outflows using \color{black}0.18$"$ resolution \color{black}near-infrared images from the {\it Hubble Space Telescope}, 10 of which are Class 0 outflows and 20 of which are Class I outflows.  To measure the opening angles they applied an edge detection algorithm to their near-infrared images in order to the detect the edges of the cavity walls seen in scattered light.  They then fit power laws to the cavity walls, \color{black}treating the power-law index as a free parameter and finding mean and median indices of 1.9 and 1.5, respectively (indices of 1 and 2 indicate conical and parabolic morphologies, respectively).  Since most of their outflows are not found to be conical, \color{black}their opening angles are defined to be the opening angles of cones that reach the same width as the power-law cavities at a vertical distance of 8000~au above the midplane.  \color{black}Their reported opening angles are the half-angles of these cones, \color{black}after applying model-derived correction factors to correct for inclination and other factors, as described below.\color{black}

\color{black}To perform an ``apples-to-apples'' comparison between our MASSES results, which are observed opening angles without inclination or other corrections, and those of \citet{habel2021:outflows}, we must "undo" their corrections.  Before attempting to undo their full model-derived corrections, we first explore the effects of inclination corrections alone.  \color{black}We first recalculate their opening angles to be those of cones that reach the same width at a vertical height of 5000~au above the midplane, rather than 8000~au above the midplane, using Equations~\ref{eq_shape1}--\ref{eq_shape3}.  This correction to a lower vertical height is chosen to be more consistent with both \citet{hsieh2017:outflows} and with the comparison of methods presented in the Appendix of \citet{myers2023:outflows}.  Next, we multiply the opening angles reported by \citet{habel2021:outflows} by a factor of two to convert from half opening angles to full opening angles.  Finally, we project these corrected opening angles onto the plane of the sky, using Equation~\ref{eq_inclination} with the source inclinations reported by \citet{furlan2016:hops} (these inclinations are derived from spectral energy distribution modeling; see Furlan et al. for details on the modeling process and associated uncertainties).  \color{black}We refer to these correction steps as correction method 1.\color{black}

\color{black}\citet{habel2021:outflows} applied model-derived correction factors to their measured cavity widths that are much larger than those derived from inclination corrections alone.  They attribute these correction factors to the combined effects of inclination, the penetration of light into and beyond the cavity edges, and other systematic biases in their measurement technique, all of which caused their measured outflow opening angles to be wider than the intrinsic opening angles.  \color{black}Since both the MASSES measurements and all other literature measurements considered in \S \ref{sec_discussion_previous} are direct measurements in the plane of the sky, without any correction for the various methodological biases present in each study, we ``undo'' their model-derived corrections to obtain their directly measured opening angles.  \color{black}In practice we do this by multiplying $x_{8000}$ (see Figure \ref{fig_corrections_shape}) by the Habel et al.~model-derived correction factor for each target in their sample\footnote{\color{black}While \citet{habel2021:outflows} do not tabulate their specific correction factors, based on the information they report in their text we apply correction factors of 1.3 (2.0) for unipolar cavities with reported half-angles greater than (less than) 15\degree, and correction factors of 1.1 (2.3) for bipolar cavities with reported half-angles greater than (less than) 15\degree.\color{black}}, \color{black}recalculating their $\alpha_{8000}$ (again, see Figure \ref{fig_corrections_shape}) using this larger width, converting from $\alpha_{8000}$ to $\alpha_{5000}$ following Appendix \ref{sec_app_corrections}, and then finally multiplying by a factor of two to convert from half to full opening angles.  We refer to these correction steps as correction method 2.\color{black}

\color{black}The bottom left and center panels of Figure \ref{fig_angletbol_compare} plot the projected outflow opening angles versus the \tbol\ of the driving sources for both our MASSES measurements and those 
presented by \citet{habel2021:outflows} after applying correction method 1 (left panel) or correction method 2 (right panel).  Overplotted are unweighted hyperbolic tangent fits to the MASSES opening angles, to the \citet{habel2021:outflows} opening angles, and to the combination of the MASSES and \citet{habel2021:outflows} opening angles, with the parameters of these fits reported in Table \ref{tab_compare}.  Also reported in Table \ref{tab_compare} are the median Class 0 and Class I opening angles in the \citet{habel2021:outflows} sample, after applying correction methods 1 and 2.  \color{black}These results show that the two samples are consistent on both the widening between Class~0 and Class~I outflows and the lack of a discernible continuation of the widening trend between the Class~0/I boundary, regardless of the specific correction method used, emphasizing that the MASSES and \citet{habel2021:outflows} results for outflow opening angle trends are in overall good agreement.  We note that correction method 2 yields opening angles that are overall more consistent with the MASSES opening angles, especially in the relative lack of very narrow Class I outflows.  Finally, comparing the two correction methods emphasizes that the large number of very narrow Class I outflows reported by \citet{habel2021:outflows} is a direct result of their large, model-derived, outflow width reductions by factors of 2--2.3 in outflows with full opening angles less than 30\degree.  

\color{black}Finally, we \color{black}note that \citet{habel2021:outflows} used the fraction of point sources (assumed to be viewed face-on through the cleared outflow cavities) to suggest a maximum \color{black}intrinsic \color{black}outflow opening angle of 70\degree\ in Class~I sources, in good quantitative agreement with our results for the existence and possible values of a maximum outflow opening angle \color{black}($\sim$66\degree\ after inclination correction; see \S \ref{sec_inclination}).\color{black}

\subsubsection{Comparison to \citet{hsieh2023:outflows}}

\color{black}\citet{hsieh2023:outflows} measured outflow opening angles, \color{black}using $\sim$1$"$ resolution ALMA \cojtwo\ data, for 17 protostellar outflows \color{black}in Orion, using a method identical to our own except implemented separately for the redshifted and blueshifted outflow lobes.  Since we fit to the wider of the two lobes in the small number of cases where the two outflow lobes are asymmetrical, for consistency with our results we select the wider of their two opening angle measurements\footnote{\color{black}The mean and median differences between the wider and narrower lobes for the \citet{hsieh2023:outflows} sample are \color{black}19.5\degree\ and 20\degree, somewhat beyond the high end of the 5--15\degree\ typical uncertainty for our method that we discussed in \S \ref{sec_method_uncertainties}.  When using the narrower lobes rather than the wider lobes, the median opening angles reported in Table \ref{tab_compare} for the \citet{hsieh2023:outflows} outflows decrease from 65\degree\ to 42\degree\ for Class 0, and from 110\degree\ to 85\degree\ for Class I.\color{black}}.  We also select their results uncorrected for inclination \color{black}for direct comparison with our results, which are also uncorrected for inclination.\color{black}

The bottom right panel of Figure \ref{fig_angletbol_compare} plots the outflow opening angles versus the \tbol\ of the driving sources for both our MASSES measurements and those presented by \citet{hsieh2023:outflows}.  Overplotted are unweighted hyperbolic tangent fits to just the MASSES opening angles, to just the \citet{hsieh2023:outflows} opening angles, and to the combination of the MASSES and \citet{hsieh2023:outflows} opening angles.  The parameters of these fits are reported in Table \ref{tab_compare}.  Also reported in Table \ref{tab_compare} are the median Class 0 and Class I opening angles in the \citet{hsieh2023:outflows} sample.

Based on their measurements, \citet{hsieh2023:outflows} concluded there is a trend of increasing opening angle with \tbol, with a notable decline in the rate of increase for more evolved sources.  Overall their results are highly consistent with our MASSES results.
\color{black}

\subsubsection{Synthesis of Comparisons to Previous Results}\label{sec_discussion_comparison_synthesis}

\begin{figure*}
    \resizebox{\hsize}{!}{\includegraphics{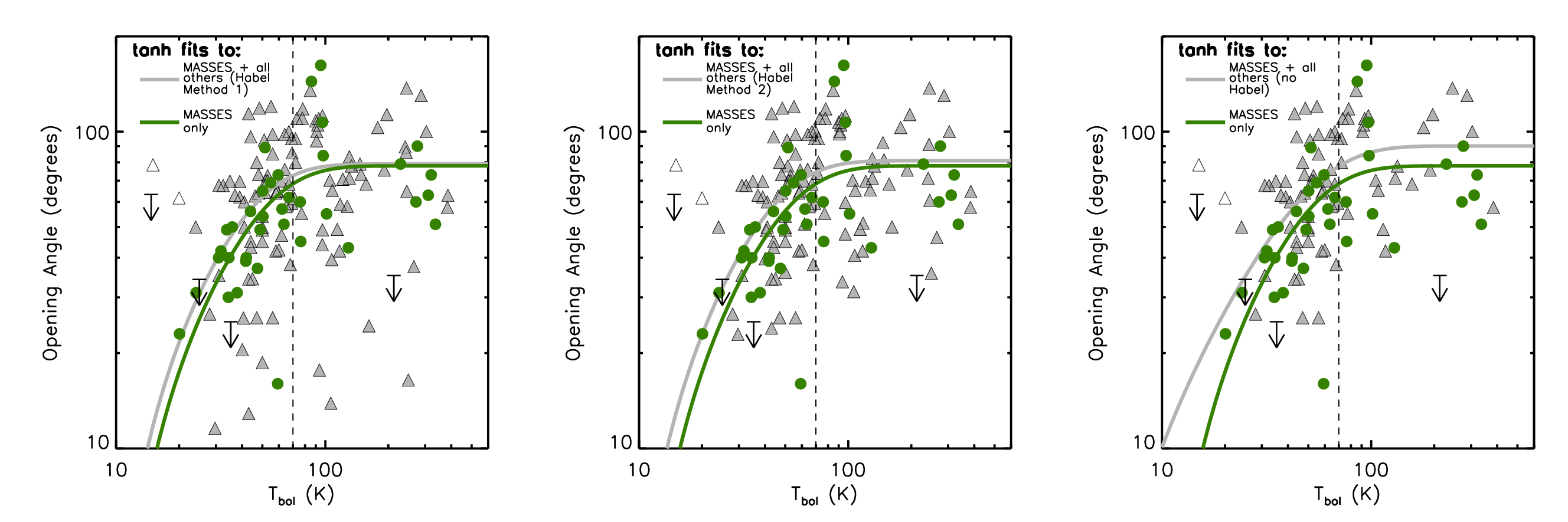}}
    \vspace{0.0in}
    \caption{\color{black}Outflow opening angles plotted versus the bolometric temperatures of their driving sources, on logarithmic axes.  In all three panels the green circles plot the MASSES high and medium confidence measurements and the downward arrows plot the MASSES upper limits.  The shaded gray triangles plot the combined measurements from the previous studies discussed in \S \ref{sec_discussion_previous}, with the \citet{habel2021:outflows} measurements corrected using method 1 in the left panel, corrected using method 2 in the middle panel, and omitted from the comparison in the right panel (see the text in \S \ref{sec_discussion_comparison_synthesis} for details on the unshaded gray triangles).  The dashed vertical line in each panel marks \tbol~$=70$~K, the Class 0/I boundary in bolometric temperature \citep{chen1995:tbol}, with Class 0 protostars to the left (\tbol~$<70$~K) and Class I protostars to the right (\tbol~$\geq 70$~K). The green line in each panel shows the same unweighted hyperbolic tangent fit to the MASSES measurements that is shown in panel (d) of Figure \ref{fig_angletbol_fits}, and the gray line in each panel shows the unweighted hyperbolic tangent fit to the combined sample of MASSES measurements and previous literature measurements.  Upper limits are not used in any of the fits presented here.\color{black}}
\label{fig_compare_summary}
\end{figure*}

\color{black}
Figure \ref{fig_compare_summary} plots the projected \color{black}outflow opening angles versus the \tbol\ of the driving sources for the combined MASSES (present work), \citet{velusamy2014:outflows}, \citet{hsieh2017:outflows}, \citet{dutta2020:almasop}, \color{black}\citet{habel2021:outflows}, \color{black}and \citet{hsieh2023:outflows} samples.  \color{black}In each panel, \color{black}the MASSES measurements are plotted with green circles and the combined sample of previous measurements are plotted with filled gray triangles.  \color{black}The \citet{habel2021:outflows} measurements are corrected using method 1 in the left panel, they are corrected using method 2 in the center panel, and they are omitted entirely in the right panel\footnote{\color{black}Including a panel where the \citet{habel2021:outflows} measurements are omitted should not be taken to signify that we distrust their results; \color{black}rather it is used to further demonstrate the robust nature of the results of this paper.\color{black}}.  Inspection of Figures \ref{fig_angletbol_compare} and \ref{fig_compare_summary} \color{black}reveals two outliers at low \tbol\ and wide opening angle, one from \citet{hsieh2017:outflows} and one from \citet{dutta2020:almasop}.  These objects are plotted as unshaded triangles in \color{black}Figure \ref{fig_compare_summary}.  \color{black}The outlier in \citet{hsieh2017:outflows}, DCE~004, is undetected in their near-infrared images and only weakly detected in their mid-infrared image, with their best-fit model image not clearly providing a good fit to their observed image.  For the outlier in \citet{dutta2020:almasop}, G203.21$-$11.20W2, Dutta et al.~report a \tbol\ of 15~K but this appears to be a mistake; recalculating \tbol\ using their tabulated photometry and stated signal-to-noise cutoffs yields a \tbol\ of 109 K.  Thus we omit these two outliers from the following analysis.

Overplotted in \color{black}each panel of Figure \ref{fig_compare_summary} \color{black} are unweighted hyperbolic tangent fits to just the MASSES opening angles and to the combined sample of MASSES and all previous results (except for those omitted as described above), \color{black}with the parameters of these fits reported in Table \ref{tab_compare}.  \color{black}Also reported in Table \ref{tab_compare} are the median Class 0 and Class I opening angles for the combined samples.  \color{black}Regardless of the specific method used to correct the \citet{habel2021:outflows} measurements (or even if the Habel et al.~measurements are included at all), the combined sample gives generally consistent results to those obtained with just the MASSES sample:  there is a moderately increasing opening angle trend in Class 0 but no clear evidence that this trend continues in Class I, with Class~0 correlation coefficients (in log-log space) of 0.39 (left panel), 0.41 (middle panel), and 0.39 (right panel), and Class~I correlation coefficients (in log-log space) of 0.06 (left panel), 0.06 (middle panel), and 0.07 (right panel).  Additionally, the maximum Class~I opening angles reached by the hyperbolic tangent fits are 79\degree\ (left panel), 81\degree\ (middle panel), and 90\degree\ (right panel), in good agreement with the value of 78\degree\ obtained from the MASSES measurements alone.  The scatter for the combined sample is larger, as expected given the use of different outflow tracers and different measurement techniques, but there is overall consistency in the results between the various studies to date.  \color{black}Larger samples using a consistent tracer and uniform \color{black}measurement technique \color{black}are needed to further quantify the exact nature of opening angle trends, and indeed such samples should become available in the coming years through the ALMA data archive.\color{black}

\subsection{Implications for the Growth of Protostars}\label{sec_discussion_implications}

Synthesizing our results and those of the previous studies described above in \S \ref{sec_discussion_previous}, Class 0 outflow opening angles show a widening trend with \tbol.  If \tbol\ is taken as a proxy for evolutionary status, outflows do appear to widen considerably throughout the Class 0 evolutionary stage.  Class I outflows, on the other hand, do not show any significant trends between opening angles and \tbol.
The transition from significant widening to no significant widening appears to occur very near the Class 0/I boundary, with approximately 90\% of the widening occurring by the Class 0/I boundary of \tbol$~=~70$~K \color{black}(based on both hyperbolic tangent and $(1-e)$ fits to the data).\color{black}

The maximum \color{black}projected \color{black}opening angle reached in the Class I stage before the widening ends remains somewhat uncertain.  \color{black}Hyperbolic tangent and $(1-e)$ fits to just our MASSES data asymptote to opening angles ranging between 65\degree\ -- 78\degree, depending on the exact details of the fit.  Unweighted hyperbolic tangent fits to various combinations of opening angles from MASSES and other previous studies \color{black}asymptote to opening angles ranging between 60\degree\ -- 108\degree\ (with mean and median values of 81\degree\ and 79\degree, respectively), and a fit to the combined measurements from MASSES and all previous surveys 
asymptote to opening angle of 79\degree\ -- 90\degree, depending on exactly which measurements are included and exactly how they are corrected for different measurement techniques.  \color{black}
Considering all of these results, we conclude that the maximum \color{black}projected \color{black}opening angle reached during the Class I stage is approximately 90\degree~$\pm$~20\degree, \color{black}where the uncertainty of 20\degree\ represents the \color{black}approximate \color{black}range of possible values rather than a statistical uncertainty.\color{black}

Bipolar conical outflow cavities with opening angles ranging from 70\degree~$-$~110\degree\ subtend between 18\%~$-$~43\% of the full $4\pi$ steradian.  Assuming spherical symmetry, and further assuming that only mass which starts within the outflow cavity can be removed by the outflow, completely cleared conical cavities of these sizes could remove no more than 18\%~$-$~43\% of the total initial core mass.  This range represents an upper limit for \color{black}two \color{black}reasons: (1) the cavities start much narrower and only reach these widths near the Class 0/I transition, allowing some mass that is within the final outflow cavity to accrete before the cavities reach their full extents; and (2) rotational flattening decreases the fraction of total core mass near the rotation axes \color{black}(and thus within the volume occupied by the outflow) \color{black}and increases it near the midplane, \color{black}compared to the fractions expected in spherically symmetric cores\color{black}.  Thus, if it is assumed that the fraction of mass removed \color{black}by outflows \color{black}is equal to the fraction of the total volume occupied by \color{black}conical outflows with these maximum \color{black}projected \color{black}opening angles\color{black}, then these outflows are incapable of removing more than, at most, a few tens of percent of the initial core mass (and even this is likely a strong upper limit).  In this regard we are in \color{black}qualitative \color{black}agreement with \citet{habel2021:outflows}, who concluded that the low star formation efficiency inferred for molecular cores can't be explained solely by outflow-induced envelope clearing.  \color{black}On the other hand, \citet{myers2023:outflows} recently showed that outflows that reach maximum \color{black}projected \color{black}opening angles of 90\degree\ can remove larger mass fractions of the initial core than those calculated here by considering the effects of conical versus curved outflow cavity shapes and/or the possibility that mass which starts outside of the outflow can still be ejected by the outflow \citep[in particular by MHD winds; e.g.,][]{machida2013:outflows}.  Thus at present the extent to which outflows are capable of setting the observed star formation efficiencies of 25\% -- 50\% on dense core scales, requiring that the outflows remove 50\% -- 75\% of the initial core mass, remains uncertain.\color{black}

\subsection{Limitations of the Current Work}\label{sec_discussion_limitations}

\subsubsection{Opacity and Chemistry Effects}

\color{black}As discussed extensively in \S \ref{sec_discussion_comparison_velusamy}, opacity and/or chemistry effects may cause outflow opening angle measurements made using \co\ emission to underestimate the intrinsic opening angles.  While such an underestimate is unlikely to change the conclusion that a maximum outflow opening angle appears to exist, it may change the quantitative result for the value of this maximum angle.  At present the full magnitude and extent of this issue is unknown due to there being too few outflows sufficiently characterized to perform a statistical comparison of the results obtained using different tracers.  Future work must be devoted to 
measuring the opening angles of the same outflows using \co, \coo, and infrared scattered light images, using consistent techniques applied to each dataset, in order to investigate the full magnitude and extent of this issue.\color{black}

\subsubsection{Possible Bias Against the Widest Outflows}\label{sec_discussion_limitations_widebias}

All investigations of outflow opening angles, whether they use molecular gas observations or scattered light images, must start with a visual identification of the outflows to be measured.  It is possible that all such results, including those presented here, suffer from a systematic bias against the widest outflows.  Outflows with opening angles that approach 180\degree\ may be difficult to identify since they may not look much like the canonical picture of an outflow, \color{black}and additionally wide outflows may suffer from incomplete recovery in our interferometric MASSES data due to limited uv-coverage, especially at the shortest uv spacings.\color{black}

\begin{figure}
    \resizebox{\hsize}{!}{\includegraphics{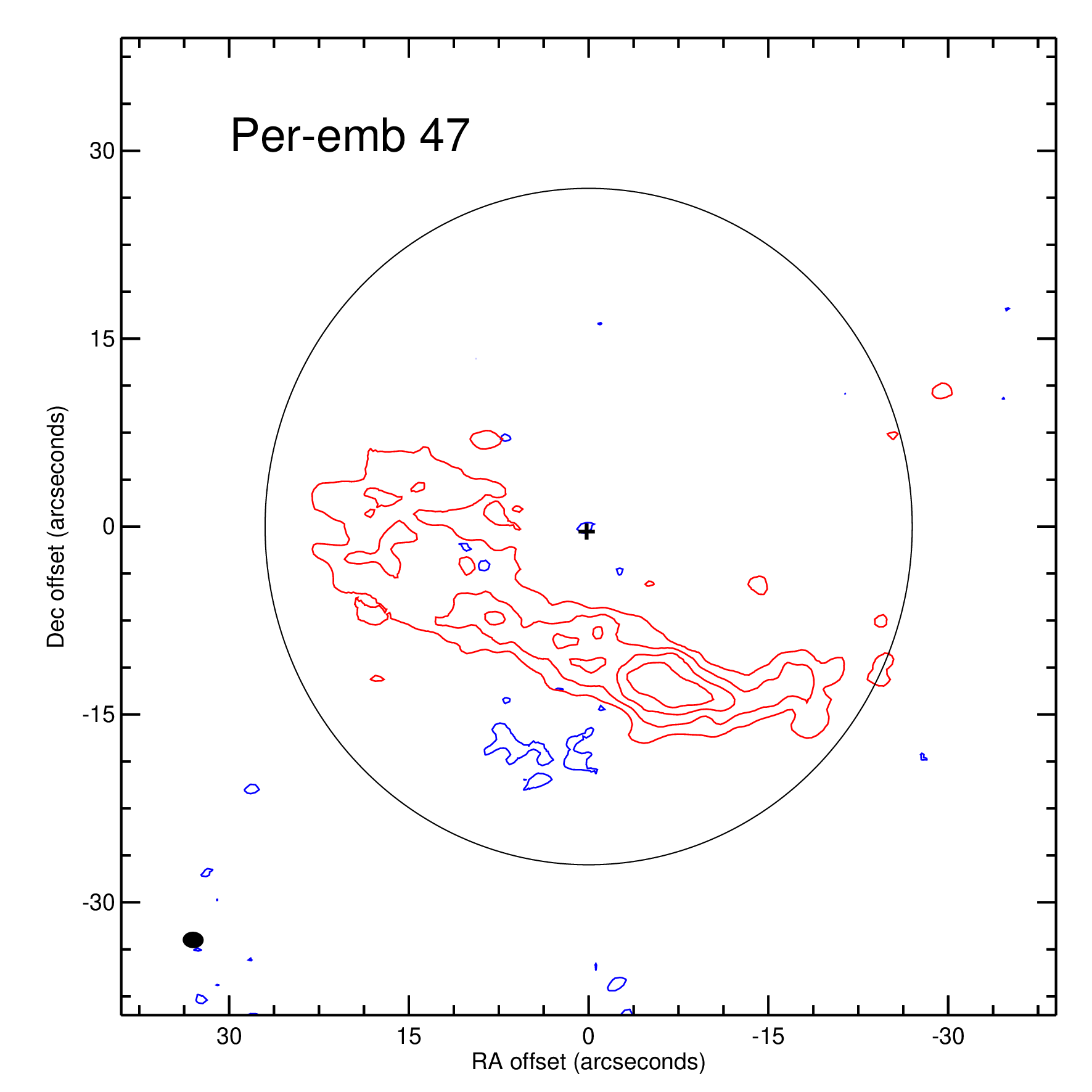}}
    \vspace{0.0in}
    \caption{Integrated (moment 0) redshifted and blueshifted \cojtwo\ emission associated with Per-emb~47.  The emission is integrated from $2-10$~\kms\ relative to the systemic velocity of 7.5~\kms.  The contours start at 3$\sigma$ and increase by 3$\sigma$, where $\sigma$ is the rms noise in the integrated maps. The large circle shows the primary beam of the SMA at 230~GHz, centered on the phase center of the observations.  The synthesized beam is shown as the filled ellipse in the lower left corner.}
\label{fig_per47}
\end{figure}

Although cases like Per-emb~36 do suggest confidence in the claim that we are able to identify outflows with very wide opening angles near their bases when they are present (see in particular the blueshifted emission associated with Per-emb~36 in Figure \ref{fig_result3}), we do find some cases in our MASSES observations where there is high-velocity gas associated with Class I protostars that we are unable to conclusively identify as an outflow.  Figure \ref{fig_per47} shows one such example for Per-emb~47.  As noted in Appendix \ref{sec_app_nofit}, we detect substantial redshifted emission in the vicinity of Per-emb~47 in our MASSES \cojtwo\ observations, extending up to $\sim$10~\kms\ from rest.  The morphology of this emission is not clearly indicative of an outflow  \color{black}However, it is difficult to account for gas moving at up to 10~\kms\ from rest without invoking outflows, and comparison to larger-scale single-dish maps does support the possibility that this emission is attributed to an outflow that is poorly recovered in our MASSES data \citep[][see in particular their Figure~5]{hatchell2009:outflows}.  \color{black}Future high-sensitivity, \color{black}high-fidelity \color{black}observations, especially in the Class I stage, must carefully revisit this question.

\subsubsection{Effects of Inclination}\label{sec_inclination}

For outflows viewed at inclination angles of $i=90$\degree, corresponding to an edge-on geometry where the outflow axis is aligned with the plane of the sky, the measured opening angle will be identical to the intrinsic opening angle.  As the inclination angle decreases toward 0\degree\ (toward a more pole-on orientation), the measured opening angle will be larger than the intrinsic opening angle.  Our results are not corrected for inclination since, in many cases, the inclinations of the systems are \color{black}either unknown or \color{black}highly uncertain.  As a result, 
\color{black}most of our measured, projected opening angles will be wider than the intrinsic, deprojected opening angles of the outflows. \color{black}While it is important to note that neither the lack of wide Class 0 outflows nor the existence of a maximum Class I outflow opening angle can be explained by inclination effects, since inclination corrections would shift our measured opening angles down rather than up, our inability to correct for inclination \color{black}will likely at least somewhat \color{black}affect our quantitative results.\color{black}

We utilize the following Monte Carlo simulation to better quantify the effects of excluding inclination corrections on our results.  For each of the 37 high and medium confidence outflows we randomly select an inclination between $i=0$\degree\ (pole-on) and $i=90$\degree\ (edge-on), weighted by solid angle so that higher inclination angles (more edge-on geometries) are more likely to be selected.  We then calculate the inclination-corrected opening angle for each of the 37 outflows \color{black}(using Equation \ref{eq_inclination}), \color{black}and then re-fit the unweighted hyperbolic tangent fit described by Equation \ref{eq_tanh}.  We then repeat this for a total of 10,000 iterations.


\begin{figure}
    \resizebox{\hsize}{!}{\includegraphics{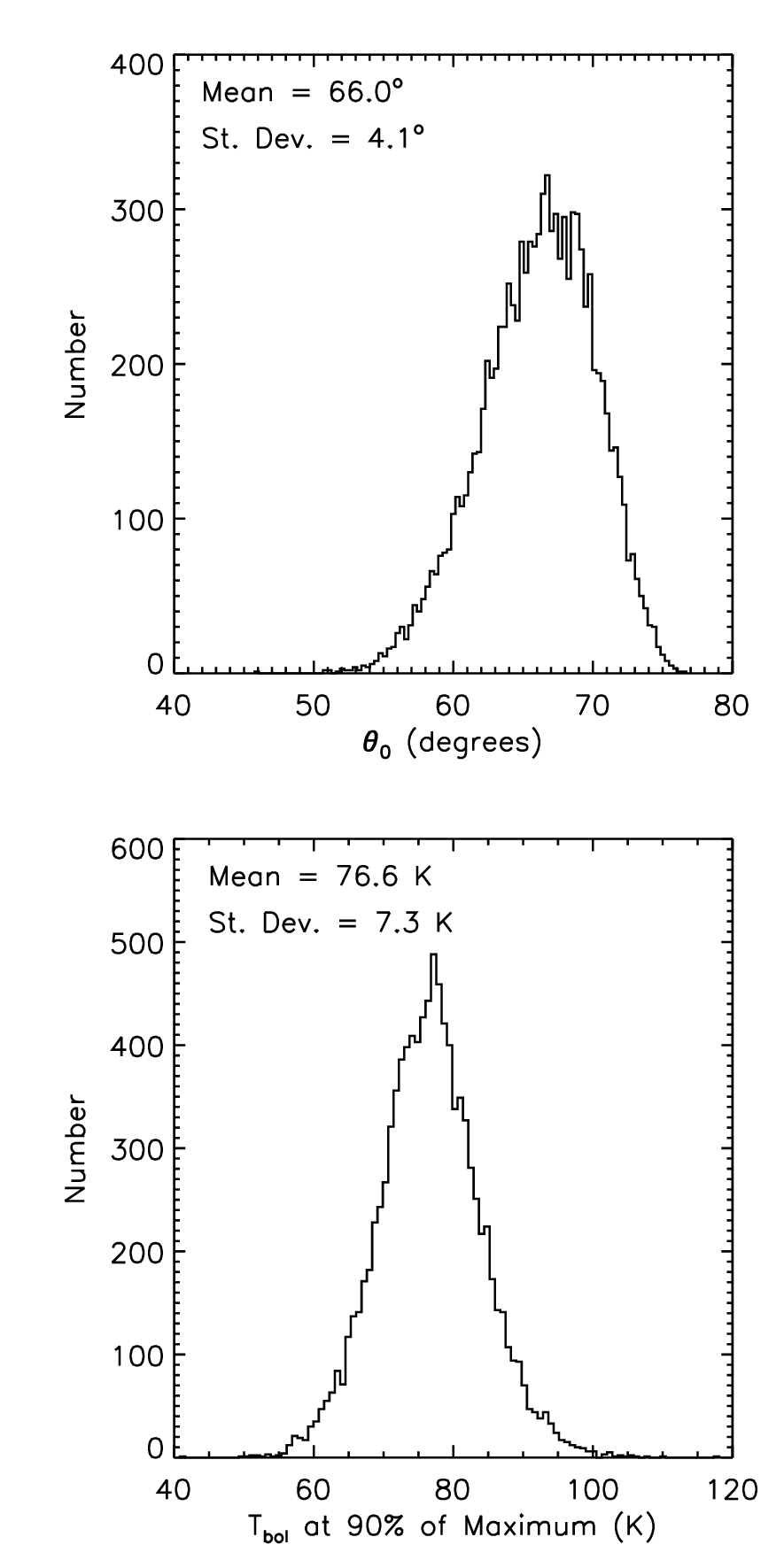}}
    \vspace{0.0in}
    \caption{Results from our Monte Carlo test of the effects of inclination corrections on our results.  The top panel shows the distribution of the parameter $\theta_0$ from the 10,000 unweighted hyperbolic tangent fits.  The bottom panel shows the distribution of \tbol\ values where each of the 10,000 hyperbolic tangent fits reach 90\% of their maximum (90\% of $\theta_0$).}
\label{fig_inc}
\end{figure}

The distribution of the parameter $\theta_0$, which gives the maximum value to which the hyperbolic tangent fit asymptotes \color{black}(and thus the maximum \color{black}projected \color{black}opening angle reached by Class I outflows), \color{black}is shown in the top panel of Figure \ref{fig_inc}.  $\theta_0$ has a mean value of \color{black}66.0\degree\ and a standard deviation of 4.1\degree\ \color{black}over the 10,000 simulation runs.  Our original value for $\theta_0$ from Table \ref{tab_fits} was \color{black}78\degree, \color{black}and the smaller mean value here makes sense given that inclination corrections will decrease the measured opening angles.  \color{black}It also agrees very well with the value of 68.7\degree\ for $\theta_0$ that is obtained by correcting $\theta_0 = 78$\degree\ after adopting the average inclination angle of 57.3\degree\ for randomly oriented objects.  \color{black}The distribution of \tbol\ values where the hyperbolic tangent fit reaches 90\% of its maximum value (90\% of $\theta_0$) from the 10,000 iterations is shown in the bottom panel of Figure \ref{fig_inc}.  The mean \tbol\ at which 90\% of the widening has occurred is \color{black}76.6 K, with a standard deviation of 7.3 K (compared to 74 K for our original fit without inclination corrections).

\color{black}A Monte Carlo simulation such as this could in theory miss the effects of the most extreme inclination corrections, although with 10,000 simulation runs the statistics are likely robust.  We can explicitly test this by considering Per-emb~44 (SVS~13A), which has a robust inclination measurement of $i = 18 \pm 3$\degree\ \citep{hodapp2014:svs13} and is thus very close to a pole-on inclination.  The large inclination correction necessary for $i = 18$\degree\ reduces the opening angle of this outflow from 144\degree\ (the second largest value in our sample) to 87\degree.  Re-performing our unweighted hyperbolic tangent fit to our full sample with the Per-emb~44 opening angle changed to 87\degree\ reduces $\theta_0$ (the maximum value to which the hyperbolic tangent fit asymptotes) from 78\degree\ to 76\degree.

Given these results, we find that the \color{black}inclusion \color{black}of inclination corrections does not affect our two main conclusions that there exists a maximum outflow opening angle reached in the Class I stage, and that the transition from widening to constant outflow opening angles occurs at or near the Class 0/I boundary.  It does reduce the value of the maximum \color{black}intrinsic \color{black}opening angle reached in the Class I stage, \color{black}compared to the projected value.  We note that this value is close to the low end of our stated range of approximately 90\degree\ $\pm$ 20\degree\ for the maximum projected opening angle.\color{black}

\subsubsection{Use of \tbol\ as an Evolutionary Indicator}\label{sec_discussion_limitations_tbol}

\color{black}As noted in \S \ref{sec_intro}, the \tbol\ of a protostar is defined to be the temperature of a blackbody with the same flux-weighted mean frequency as the SED of that protostar \citep{myers1993:tbol}, and \tbol\ is commonly used as an evolutionary indicator \citep[e.g.,][]{chen1995:tbol,enoch2009:protostars,evans2009:c2d,dunham2014:ppvi,frimann2016:simulations}.  However, in addition to evolutionary status, \tbol\ can also be affected by geometry and accretion history, thus assuming a monotonic increase in \tbol\ with evolutionary status (as is done when using plots of opening angle versues \tbol\ as proxies for plots of opening angle versus evolutionary status) adds potential uncertainties and biases to the results presented here that are difficult to fully quantify.

Several authors have argued that alternative classification methods, such as the ratio of submillimeter (longward of 350~\um) to bolometric luminosity (\lsmmbol), better trace the underlying evolutionary status of a protostar \citep[e.g.,][]{dunham2014:ppvi,frimann2016:simulations}.  Our choice to use \tbol\ is motivated primarily by a desire to remain consistent with previous studies (for ease of comparison) that also used \tbol.  Future work must examine the extent to which the results of this paper hold up to alternative evolutionary indicators.\color{black}

\section{Summary}\label{sec_summary}

In this paper we use \cojtwo\ observations from the MASSES (Mass Assembly of Stellar Systems and their Evolution with the SMA) project to measure the \color{black}projected \color{black}opening angles of protostellar outflows in the Perseus Molecular Cloud.  We summarize our results as follows:

\begin{enumerate}
    \item We measure the opening angles for a total of 46 outflows, 37 of which are measured with sufficiently high confidence to use for further analysis.  Out of these 37 outflows, 23 are driven by Class 0 protostars and 14 are driven by Class I protostars.
    \item We measure the outflow opening angles using a fitting method adapted from \citet{offner2011:outflows}.  This method shows excellent agreement with the cones that would be drawn by hand for conical outflows, and it accurately measures the opening angles of the conical (or approximately conical) bases of outflows with \color{black}curved \color{black}morphologies. 
    \item There is a statistically significant difference in the distributions of outflow opening angles for Class 0 and Class I outflows.  In particular there is a distinct lack of both wide-angle Class 0 outflows and highly collimated Class I outflows, and the median Class I outflow (73\degree) is wider than the median Class 0 outflow (49\degree).
    \item Class 0 outflows have opening angles that exhibit a moderate positive linear correlation with \tbol, whereas Class I outflows have opening angles that exhibit either no, or a very weak negative, linear correlation with \tbol.  If \tbol\ is taken to be an evolutionary tracer, Class 0 outflows appear to widen with age whereas Class I outflows do not.
    \item Multiple types of fits to the opening angles versus \tbol, including power-law fits (linear fits in log-log space), weighted and unweighted hyperbolic tangent fits in linear space, and weighted and unweighted exponential fits in linear space, all demonstrate that a transition between outflow opening angles that widen with \tbol\ (widen with age) and outflow opening angles that remain approximately constant with \tbol\ (approximately constant with age) occurs at or near the Class 0/I boundary.  These results are seen to be robust even when inclination corrections are taken into account.
    \item \color{black}The results from this survey are generally consistent with those from all previously published studies of outflow opening angles over the last 15 years.\color{black}
    \item Outflows do not widen all the way to 180\degree, at which point they would fully disperse the surrounding core.  Instead they appear to reach maximum \color{black}projected \color{black}opening angles of approximately  90\degree~$\pm$~20\degree.  The volume fractions occupied by these outflows are incapable of removing more than a few tens of percent of the initial core mass, at most, assuming the fraction of initial core mass removed by the outflow matches the volume fraction of the core occupied by these conical outflows. 
\end{enumerate}

While these results challenge the notion that outflows widen all the way to 180\degree, at which point they disperse the surrounding cores and terminate the protostellar mass accretion process, \color{black}recent theoretical work suggests outflows may still be capable of playing a central role in setting the low star formation efficiencies of 25\% -- 50\% observed on core scales.  Future work must concentrate on building larger samples with improved statistics, especially in the Class I stage, on understanding the systematic biases introduced by measuring outflow opening angles with different datasets and methodologies, and on testing the effects of using alternative evolutionary indicators to measure the relative evolutionary status of the driving sources.\color{black}

\section*{Acknowledgements}

The authors acknowledge support from NASA grant NNX14AG96G, and HGA acknowledges support from NSF award AST-1714710.  We gratefully acknowledge helpful discussions with Stella Offner, Lars Kristensen, and Katherine Lee during the early implementaton of this project, as well as intellectual contributions provided by Forest Mathieu and Oscar DeLaRosa.  We also acknowledge helpful discussions with Sarah Sadavoy and comments from Cheng-Han Hsieh.  \color{black}We thank the referee for extremely helpful comments that have greatly improved the quality of this publication, and we express our gratitude to Adele Plunkett for allowing us to overplot contours from her 2013 survey of outflows in NGC~1333.  \color{black}This work is based on data obtained by the MASSES project.  This large-scale project would not have been successful without the assistance and support provided by Charlie Qi, Glen Petitpas, Qizhou Zhang, Garrett ``Karto'' Keating, Eric Keto, and Ray Blundell.

The Submillimeter Array is a joint project between the Smithsonian Astrophysical Observatory and the Academia Sinica Institute of Astronomy and Astrophysics and is funded by the Smithsonian Institution and the Academia Sinica.  The authors wish to recognize and acknowledge the very significant cultural role and reverence that the summit of Mauna Kea has always had within the indigenous Hawaiian community.  We are most fortunate to have the opportunity to conduct observations from this mountain.  This research has made use of NASA's Astrophysics Data System (ADS) Abstract  Service and the IDL Astronomy Library hosted by the NASA Goddard Space Flight  Center.  This research has made use of the SIMBAD database, operated at CDS, Strasbourg, France.

\section*{Data Availability}

{\it  
All MASSES data are publicly available through the Harvard Dataverse at https://dataverse.harvard.edu/dataverse/full\_MASSES/.
}



\bibliographystyle{mnras}
\bibliography{dunham_citations} 




\appendix

\section{Outflows in Clusters and Groups}\label{sec_app_clusters}

\begin{figure*}
    \resizebox{7in}{!}{\includegraphics{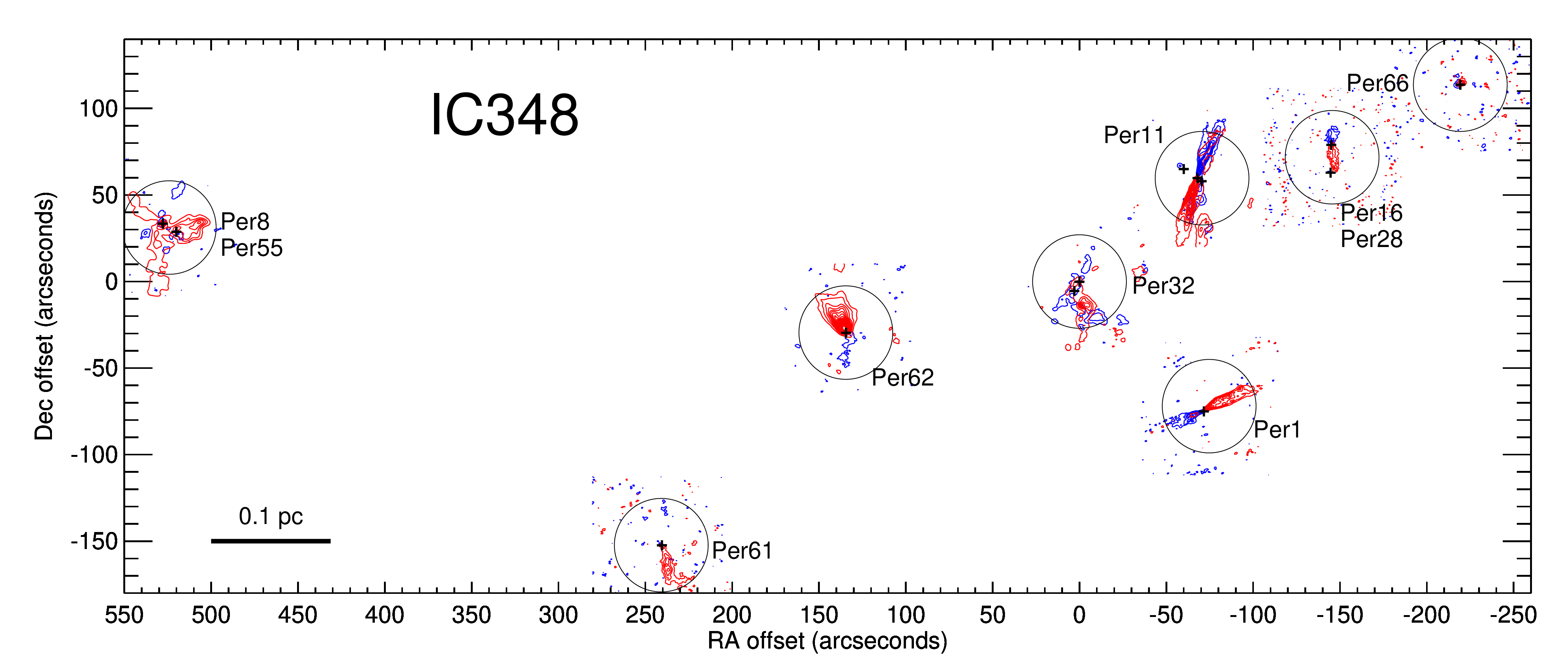}}
    \caption{\cojtwo\ outflow maps for each of the MASSES pointings in the IC348 region of Perseus, with the protostars located within each pointing labeled in the figure.  The blue contours show blueshifted emission integrated from $3-8$~\kms\ relative to the systemic velocity of 8.5~\kms, and the red contours show redshifted emission integrated over the same velocity range relative to the systemic velocity.  The contours start at 2.5$\sigma$ and increase by 5$\sigma$, where $\sigma$ is the rms noise in each integrated map.  The large circles show the primary beam of the SMA at 230~GHz, centered on each of the pointings. The crosses mark the positions of the protostars.  See Figures~\ref{fig_result1}~--~\ref{fig_result3} for labels of the individual protostars within each pointing, as well as ellipses showing the size and orientation of each synthesized beam (which would be too small to be visible in this figure).  The axes are plotted using coordinate offsets from the phase center of Per-emb~32 (R.A.~03:44:02.40, Decl.~$+$32:02:04.90).}
    \label{fig_mosaic_ic348}
\end{figure*}

\begin{figure*}
    \resizebox{4in}{!}{\includegraphics{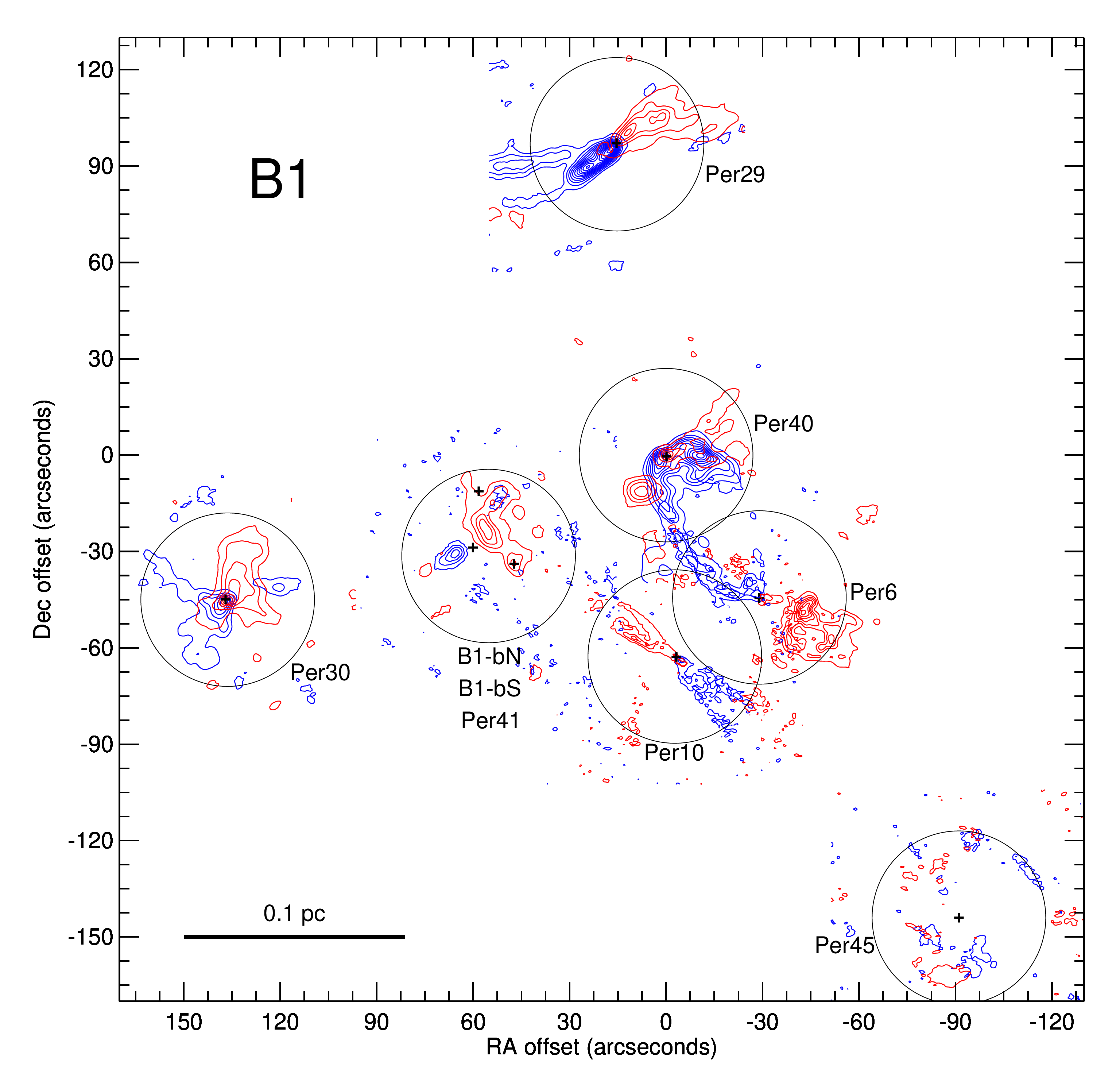}}
    \caption{\cojtwo\ outflow maps for each of the MASSES pointings in the B1 region of Perseus, with the protostars located within each pointing labeled in the figure.  The blue contours show blueshifted emission integrated from $3-8$~\kms\ relative to the systemic velocity of 6.5~\kms, and the red contours show redshifted emission integrated over the same velocity range relative to the systemic velocity.  The contours start at 2$\sigma$ and increase by 5$\sigma$, where $\sigma$ is the rms noise in each integrated map.  The large circles show the primary beam of the SMA at 230~GHz, centered on each of the pointings. The crosses mark the positions of the protostars.  See Figures~\ref{fig_result1}~--~\ref{fig_result3} for labels of the individual protostars within each pointing, as well as ellipses showing the size and orientation of each synthesized beam (which would be too small to be visible in this figure).  The axes are plotted using coordinate offsets from the phase center of Per-emb~40 (R.A.~03:33:16.66, Decl.~$+$31:07:55.20).}
    \label{fig_mosaic_b1}
\end{figure*}

\begin{figure*}
    \resizebox{\hsize}{!}{\includegraphics{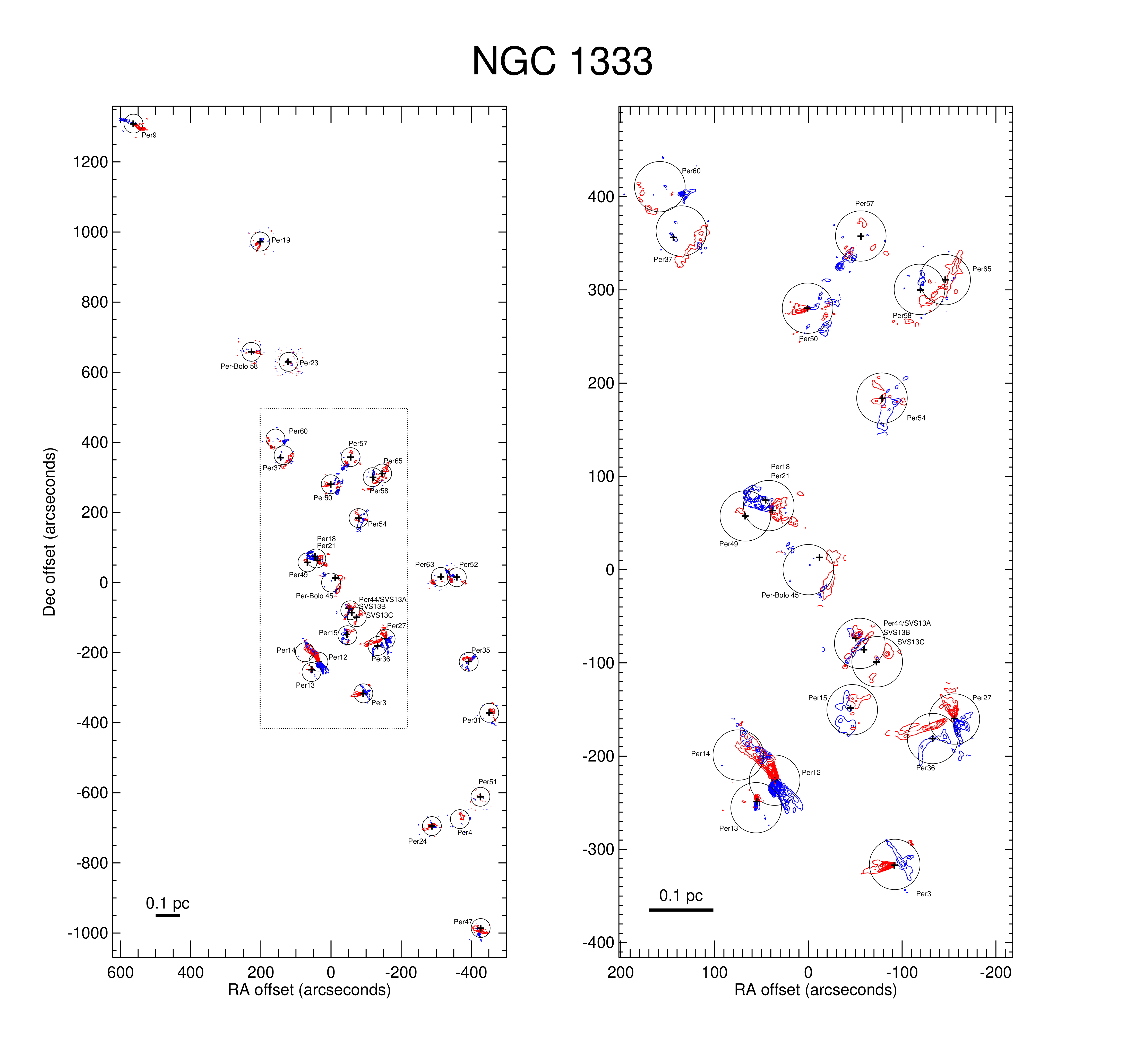}}
    \caption{\cojtwo\ outflow maps for each of the MASSES pointings in the NGC~1333 region of Perseus, with the protostars located within each pointing labeled in the figure.  The blue contours show blueshifted emission integrated from $3-8$~\kms\ relative to the systemic velocity of 7.5~\kms, and the red contours show redshifted emission integrated over the same velocity range relative to the systemic velocity.  The contours start at 3$\sigma$ and increase by 5$\sigma$, where $\sigma$ is the rms noise in each integrated map.  The large circles show the primary beam of the SMA at 230~GHz, centered on each of the pointings. The crosses mark the positions of the protostars.  See Figures~\ref{fig_result1}~--~\ref{fig_result3} for labels of the individual protostars within each pointing, as well as ellipses showing the size and orientation of each synthesized beam (which would be too small to be visible in this figure).  The axes are plotted using coordinate offsets from the phase center of Per-Bolo~45 (R.A.~03:29:07.70, Decl.~$+$31:17:16.80).  The right panel zooms in on the region outlined with a dotted rectangle in the left panel.}
    \label{fig_mosaic_ngc1333}
\end{figure*}

\begin{figure*}
    \resizebox{3.7in}{!}{\includegraphics{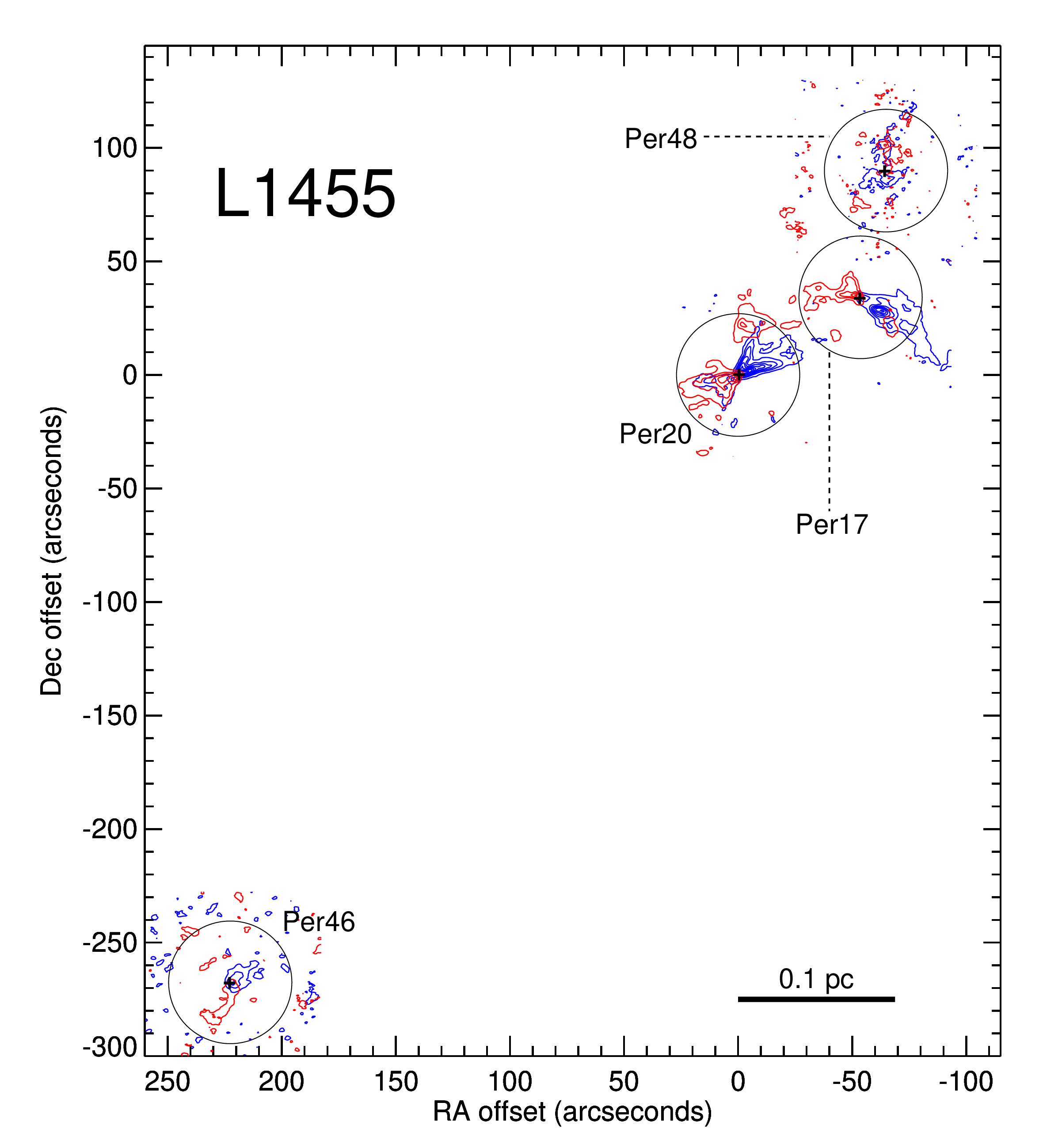}}
    \caption{\cojtwo\ outflow maps for each of the MASSES pointings in the L1455 region of Perseus, with the protostars located within each pointing labeled in the figure.  The blue contours show blueshifted emission integrated from $3-8$~\kms\ relative to the systemic velocity of 5.5~\kms, and the red contours show redshifted emission integrated over the same velocity range relative to the systemic velocity.  The contours start at 2$\sigma$ and increase by 5$\sigma$, where $\sigma$ is the rms noise in each integrated map.  The large circles show the primary beam of the SMA at 230~GHz, centered on each of the pointings. The crosses mark the positions of the protostars.  See Figures~\ref{fig_result1}~--~\ref{fig_result3} for labels of the individual protostars within each pointing, as well as ellipses showing the size and orientation of each synthesized beam (which would be too small to be visible in this figure).  The axes are plotted using coordinate offsets from the phase center of Per-emb~20 (R.A.~03:27:43.23, Decl.~$+$30:12:28.80).}
    \label{fig_mosaic_l1455}
\end{figure*}

\begin{figure*}
    \resizebox{5.0in}{!}{\includegraphics{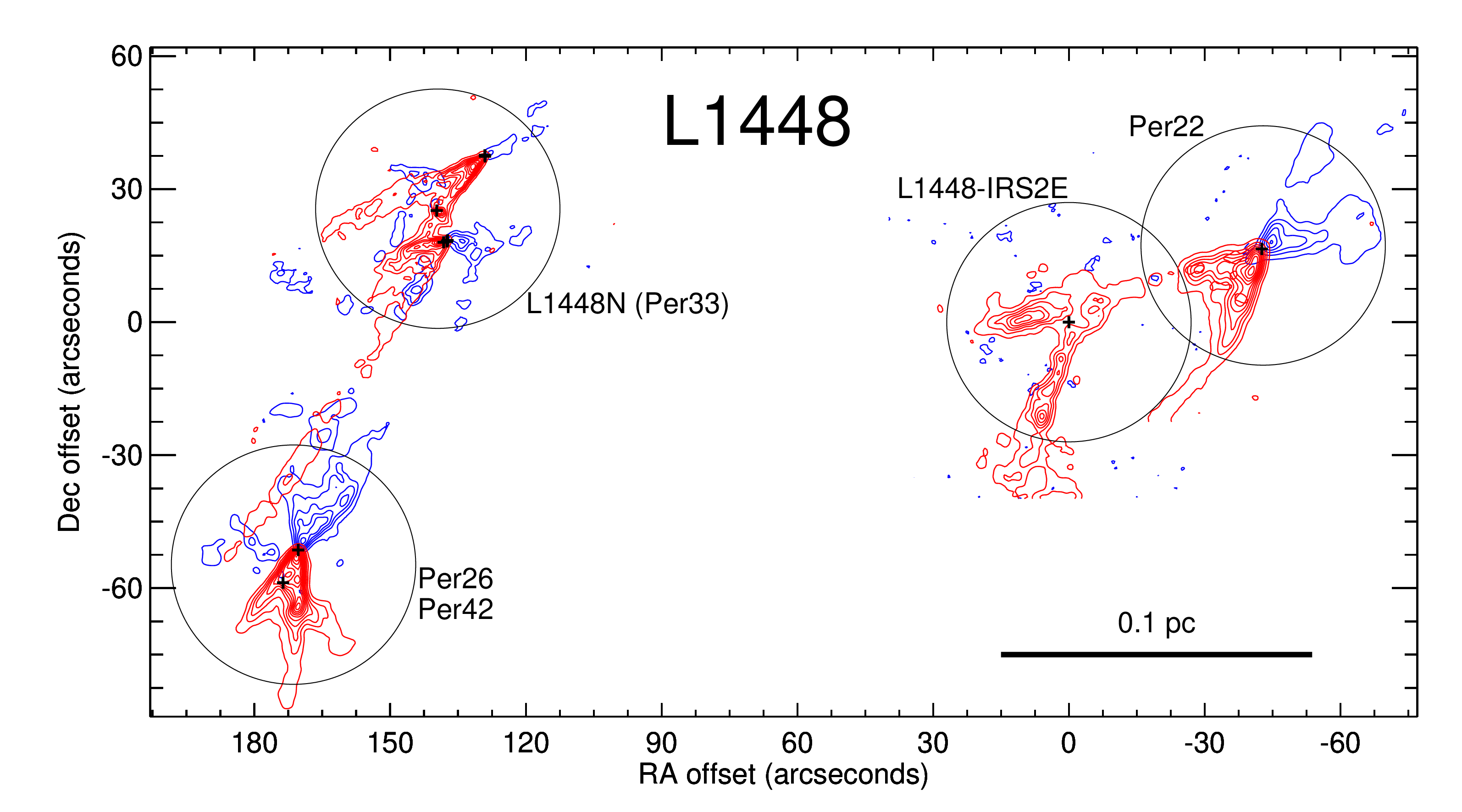}}
    \caption{\cojtwo\ outflow maps for each of the MASSES pointings in the L1448 region of Perseus, with the protostars located within each pointing labeled in the figure.  The blue contours show blueshifted emission integrated from $2-6$~\kms\ relative to the systemic velocity of 5~\kms, and the red contours show redshifted emission integrated over the same velocity range relative to the systemic velocity.  The contours start at 2.5$\sigma$ and increase by 5$\sigma$, where $\sigma$ is the rms noise in each integrated map.  The large circles show the primary beam of the SMA at 230~GHz, centered on each of the pointings. The crosses mark the positions of the protostars.  See Figures~\ref{fig_result1}~--~\ref{fig_result3} for labels of the individual protostars within each pointing, as well as ellipses showing the size and orientation of each synthesized beam (which would be too small to be visible in this figure).  The axes are plotted using coordinate offsets from the phase center of L1448-IRS2E (R.A.~03:25:25.66, Decl.~$+$30:44:56.70).}
\label{fig_mosaic_l1448}
\end{figure*}

In this appendix we present we present large-scale maps showing the results of our individual MASSES pointings in various regions of Perseus. Figures~\ref{fig_mosaic_ic348}~--~\ref{fig_mosaic_l1448} show the  IC348, B1, NGC~1333, L1455, and L1448 regions of Perseus, respectively. \color{black}Individual protostars are labeled as PerXX instead of Per-emb~XX for maximum clarity in the figures.  \color{black}We emphasize that these are {\it not} mosaics.  Instead, these figures plot the individual pointings in each region offset from each other by the proper amounts to reflect their positions relative to each other.  The choice of whether to display the robust~=~$+1$ or robust~=~$-1$ map for each pointing matches the rest of the paper, as summarized in Table \ref{tab_sample}. For pointings with undetected or unclear outflows that are not considered in this work, we default to the robust~=~$+1$ map.  \color{black}By plotting in offset coordinates there are technically projection errors in these maps in that the decrease in the total distance across the Right Ascension axis by a factor of ${\rm cos}(\delta)$ with increasing $\delta$ (where $\delta$ represents the Declination) is not accounted for.  However, the total change in distance across the Right Ascension axis between the bottom and top of these figures is less than 1$"$, and thus negligible, for all but NGC~1333.  This total change is 14$"$ for NGC~1333; while still small compared to the 57$"$ FWHM of the primary beam of our observations, it should nevertheless be considered when comparing to maps plotted in absolute coordinates.

In these figures \color{black}the velocity ranges for integration and the contour levels plotted do not match those adopted for Figures~\ref{fig_result1}~--~\ref{fig_result3}, \color{black}where they \color{black}were both customized on a source-by-source basis to best display the outflows in each pointing.  Here we adopt the same velocity range and contour levels for each pointing in a given region, since these figures are primarily intended to give context on the broader region in which each pointing is located, and we refer to Figures~\ref{fig_result1}~--~\ref{fig_result3} for maps optimized to show each detected outflow.

\section{Notes on Individual Sources with Measured Opening Angles}\label{sec_app_fit}

In this appendix we provide relevant notes on each outflow for which we obtain an opening angle measurement.  We note here that we applied our fitting method to both outflow lobes simultaneously in all cases except for those specifically noted below.

\begin{figure}
    \resizebox{\hsize}{!}{\includegraphics{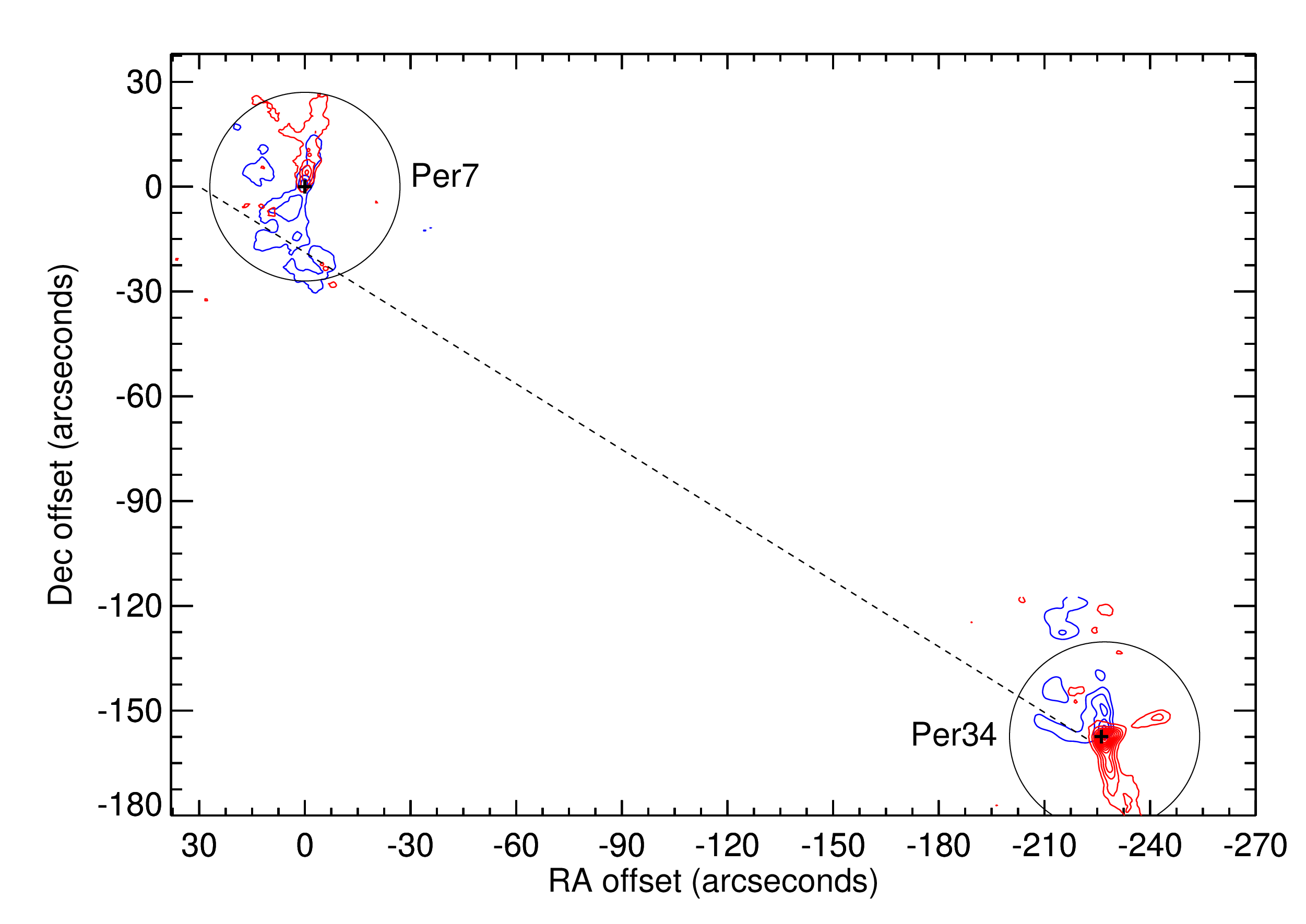}}
    \caption{Integrated redshifted and blueshifted \cojtwo\ emission associated with Per-emb~7 and Per-emb~34, integrated over the velocity ranges listed in Table \ref{tab_sample}.  The contours start at 2.5$\sigma$ and increase by 3$\sigma$, where $\sigma$ is the rms noise in each integrated map. The large circles show the primary beam of the SMA at 230~GHz, centered on each of the two pointings. The dashed line is drawn to guide the eye, and indicates that the blueshifted outflow lobe from Per-emb~34 is directed toward Per-emb~7.  The synthesized beams for each pointing are shown in Figures \ref{fig_result1} and \ref{fig_result3} (they are so small they would be nearly invisible on the scale of this figure).  The axes are plotted using coordinate offsets from the phase center of Per-emb~7.}
    \label{fig_per7per34}
\end{figure}

\vspace{0.1in}
\noindent \textbf{Per-emb~1: }
The Per-emb~1 outflow is well-detected in our MASSES \cojtwo\ observations.  While it is not well-resolved in our robust $+1$ image, it is well-resolved in our robust $-1$ image.  This outflow is well-fit by our method, thus we place it in the high confidence group.

\vspace{0.1in}
\noindent \textbf{Per-emb~2: }
Per-emb~2 is well-fit by our method, thus we place this object in the high confidence group.  While Per-emb~2 is a binary system with a projected separation of 24~au, as revealed by the \underline{V}L\underline{A} \underline{N}ascent \underline{D}isk \underline{A}nd \underline{M}ultiplicity Survey of Perseus Protostars \citep[VANDAM;][]{tobin2016:vandam,tobin2018:vandam}, we see no evidence of multiple outflows and thus assume that one component dominates both the outflow and the observed SED.

\vspace{0.1in}
\noindent \textbf{Per-emb~3: }
The Per-emb~3 outflow is well-detected in our MASSES \cojtwo\ observations.  While some of the low-velocity blueshifted channels (notably those at $\sim$3~\kms\ from rest) are affected by striping due to poor $uv$ coverage, we are able to obtain a reliable fit to by integrating over higher velocities. Given the clear outflow signature and the extent to which our fit agrees visually with the morphology of this outflow, we place this object in the high confidence group.

\vspace{0.1in}
\noindent \textbf{Per-emb~5: }
Per-emb~5 is is well-fit by our method, thus we place this object in the high confidence group.  As with Per-emb~2, Per-emb~5 is a binary system with a projected separation of 29~au
\citep{tobin2016:vandam,tobin2018:vandam}. With no evidence of multiple outflows in our data, we again assume that one component dominates both the outflow and observed SED.

\vspace{0.1in}
\noindent \textbf{Per-emb~6: }
The Per-emb~6 outflow is well-detected in our MASSES \cojtwo\ observations.  Despite being located in the B1 region where the CO emission is quite complex, this outflow is sufficiently isolated for identification and fitting (see Figure \ref{fig_mosaic_b1}).  While our fitting method provides a good visual match to the redshifted lobe, the match to the blueshifted lobe is somewhat more questionable, possibly due to contamination by other outflows in the B1 complex.  We thus place this object in the medium confidence group.

\vspace{0.1in}
\noindent \textbf{Per-emb~7: }
Per-emb~7 is associated with a highly collimated, redshifted outflow lobe that extends to the northwest of the protostar, but the blueshifted emission is significantly more complex.  While
there are some hints of a similarly collimated blueshifted lobe extending to the southeast, the emission is dominated by a diagonal, southwest-northeast stripe that is not clearly associated with Per-emb~7 itself.  As seen in Figure \ref{fig_per7per34}, this stripe is aligned with the blueshifted lobe of the outflow driven by Per-emb~34, located 4.5$\arcmin$ to the southwest.  Thus it is likely that the blueshifted emission in the vicinity of Per-emb~7 is contaminated by the outflow from Per-emb~34, as also noted by \citet{hatchell2009:outflows}.  As a result, we fit to only the redshifted outflow lobe.  Given that this lobe is only marginally resolved in our data, our measured opening angle is an upper limit to the actual opening angle (we are unable to use the higher-resolution robust $-1$ image for this outflow due to insufficient sensitivity).

\vspace{0.1in}
\noindent \textbf{Per-emb~9: }
The Per-emb~9 outflow is well-detected in our MASSES \cojtwo\ observations, where it displays an S-shaped symmetry in both lobes suggesting possible outflow precession.  While Per-emb~9 is single down to the VANDAM resolution limit of 20~au \citep{tobin2016:vandam,tobin2018:vandam}, it's possible there is an undetected close companion responsible for this apparent precession.  Additionally, the redshifted lobe features a northwestern extension of unknown origin approximately halfway down its length, but since this extension does not clearly point back to Per-emb~9, and is only marginally spatially connected to the southwestern redshifted outflow lobe, it is unlikely but ultimately unclear if it is indeed part of the Per-emb~9 outflow. In fitting the opening and position angles, we only fit close to the source (within 15$''$) where the curvature in each lobe has not yet become apparent, and we use the higher-resolution robust $-1$ image as the collimated outflow lobes are better resolved in this image.  While our fitting method provides a good visual match to the observed outflow morphology, given the uncertainties associated with this outflow, including the possible precession and the redshifted extension of unknown origin, we place Per-emb~9 in the medium confidence group.

\vspace{0.1in}
\noindent \textbf{Per-emb~10: }
The Per-emb~10 outflow is well-detected in our MASSES \cojtwo\ observations.  Like Per-emb~6, despite being located in the B1 region where the CO emission is quite complex, this outflow is sufficiently isolated for identification and fitting (see Figure \ref{fig_mosaic_b1}).  A companion object detected to the southeast by both \citet{pokhrel2018:masses} and \citet{stephens2018:masses}, and shown to be driving an outflow in the latter study, is in fact an artifact due to phase errors in the original dataset that have now been corrected \citep{stephens2019:masses}.  While the outflow is not well-resolved in our robust $+1$ image it is well-resolved and well-fit in our robust $-1$ image, thus we place this object in the high confidence group.

\vspace{0.1in}
\noindent \textbf{Per-emb~11: }
Per-emb~11 is a binary system with a projected angular separation of 2.95$\arcsec$, corresponding to a projected linear separation of 885~au \citep{tobin2016:vandam,tobin2018:vandam}. Both components in this binary system drive nearly aligned, overlapping, bipolar outflows, except with reversed redshifted and blueshifted lobes (the blueshifted lobe extends to the northwest [southeast] for the primary [secondary] source, and the redshifted lobe extends to the southeast [northwest] for the primary [secondary] source). It is worth pointing out that \citet{pech2012:jetrotation} interpreted these parallel outflows as evidence of jet rotation in lower resolution observations, although they did explicitly acknowledge the possibility of parallel outflows driven by unresolved (at the time) binary components.

Despite the spatial overlap the two outflows are well-separated in velocity, allowing us to apply our fitting procedure to the primary outflow (assumed to be driven by the same source that dominated the observed SED and thus dominates the measured \tbol).  Since our measured opening angle accurately reproduces the extent of the primary outflow, we place this source in the high confidence bin.  We do not fit to the secondary outflow since we do not have a reliable measurement of \tbol\ for this outflow's driving source.

A third protostar, IC348-SMM2E, is located at a projected separation of 9.47$\arcsec$ to the northeast of the primary component of the Per-emb~11 binary system, corresponding to a projected linear separation of 2841~au \citep{palau2014:ic348smm2e,tobin2016:vandam,tobin2018:vandam}. It drives a compact outflow first detected by \citet{palau2014:ic348smm2e}, and indeed the blueshifted lobe is visible in Figure \ref{fig_result1}.  However, as this outflow is unresolved in our data and contained entirely within 4$\arcsec$ of its driving source, we are unable to reliably apply our fitting procedure.

\vspace{0.1in}
\noindent \textbf{Per-emb~13: }
A compact outflow associated with Per-emb~13 is well-detected in our MASSES \cojtwo\ observations.  The northern redshifted lobe is contaminated by sidelobe emission associated with the nearby Per-emb~12 outflow, thus we fit to only the blueshifted emission.  The width of ths outflow is unresolved in both the robust~$+1$ and robust~$-1$ images, thus our measured opening angle is an upper limit to the true opening angle.

An additional Class 0 protostar, referred to by various names in the literature including IRAS4C \citep{looney2000:multiplicity}, IRAS4B2 \citep{hull2014:tadpol}, and IRAS4B$^\prime$ \citep{tobin2016:vandam}, is located 10.66$\arcsec$ east of Per-emb~13, corresponding to a projected linear separation of 3196~au \citep{looney2000:multiplicity,tobin2016:vandam,tobin2018:vandam}. It drives a weak, compact outflow whose redshifted lobe is weakly detected in our data, but as this outflow is unresolved in both dimensions and contaminated by sidelobe emission from the Per-emb~12 outflow, we are unable to measure its opening and position angles.

\vspace{0.1in}
\noindent \textbf{Per-emb~15: }
A bipolar displaying outflow a wide, shell-like morphology, consistent with the \cojone\ map presented by \citet{plunkett2013:ngc1333}, is clearly detected from Per-emb~15.  While the outflow morphology is clearly not conical, the opening angle near the source is very wide and our fitting method accurately reproduces this wide angle, leading us to place this object in the high confidence group.

\vspace{0.1in}
\noindent \textbf{Per-emb~16: }
Per-emb~16 is part of the Per-emb~16/28 wide binary system, with a projected separation of 16$\arcsec$ \citep[corresponding to a projected linear separation of  4819~au;][]{tobin2016:vandam,tobin2018:vandam}. The two components drive approximately perpendicular outflows \citep[e.g.,][]{lee2016:masses}, allowing us to easily separate their \cojtwo\ emission.  We use the robust~$-1$ image to best separate the two outflows, and we restrict our fitting method to only those pixels within 13$\arcsec$ of the driving source to avoid contamination from the Per-emb~28 outflow.  Our fitting method consistently gives a rather wide opening angle that provides a good fit very close to the driving source, but appears to be a poor fit farther from the source. This fact, coupled with possible asymmetries between the two lobes and contamination from the Per-emb~28 outflow, lead us to place this object into the low confidence group.

\vspace{0.1in}
\noindent \textbf{Per-emb~17: }
The outflow driven by Per-emb~17 is clearly detected and well-fit by our method, thus we place this object in the high confidence group.  Per-emb~17 is actually a binary system with a projected angular separation of 0.28$\arcsec$, corresponding to a projected linear separation of 83~au \citep{tobin2016:vandam,tobin2018:vandam}. As we see no evidence of multiple outflows, we assume that the same single component dominates both the outflow and the observed SED.   

\vspace{0.1in}
\noindent \textbf{Per-emb~19: }
While an outflow driven by Per-emb~19 is clearly detected, the two lobes display somewhat different morphologies.  The southeast, redshifted lobe appears to trace two cavity walls with a cleared interior, with an initially wide angle near the protostar and a more collimated appearance at larger distances along the lobe.  The northwest blueshifted lobe does not exhibit a clear cavity structure, and is instead dominated by a bright peak near the protostar.  As Per-emb~19 is located to the north of NGC~1333, it is possible that the blueshifted lobe is propagating into a less dense part of the cloud where there is less gas to entrain into the outflow, possibly explaining the lack of emission farther along the blue lobe.  While we fit to both lobes simultaneously and our fit accurately traces the wide morphology near the driving source (especially prevalent in the redshifted lobe), the asymmetry in the two lobes leads us to place this source in the medium confidence group.

\vspace{0.1in}
\noindent \textbf{Per-emb~20: }
An bipolar outflow driven by Per-emb~20 is clearly detected.  There is a redshifted emission peak detected to the north of the blueshifted lobe, visible in Figures \ref{fig_result2} and \ref{fig_mosaic_l1455}, that appears unrelated to the outflow and is thus excluded from our fit (it may be related to the Per-emb~17 outflow, but we are unable to conclusively determine its origin).  As both lobes of this outflow are well-fit by our method, we place this source in the high confidence group.

\vspace{0.1in}
\noindent \textbf{Per-emb~21: }
Per-emb~21 is part of the Per-emb~18/21 wide binary system, with a projected angular separation of 13.3$\arcsec$ \citep[corresponding to a projected linear separation of 3976~au;][]{tobin2016:vandam,tobin2018:vandam}. The MASSES data clearly detects a relatively wide-angle, bipolar outflow driven by Per-emb~21 \citep[see also][]{lee2016:masses}, but no outflow is detected from Per-emb~18.  The Per-emb~21 outflow displays a substantial asymmetry, with the southwestern, redshifted lobe appearing wider and at a somewhat different position angle than the northeastern, blueshifted lobe.  We choose to fit to the blueshifted lobe as our method obtains qualitatively better fits for this lobe.  Due to the substantial uncertainties associated with this asymmetrical outflow, along with possible confusion from an undetected Per-emb~21 outflow, we place this source in the low confidence group.

\vspace{0.1in}
\noindent \textbf{Per-emb~22: }
The outflow driven by Per-emb~22 is clearly detected and well-fit by our method, thus we place this object in the high confidence group.  Per-emb~22 is actually a binary system with a projected angular separation of 0.75$\arcsec$, corresponding to a projected linear separation of 225~au \citep{tobin2016:vandam,tobin2018:vandam}. As we see no evidence of multiple outflows, we assume that the same single component dominates both the outflow and the observed SED.   

\vspace{0.1in}
\noindent \textbf{Per-emb~23: }
The outflow driven by Per-emb~23 is clearly detected and well-fit by our method, thus we place this object in the high confidence group.  We use the robust~$-1$ image since this outflow is quite collimated and better resolved in this image.  The detection of both redshifted and blueshifted emission in each lobe suggests that this source may be viewed at an inclination close to edge-on.

\vspace{0.1in}
\noindent \textbf{Per-emb~24: }
The outflow driven by Per-emb~24 is clearly detected and well-fit by our method, thus we place this object in the high confidence group.

\vspace{0.1in}
\noindent \textbf{Per-emb~25: }
The outflow driven by Per-emb~25 is clearly detected and well-fit by our method, thus we place this object in the high confidence group.

\vspace{0.1in}
\noindent \textbf{Per-emb~26: }
Per-emb~26 is part of the Per-emb~26/42 wide binary system, with a projected separation of 8.1$\arcsec$ \citep[corresponding to a projected linear separation of $\sim$2431~au;][]{tobin2016:vandam,tobin2018:vandam}. The two components drive approximately perpendicular outflows that are thus able to be separated spatially.  Per-emb~26 drives the dominant outflow in this system, with a northwest blueshifted lobe and a southeast redshifted lobe.  As Per-emb~42 is located within the redshifted lobe of the Per-emb~26 outflow, at least in projection, and the outflow driven by Per-emb~42 overlaps spatially with this redshifted lobe, we fit to only the blueshifted lobe of the Per-emb~26 outflow.  Redshifted emission northeast of the Per-emb~26 protostar arises from one of the outflows driven by the Per-emb~33 system, as seen in Figure \ref{fig_mosaic_l1448}, and is thus omitted from our fit.  Given that the two lobes of the Per-emb~26 outflow are symmetrical and the blueshifted lobe is well-fit by our method, we place this object in the high confidence group.

\vspace{0.1in}
\noindent \textbf{Per-emb~27: }
Per-emb~27 is a binary system with a projected separation of 0.62$\arcsec$ \citep[corresponding to a projected linear separation of 186~au;][]{tobin2015:iras2a,tobin2016:vandam,tobin2018:vandam}.  The two components drive perpendicular outflows \citep[north-south for the primary and east-west for the secondary;][]{plunkett2013:ngc1333,tobin2015:iras2a}.  Since the SEDs of these components are unresolved at most wavelengths we fit only the primary outflow, under the assumption that the primary source dominates the observed SED (and thus the measured \tbol).  Inspection of our data, as well as the outflow maps presented by \citet{plunkett2013:ngc1333} and \citet{tobin2015:iras2a}, reveals an asymmetry between the two lobes, especially in terms of position angle.  However, the blueshifted lobe is contaminated by blueshifted emission from both the secondary outflow and the wide-angle outflow driven by Per-emb~36 \citep[see below, as well as][]{plunkett2013:ngc1333}, thus we only fit to the northern, redshifted lobe.  Given the uncertainties associated with both the asymmetry and the contamination from nearby outflows, we place this object in the medium confidence group.

\vspace{0.1in}
\noindent \textbf{Per-emb~28: }
Per-emb~26 is part of the Per-emb~16/28 wide binary system, with a projected separation of 16$\arcsec$ \citep[corresponding to a projected linear separation of  4819~au;][]{tobin2016:vandam,tobin2018:vandam}. The two components drive approximately perpendicular outflows \citep[e.g.,][]{lee2016:masses}, allowing us to easily separate their \cojtwo\ emission.  For Per-emb 28 we remove the central pixels where the two outflows overlap to avoid contamination from the Per-emb 16 outflow.  There is a change in position angle between the two lobes of this outflow, as also seen in the map presented by \citet{lee2016:masses}, possibly due to interaction between the two outflows.  We thus fit to only the northwestern outflow lobe, and note here that a fit to the southeastern lobe yields a different position angle but a consistent opening angle.  Due to the uncertainties from asymmetry and potential confusion, we place Per-emb~28 in the medium confidence group.

\vspace{0.1in}
\noindent \textbf{Per-emb~29: }
The Per-emb~29 outflow is well-detected in our MASSES \cojtwo\ observations, where it displays an S-shaped symmetry in both lobes suggesting possible outflow precession.  While Per-emb~29 is single down to the VANDAM resolution limit of 20~au \citep{tobin2016:vandam,tobin2018:vandam}, it's possible there is an undetected close companion causing this apparent precession.
As our method is incapable of fitting an outflow with a variable position angle, we restrict our fit to only the emission located within 15$\arcsec$ of the protostar in Right Ascension.  Given the uncertainties associated with this choice, we place Per-emb~29 in the medium confidence group. 

\vspace{0.1in}
\noindent \textbf{Per-emb~30: }
The outflow driven by Per-emb~30 is clearly detected and well-fit by our method, thus we place this object in the high confidence group.

\vspace{0.1in}
\noindent \textbf{Per-emb~31: }
The outflow driven by Per-emb~31 is clearly detected in our data. The northwest lobe is associated with redshifted emission that appears to be tracing resolved outflow cavity walls and spatially coincident blueshifted emission that likely indicates a nearly edge-on inclination for this source.  While there is blueshifted emission to the southeast at the expected position angle for this outflow, there is no matching redshifted emission.  Since the redshifted emission displays the widest morphology, with resolved cavity walls clearly evident, we only fit to the redshifted emission, and we place this source in the medium confidence group due to the uncertainties associated with the asymmetry.  We also note that Per-emb~31 is not detected in our MASSES 1.3~mm continuum observations, but spatially coincident emission is detected at both 8~mm in the VANDAM survey \citep{tobin2016:vandam} and in the infrared by \citet{enoch2009:protostars}.

\vspace{0.1in}
\noindent \textbf{Per-emb~33: }
The Per-emb~33 system (also referred to as the L1448N system or the L1448~IRS~3 system) is a hierarchical system of six protostars distributed in a close triple system, a close binary system, and a single system \citep{lee2015:masses,tobin2016:vandam}.  We are able to identify, separate, and reliably measure opening angles for two outflows in this system: L1448N-NW and L1448N-B.  These outflows are discussed in the appropriate paragraphs below.

\vspace{0.1in}
\noindent \textbf{Per-emb~34: }
The outflow driven by Per-emb~30 is clearly detected and generally well-fit by our method, although the morphology differs somewhat between the two lobes, with a more conical appearance in the redshifted lobe and a more parabolic appearance in the blueshifted lobe.  While our fit provies a good visual match to the image, to be conservative we place this object in the medium confidence group to account for uncertainties due to the asymmetry.

\vspace{0.1in}
\noindent \textbf{Per-emb~36: }
Per-emb~36 drives a very wide outflow with a blueshifted lobe clearly detected in our MASSES data that extends to the southwest and exhibits a shell-like morphology \citep[see also][]{plunkett2013:ngc1333}.  There is redshifted emission to the north, but it is associated with the secondary outflow from Per-emb~27 rather than the outflow driven by Per-emb~36.  Farther to the north, outside of the extent of our SMA observations, there is redshifted emission that likely arises from both the Per-emb~27 and Per-emb~36 outflows\citep[see][]{plunkett2013:ngc1333}. While our method provides a good visual match to the width of the outflow near the driving source, which indeed appears to span nearly the full 180\degree, the lack of a detected redshifted lobe with which to validate our results leads us to place this object in the medium confidence group. Finally, we note that this object is actually a binary system unresolved in our observations, with a projected linear separation of 93~au \citep{tobin2016:vandam,tobin2018:vandam}. As we see no evidence of multiple outflows, we assume that one component dominates both the outflow and the observed SED.

\vspace{0.1in}
\noindent \textbf{Per-emb~40: }
The outflow driven by Per-emb~40 is most apparent in our MASSES data as a conical redshifted lobe extending to the northwest.  Blueshifted emission from the Per-emb~6 outflow overlaps with this object (see Figure \ref{fig_mosaic_b1}, causing difficulty in identifying a clear blueshifted component to the Per-emb~40 outflow.  We thus fit to only the redshifted emission after manually cutting out the additional redshifted emission visible to the southeast of the protostar, as we are unable to determine the origin of this emission.  Finally, we also note that this object is actually a binary system unresolved in our observations, with a projected linear separation of 117~au \citep{tobin2016:vandam,tobin2018:vandam}.  As we see no evidence of multiple outflows, we assume that one component dominates both the outflow and the observed SED.  Given the various uncertainties associated with this source, we place it in the medium confidence group.

\vspace{0.1in}
\noindent \textbf{Per-emb~41: }
Per-emb~41 is part of the B1-b system, located approximately 14$\arcsec$ (4200~au) southwest of B1-bS, and approximately 25$\arcsec$ (7500~au) southwest  of B1-bN \citep[][]{tobin2016:vandam}. Per-emb~41 itself is a single system down to a projected separation of at least 20~au \citep{tobin2016:vandam,tobin2018:vandam}. As previously identified by both \citet{hirano2014:b1b} and \citet{lee2016:masses}, Per-emb~41 drives a northeast-southwest outflow, with blueshifted emission extending to the southwest and redshfited emission extending to the northeast.  Both of these lobes are detected in our MASSES data.  While the redshifted lobe eventually merges with confused redshifted emission from the other components of the B1-b system, a clear, compact outflow can be identified in both the redshifted and blueshifted lobes.  As both lobes are only marginally resolved in their lengths, and essentially unresolved in their widths, we are only able to measure an upper limit to the opening angle of this outflow.

\vspace{0.1in}
\noindent \textbf{Per-emb~42: }
Per-emb~42 is part of the Per-emb~26/42 wide binary system, with a projected separation of 8.1$\arcsec$ \citep[corresponding to a projected linear separation of $\sim$2431~au;][]{tobin2016:vandam,tobin2018:vandam}. The two components drive approximately perpendicular outflows that are thus able to be separated spatially.  A northeast-southwest outflow driven by Per-emb~42 is weakly detected in our MASSES data.  As the southwestern lobe overlaps substantially with the Per-emb~26 outflow, we only fit to the blueshifted emission in the northeastern lobe (as this lobe overlaps with redshifted emission from outflows driven by the Per-emb~33 system [see Figure \ref{fig_mosaic_l1448}], we do not include redshifted emission in our fit).  While our fit provides an excellent visual match to the northeastern lobe, the uncertainties associated with the overlapping emission from the multiple outflows lead us to place this object in the medium confidence group.

\begin{figure}
    \resizebox{\hsize}{!}{\includegraphics{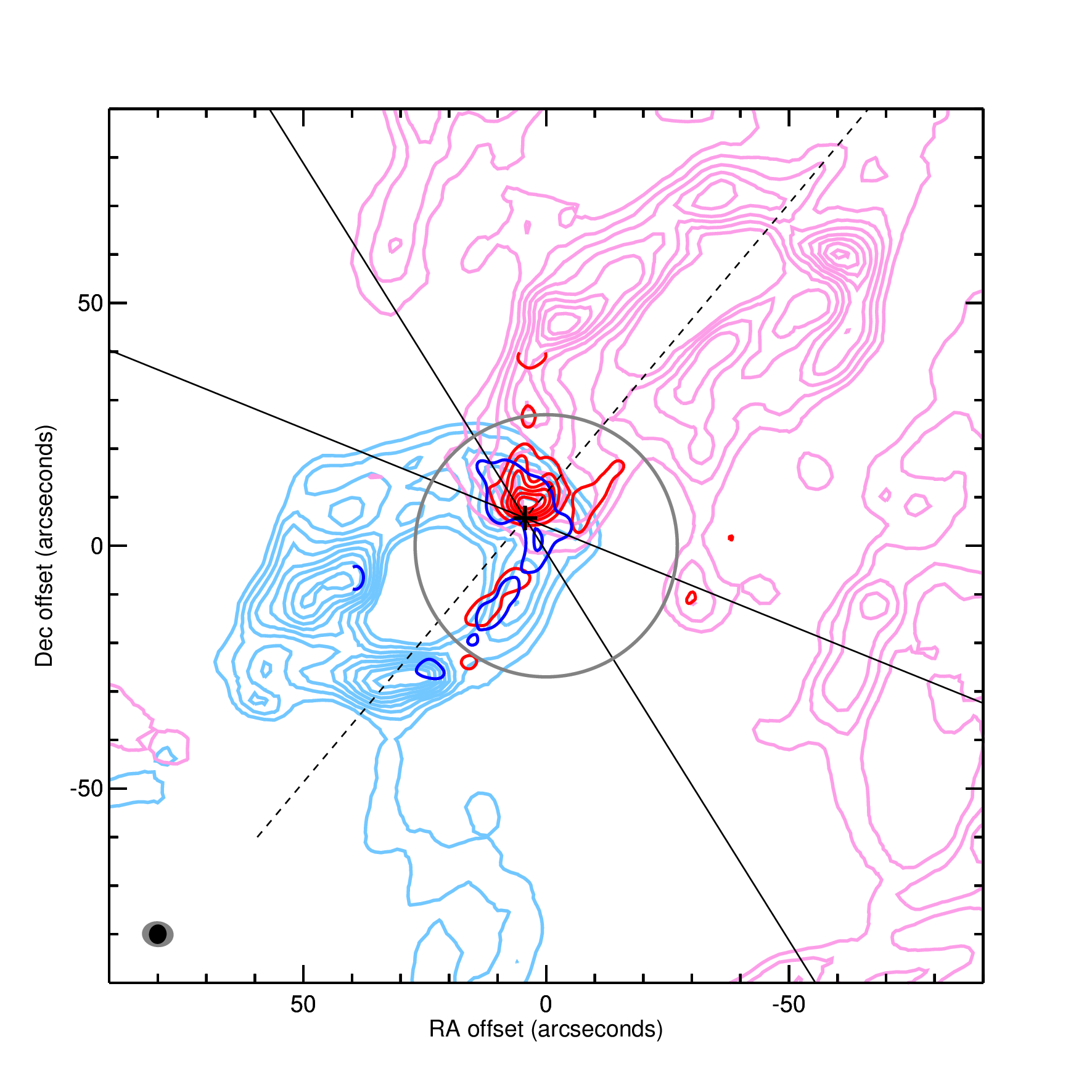}}
    \caption{\color{black}Integrated redshifted (red contours) and blueshifted (dark blue contours) MASSES \cojtwo\ emission associated with Per-emb~44, with the same velocity ranges and contour levels as in Figure \ref{fig_result2}.  The large gray circle shows the primary beam of the SMA, the cross shows the position of the Per-emb~44 protostar, and the dashed and solid lines show the position and opening angles of the Per-emb~44 outflow from our fit (these are the same lines shown in Figure \ref{fig_result2}).  Overplotted are redshifted (pink contours) and blueshifted (light blue contours) \cojone\ emission from \citet{plunkett2013:ngc1333}.  The synthesized beams for the  MASSES (black) and \citet{plunkett2013:ngc1333} (gray) data are plotted as filled ovals in the lower left corner.  The axes are plotted using coordinate offsets from the phase center of the MASSES observations of Per-emb~44.\color{black}}    
    \label{fig_per44}
\end{figure}

\vspace{0.1in}
\noindent \textbf{Per-emb~44: }
Per-emb~44 drives a wide, bipolar, shell-like outflow, with blueshifted emission extending to the southeast and redshifted emission extending to the northwest \color{black}\citep[][see, in particular, their Figures 4 and 5]{plunkett2013:ngc1333}.  \color{black}Only \color{black}patchy emission from \color{black}the inner portion of this outflow is \color{black}detected in our MASSES data.  We \color{black}fit only to the blueshifted lobe since the \color{black}very wide, \color{black}shell-like structure is more \color{black}apparent \color{black}in this lobe, \color{black}but we note that \color{black}a fit to the redshifted lobe gives a similar opening angle.  \color{black}Figure \ref{fig_per44} repeats the Per-emb~44 panel from Figure \ref{fig_result2} except with modified axes ranges to show a larger region and with the \cojone\ contours from \citet{plunkett2013:ngc1333} overplotted.  This figure emphasizes that the base of each lobe is well-resolved compared to the synthesized beams of the MASSES and \citet{plunkett2013:ngc1333} datasets, confirming that this outflow is indeed very wide at its base.  Due to the uncertainties associated with fitting to the very patchy MASSES \cojtwo\ emission, \color{black}we place this object in the medium confidence group.  Finally, we note that Per-emb~44 is actually a binary system with a projected separation of 0.3$\arcsec$ \citep[corresponding to a projected linear separation of 90~au][]{tobin2016:vandam,tobin2018:vandam}.  As we see no evidence of multiple outflows, we assume that one component dominates both the outflow and the observed SED.  

\vspace{0.1in}
\noindent \textbf{Per-emb~46: }
An outflow driven by Per-emb~46 is clearly detected in our MASSES data, with the blueshifted lobe exhibiting a resolved cavity morphology but only one of the two cavity walls robustly detected in the redshfited lobe.  As the ``missing'' redshifted cavity wall is faintly present in individual velocity channels, and our method provides a good visual fit to the blueshifted lobe, we place this object in the high confidence group.

\begin{figure}
    \resizebox{\hsize}{!}{\includegraphics{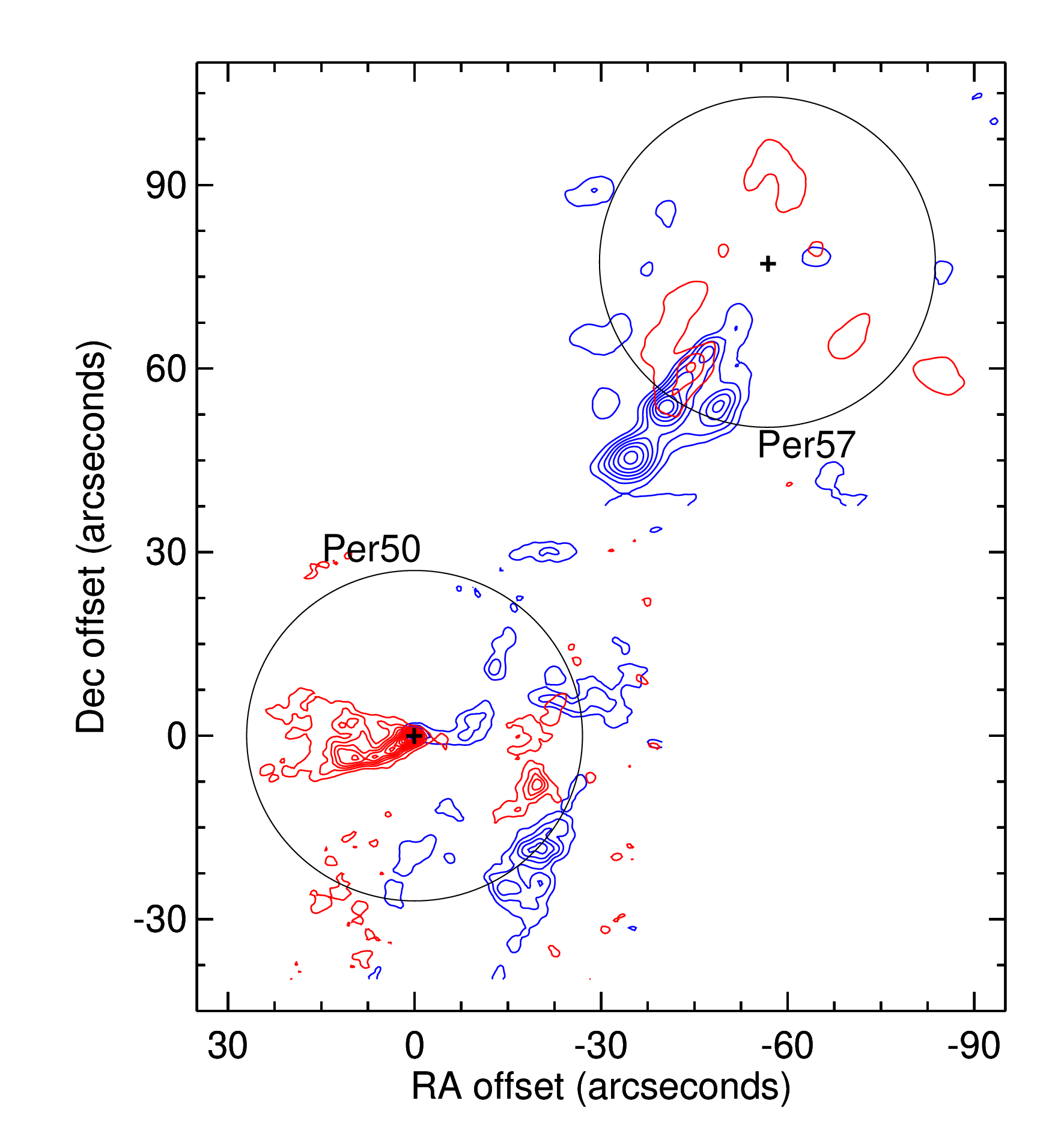}}
    \caption{Integrated redshifted and blueshifted \cojtwo\ emission associated with Per-emb~50 and Per-emb~57, integrated over the velocity ranges listed in Table \ref{tab_sample}.  The contours start at 2$\sigma$ and increase by 3$\sigma$, where $\sigma$ is the rms noise in each integrated map. The large circles show the primary beam of the SMA at 230~GHz, centered on each of the two pointings. The synthesized beams for each pointing are shown in Figure \ref{fig_result3} (they are so small they would be nearly invisible on the scale of this figure).  The axes are plotted using coordinate offsets from the phase center of Per-emb~50.}
    \label{fig_per50per57}
\end{figure}

\vspace{0.1in}
\noindent \textbf{Per-emb~50: }
An outflow driven by Per-emb~50 is clearly detected in our MASSES data.  Close to the driving source a classic bipolar outflow oriented east-west is clearly visible.  Farther to the south and west there is additional blueshifted emission detected that is not clearly part of this outflow. A larger map reveals that most or all of this additional emission is likely associated with the 
Per-emb~57 outflow (see Figure \ref{fig_per50per57}), and is thus excluded from our fit.  Given the clear outflow signature and the extent to which our fit agrees visually with the morphology of this outflow, we place this object in the high confidence group.

\vspace{0.1in}
\noindent \textbf{Per-emb~52: }
A fairly colllimated, bipolar, northeast-southwest outflow driven by Per-emb~52 is detected in our MASSES data, with the northeast lobe robustly detected but the southwest lobe only marginally detected in a few individual velocity channels.  While our method does provide a reliable fit to the blueshifted emission, given the overall weak emission and the near total absence of the redshifted lobe, we place this object in the medium confidence group.

\vspace{0.1in}
\noindent \textbf{Per-emb~53: }
Per-emb~53 is clearly associated with a bipolar molecular outflow.  Our method accurately recovers the opening angle of this outflow close to the driving source, where the outflow is widest.  We thus place this object in the high confidence group.

\vspace{0.1in}
\noindent \textbf{Per-emb~56: }
The outflow driven by Per-emb~56 is clearly detected and well-fit by our method, thus we place this object in the high confidence group.

\vspace{0.1in}
\noindent \textbf{Per-emb~57: }
A collimated, jet-like outflow feature is detected south of the Per-emb~57 protostar, extending on a line that traces back to the protostar.  Assuming this emission is indeed associated with an outflow driven by Per-emb~57, our method accurately fits the morphology of this jet-lie feature.  However, the fact that this feature does not extend all the way back to the protostar, coupled with the presence of only weak emission to the northwest that is not consistent with being the counterpart to this feature, lead us to place this object in the low confidence group.

\vspace{0.1in}
\noindent \textbf{Per-emb~61: }
A bipolar north-south outflow driven by Per-emb~61 is detected in our MASSES data.  While each lobe appears fairly collimated at first glance, the position angles of the two lobes appear misaligned.  An alternative interpretation is that the outflow is wider than implied by either lobe individually, with only one of the two cavity walls detected in each lobe.  We favor this second interpretation based on the detection, in individual velocity channels, of weak, patchy emission where the other cavity walls would be present.  Since we can't fully rule out a more collimataed outflow with misaligned lobes, we place this object in the medium confidence group.

\vspace{0.1in}
\noindent \textbf{Per-emb~62: }
The outflow driven by Per-emb~62 is clearly detected, although only one of the two cavity walls appears to be deteted in the southern lobe.  As this outflow is well-fit by our method, we place this object in the high confidence group.  

\vspace{0.1in}
\noindent \textbf{B1-bS: }
B1-bS, either a very young Class 0 protostar or a first hydrostatic core, is located in the B1-b system \citep{pezzuto2012:fhsc} and drives a southeast (blueshifted) -- northwest (redshifted) bipolar outflow, with the redshifted outflow not detected in CO due to confusion with emission from other components in the B1-b system but detected in other tracers \citep{hirano2014:b1b,lee2016:masses,gerin2015:b1b}.  Since only the blueshifted lobe can be clearly identified and separated in our \cojtwo\ observations, we only apply our fitting method to this lobe.  While the robust $+1$ image only marginally resolves the length of this outflow and does not resolve its width, the  sensitivity in the robust $-1$ image is too low to obtain a reliable fit.  Thus we apply our fitting method to the robust $+1$ image and present our opening angle as an upper limit.

\vspace{0.1in}
\noindent \textbf{L1448N-NW: }
L1448N-NW is part of the Per-emb~33 system (also referred to as the L1448N system or the L1448~IRS~3 system), a hierarchical system of six protostars distributed in a close triple system (L1448N-B), a close binary system (L1448N-A), and a single system (L1448N-NW) \citep{lee2015:masses,tobin2016:vandam}.  The MASSES \cojtwo\ observations show a complicated tangle of outflow emission from the six protostars in this system, as well as from the nearby Per-emb~26 outflow \citep[see Figure \ref{fig_mosaic_l1448} and][]{lee2015:masses}.  Despite the confusion in this region the outflow from L1448N-NW can be clearly identified and separated from the other emission components, with a southeast redshifted lobe and a northwest blueshifted lobe.  As the driving source is near the edge of the primary beam the blueshifted lobe (which extends away from the phase center) is not well-detected, thus we only fit to the redshifted lobe.  We still include this outflow in the high confidence group due to its simply morphology and the excellent visual match between the fit and the image.  We note that L1448N-NW is actually a binary system with a projected separation of 0.25$\arcsec$ \citep[corresponding to a projected linear separation of 75~au ][]{tobin2016:vandam,tobin2018:vandam}. As we see no evidence of multiple outflows, we assume that one component dominates both the outflow and the observed spectral energy distribution. 

\vspace{0.1in}
\noindent \textbf{L1448N-B: }
L1448N-B is part of the Per-emb~33 system (also referred to as the L1448N system or the L1448~IRS~3 system), a hierarchical system of six protostars distributed in a close triple system (L1448N-B), a close binary system (L1448N-A), and a single system (L1448N-NW) \citep{lee2015:masses,tobin2016:vandam}.  The MASSES \cojtwo\ observations show a complicated tangle of outflow emission from the six protostars in this system, as well as from the nearby Per-emb~26 outflow \citep[see Figure \ref{fig_mosaic_l1448} and][]{lee2015:masses}.  Despite this complexity, we are able to clearly identify an outflow driven by the L1448N-B triple system.  The southeastern redshifted lobe is particularly clear, whereas the northwestern blueshifted lobe is likely contaminated by blueshifted emission from the Per-emb~26 outflow (see Figure \ref{fig_mosaic_l1448}.  Thus we fit only the redshifted lobe, and assume that both the outflow and the observed SED are dominated by the same single member of this triple system.  Given the uncertainties in these assumptions, we place this outflow in the medium confidence group.

\vspace{0.1in}
\noindent \textbf{Per-Bolo~45: }
Per-Bolo~45 was identified as a candidate first hydrostatic core based first on the detection of compact 3~mm continuum emission \citep{schnee2010:starless} and then on the tentative detection of an outflow in SiO~(2--1) emission \citep{schnee2012:starless}.  
We detect blueshifted \cojtwo\ emission to the southeast of the compact continuum source that is broadly consistent with the location, morphology, and velocities of the outflow detection claimed by \citet{schnee2012:starless}.  While this blueshifted feature does not extend all the way back to the position of the continuum source, the long axis of the extended morphology does trace back to the continuum source. We do not detect a corresponding redshifted feature to the northwest, but this could possibly be due to the significant offset between the phase center of the observations and the position of the compact continuum source. While the general agreement with the claimed outflow by \citet{schnee2012:starless} is encouraging, the highly tentative nature of this detection leads us to place this object in the low confidence group, a point reinforced by the fact that \citet{maureira2020:fhsc} argued in favor of a prestellar nature for this object.

\vspace{0.1in}
\noindent \textbf{Per-Bolo~58: }
Our MASSES data detects a known bipolar outflow driven by Per-Bolo~58, first identified as a candidate first hydrostatic core \citep{enoch2010:fhsc} but currently believed to be a very young Class 0 protostar \citep{maureira2020:fhsc}.  Our outflow detection matches that of \citet{dunham2011:fhsc}.  Additional blueshifted emission detected to the north of the redshifted lobe, and to the south of the blueshifted lobe, likely arises from the outflow driven by Per-emb~23 and is excluded from our analysis.  Given the extent to which our fit agrees visually with the morphology of this outflow, we place this object in the high confidence group.

\vspace{0.1in}
\noindent \textbf{SVS~13C: }
SVS~13C is part of the SVS~13 system, located approximately 20$\arcsec$ (6000~au) southwest of SVS~13A and approximately 9$\arcsec$ (2700~au) southwest of SVS~13B \citep{looney2000:multiplicity}.  \citet{plunkett2013:ngc1333} identified a tentative outflow extending approximately north-south, with the southern lobe associated with weak blueshifted emission that aligns with a scattered light cavity seen in a 4.5 \um\ image from the {\it Spitzer Space Telescope}.  The counter lobe extending to the north is less clearly detected, although there is substantial confusion in the CO emission in this region \citep[see][for details]{plunkett2013:ngc1333}.  In our MASSES data we detect blueshifted emission extending north and south of SVS~13C.  To the south, this emission aligns well with the scattered light cavity seen in the \spitzer\ data \citep[see, in particular, Figure 8 of][]{plunkett2013:ngc1333}, and with the axis of the much larger-scale outflow identified by Plunkett et al.  We also detect redshifted emission to the northwest of the protostar, but like Plunkett et al.~are unable to conclusively identify the origin of this emission.  We fit only the blueshifted emission, but given the substantial uncertainties in interpreting this complicated environment, we place this object in the low confidence group.

\begin{figure*}
    \resizebox{\hsize}{!}{\includegraphics{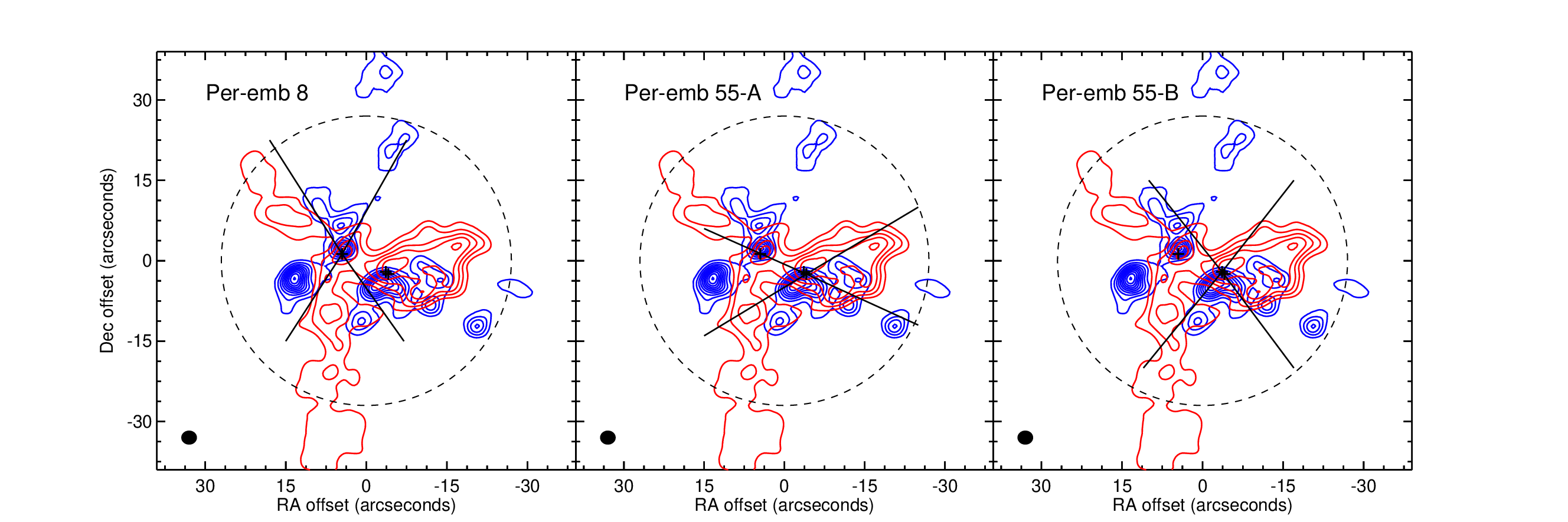}}
    \caption{MASSES \cojtwo\ integrated emission contours for the Per-emb~8/55 protostellar system.  The contours, which are the same in all three panels, show blueshifted (blue contours) and redshifted (red contours) integrated over the LSR velocity ranges of 5.5--7 (blue) and 11.5--13 (red) \kms.  The contours start at 3$\sigma$ and increase by 3$\sigma$, where $\sigma$ is the rms noise in each integrated map.  The large dashed circle shows the primary beam of the SMA at 230~GHz, centered on the phase center.  The synthesized beam is shown in the lower left corner of each panel, and the crosses mark the positions of the protostars.  The solid lines in each panel indicate the approximate morphologies of the outflows we have identified as being driven by each of the three protostars.  These lines are not fits, they are drawn by hand and are only meant to serve as a rough guide.}
    \label{fig_per8per55}
\end{figure*}

\section{Notes on Individual Sources without Measured Opening Angles}\label{sec_app_nofit}

In this appendix we provide relevant notes on each source in the MASSES sample for which we are unable to obtain opening angle measurements. In general we do not show images for these objects and instead refer to \citet{stephens2018:masses} and \citet{stephens2019:masses}, where the full set of MASSES \cojtwo\ images are displayed.

\vspace{0.1in}
\noindent \textbf{Per-emb~4: }
Per-emb~4 was identified as a protostar by \citet{enoch2009:protostars} based on the association of a mid-infrared \spitzer\ source with a 1.1~mm Bolocam continuum detection.  However, as first noted by \citet{stephens2018:masses}, the protostellar status of this object is questionable due to its non-detection in our MASSES 230~GHz continuum observations \citep{pokhrel2018:masses}, our MASSES \cooojtwo\ observations \citep{stephens2018:masses}, and in the VANDAM 8~mm continuum observations \citep{tobin2016:vandam}.  While we detect some redshifted emission in the vicinity of Per-emb~4, up to velocities of at least 5~\kms\ from the systemic cloud velocity, we do not detect any clear signatures of an outflow associated with this object.  We thus tentatively conclude that there is no outflow driven by this object detected in our data.

\vspace{0.1in}
\noindent \textbf{Per-emb~8/55: }
Per-emb~8 and Per-emb~55 form a wide binary system, with a projected separation of 9.6$\arcsec$ \citep[corresponding to a projected linear separation of 2867~au;][]{tobin2016:vandam,tobin2018:vandam}. Per-emb~55 is additionally a close binary system, with a projected linear separation of 185~au.  Figure \ref{fig_per8per55} plots contours displaying the integrated blueshifted and redshifted \cojtwo\ emission over LSR velocity ranges of 5.5--7 and 11.5--13 \kms, respectively.  Careful inspection of these contours, along with individual channel maps (not shown), reveals substantial but ultimately inconclusive evidence for three distinct outflows in this system \color{black}\citep[see also the higher resolution \cojtwo\ ALMA maps presented by][]{tobin2018:vandam}.  \color{black}Lines are drawn in Figure \ref{fig_per8per55} to indicate the approximate opening angles of these outflows.  These lines are drawn by eye and should not be interpreted as fits to the data.  While these outflows do generally agree with the detections reported previously by \citet{walawender2006:outflows}, we are unable to obtain even low confidence fits given the extensive degree to which they overlap with each other.  Thus we do not include these outflows in our analysis.

\vspace{0.1in}
\noindent \textbf{Per-emb~12: }
Per-emb~12 (NGC1333~IRAS~4A) is a binary system with a projected angular separation of 1.8$\arcsec$, corresponding to a projected linear separation of 549~au \citep[e.g.,][]{tobin2016:vandam,tobin2018:vandam}. Both components drive outflows that overlap and likely interact with each other \citep[e.g.,][]{santangelo2015:iras4a,ching2016:iras4a}.  As these two components are impossible to fully disentangle due to their overlapping nature, and furthermore given that the component that dominates the large-scale outflow emission is not the component that dominates the observed infrared and (sub)millimeter spectral energy distribution \citep{looney2000:multiplicity,reipurth2002:multiplicity,enoch2009:protostars,santangelo2015:iras4a}, we are unable to reliably measure opening angles and bolometric temperatures for either component.

\vspace{0.1in}
\noindent \textbf{Per-emb~14: }
No outflow was detected from Per-emb~14.  Similarly, no outflow was detected from this target in the large-scale map of NGC~1333 outflows presented by \citet{plunkett2013:ngc1333}.

\vspace{0.1in}
\noindent \textbf{Per-emb~18: }
Per-emb~18 is part of the Per-emb~18/21 wide binary system, with a projected angular separation of 13.3$\arcsec$ \citep[corresponding to a projected linear separation of 3976~au;][]{tobin2016:vandam,tobin2018:vandam}. We do detect a clear bipolar outflow driven by Per-emb~21
\citep[see also][]{lee2016:masses}, but not for Per-emb~18.  While there is some evidence for such an outflow in the MASSES \cojthree\ data and {\it Spitzer--IRAC} 4.5~\um\ image \citep{lee2016:masses}, \color{black}as well as in higher-resolution \cojtwo\ ALMA maps presented by \citet{tobin2018:vandam}, \color{black}we are unable to clearly identify, separate, and resolve this outflow.

\vspace{0.1in}
\noindent \textbf{Per-emb~32: }
Per-emb~32, located within IC348, is a binary system with a projected separation of 6$\arcsec$ \citep[corresponding to a projected linear separation of 1820~au;][]{evans2009:c2d,dunham2015:gb,young2015:perseus,tobin2016:vandam,tobin2018:vandam}.  Per-emb~32 is associated with both blueshifted and redshifted \cojtwo\ emission at velocities exceeding 3~\kms\ relative to the systemic cloud velocity, and while this emission is not clearly associated with other outflows in IC348 (see Figure \ref{fig_mosaic_ic348}), we are unable to clearly identify outflows from one or both components of this binary system. 

\vspace{0.1in}
\noindent \textbf{Per-emb~35: }
Per-emb~35, located near the western edge of NGC~1333, is a binary system with a projected separation of 1.9$\arcsec$ \citep[corresponding to a projected separation of 572~au;][]{tobin2016:vandam,tobin2018:vandam}. \color{black}\citet{tobin2018:vandam} present evidence for approximately parallel southeast-northwest outflows driven by the two components of this binary system using higher-resolution \cojtwo\ ALMA data, but we are unable to adequately resolve these separate outflows in our MASSES data.  \color{black}Thus, despite the likely presence of one or more outflows, we are unable to conclusively measure position and opening angles for this system.

\vspace{0.1in}
\noindent \textbf{Per-emb~37: }
Per-emb~37, located in the northeastern portion of NGC~1333, is a wide binary system with a projected separation of 10.6$\arcsec$ \citep[corresponding to a projected linear separation of 3167~au;][]{young2015:perseus,tobin2016:vandam,tobin2018:vandam}. Both components are detected in the VANDAM 8~mm continuum observations \citep{tobin2016:vandam}, but only one is detected in the MASSES 1.3~mm continuum observations \citep{pokhrel2018:masses,stephens2018:masses}.  While there is clearly high-velocity emission detected in the MASSES \cojtwo\ observations, the complexity of this region prevents us from clearly identifying individual outflows.

\vspace{0.1in}
\noindent \textbf{Per-emb~38: }
Per-emb~38, located 11$\arcmin$ (0.96~pc) southwest of the B1 region, is single down to a projected separation of at least 20~au \citep{tobin2016:vandam,tobin2018:vandam}.  This object is asociated with high-velocity \cojtwo\ emission in our MASSES data, and given the isolated nature of this protostar it is highly likely that the emission originates in an outflow driven by Per-emb~38.  However, we are unable to clearly identify distinct outflow lobes.

\vspace{0.1in}
\noindent \textbf{Per-emb~39: }
Per-emb~39 is not detected in the 1.3~mm continuum or \cooojtwo\ MASSES observations \citep{stephens2018:masses}, nor is it detected in the VANDAM 8~mm continuum \citep{tobin2016:vandam}.  It also has an extremely low and highly uncertain bolometric luminosity \citep{enoch2009:protostars,stephens2018:masses}.  For these reasons, \citet{stephens2018:masses} questioned the protostellar status of this object.  While we detect \cojtwo\ emission in the vicinity of of Per-emb~39 in our MASSES data up to $\sim$5~\kms\ from rest, particularly at redshifted velocities, we are unable to conclusively identify the origin of this emission  Furthermore, while \citet{hatchell2009:outflows} claim the detection of an outflow associated with this object, the claim is based only on the presence of high-velocity emission, and indeed their Figure~3 shows no clear morphological evidence for an outflow.  We are thus unable to clearly identify and fit to an outflow.

\vspace{0.1in}
\noindent \textbf{Per-emb~43: }
Per-emb~43 was identified as a protostar by \citet{enoch2009:protostars} based on the association of a mid-infrared \spitzer\ source with a 1.1~mm Bolocam continuum detection.  However, as first noted by \citet{stephens2018:masses}, the protostellar status of this object is questionable due to its non-detection in our MASSES 230~GHz continuum observations \citep{pokhrel2018:masses}, our MASSES 230~GHz \cooojtwo\ observations \citep{stephens2018:masses}, and in the 8~mm VANDAM observations \citep{tobin2016:vandam}.  We see no evidence for an outflow in our MASSES \cojtwo\ observations \citep[see also Fig.~3 of ][for a similar result]{stephens2018:masses}. 

\vspace{0.1in}
\noindent \textbf{Per-emb~45: }
Per-emb~45 was identified as a protostar by \citet{enoch2009:protostars} based on the association of a mid-infrared \spitzer\ source with a 1.1~mm Bolocam continuum detection. However, similar to Per-emb~43, the status of Per-emb~45 as a protostar was noted as questionable by \citet{stephens2018:masses} due to its non-detection in our MASSES 230~GHz continuum observations
\citep{pokhrel2018:masses}, our MASSES 230~GHz \cooojtwo\ observations \citep{stephens2018:masses}, and in the 8~mm VANDAM observations \citep{tobin2016:vandam}.  We see no evidence for an outflow in our MASSES \cojtwo\ observations, as seen in Figure \ref{fig_mosaic_b1}. 

\vspace{0.1in}
\noindent \textbf{Per-emb~47: }
We detect substantial redshifted emission in the vicinity of Per-emb~47 in our MASSES \cojtwo\ observations, extending up to $\sim$10~\kms\ from rest.  \color{black}While this emission is likely associated with an outflow, as seen by comparing to the single-dish maps presented by \citet[][see in particular their Figure~5]{hatchell2009:outflows}, the morphology of the emission detected by MASSES is not clearly indicative of an outflow (see \S \ref{sec_discussion_limitations_widebias} for further discussion of this object).  We thus do not include this object in our analysis.\color{black}

\vspace{0.1in}
\noindent \textbf{Per-emb~48: }
Per-emb~48, located in the L1455 region of Perseus (see Figure \ref{fig_mosaic_l1455}), is a binary system with a projected angular separation of 0.35$\arcsec$ \citep[corresponding to a projected linear separation of 104~au;][]{tobin2016:vandam,tobin2018:vandam}. While we detect \cojtwo\ emission at velocities exceeding 3~\kms\ from the systemic velocity, we are unable to identify clear signatures of one or more outflows driven by this system.  There is also no outflow clearly associated with this system in the single-dish maps presented by \citet{curtis2010:outflows}.  Furthermore, we note that the blueshifted lobe of the Per-emb~20 outflow is pointed toward Per-emb~48 (see Figure \ref{fig_mosaic_l1455}), rendering it unclear if the blueshifted emission in the vicinity of Per-emb~48 is actually associated with this system or with Per-emb~20.

\vspace{0.1in}
\noindent \textbf{Per-emb~49: }
Per-emb~49 is a binary system with a projected angular separation of 0.313$\arcsec$ \citep[corresponding to a projected linear separation of 94~au;][]{tobin2016:vandam,tobin2018:vandam}.  While undetected in our MASSES 1.3~mm continuum observations, we do detect evidence for a northeast-southwest outflow driven by this sytem.  However, as our map is dominated by sidelobe emission from the nearby Per-emb~21 outflow (see Figure \ref{fig_mosaic_ngc1333}), we are unable to reliably fit for the opening and position angles of this outflow.

\vspace{0.1in}
\noindent \textbf{Per-emb~51: }
We do not detect any evidence of an outflow driven by Per-emb~51 in our MASSES \cojtwo\ observations.

\vspace{0.1in}
\noindent \textbf{Per-emb~54: }
Per-emb~54, located in NGC~1333, is detected as a single source by the VANDAM survey at 8~mm \citep{tobin2016:vandam} but is not associated detected in our MASSES 230~GHz continuum observations \citep{pokhrel2018:masses,stephens2018:masses}. We detect high-velocity \cojtwo\ emission in our MASSES observations, extending up to $\sim$10~\kms\ from the systemic velocity.  This emission, which is also detected at much lower resolution by \citet{hatchell2007:outflows} and
\citet{curtis2010:outflows}, does not clearly originate from any other object in NGC~1333 and is thus likely indicative of an outflow driven by Per-emb~54.  However, we are unable to identify any clear morphological signatures of an outflow.

\vspace{0.1in}
\noindent \textbf{Per-emb~58/65: }
Per-emb~58 and Per-emb~65, located in NGC~1333, are separated by 29$\arcsec$ \citep[corresponding to a projected linear separation of 8700~au;][]{tobin2016:vandam,tobin2018:vandam}. Additionally, there are several Class II YSOs in the vicinity of these two protostars \citep[e.g.,][]{young2015:perseus}. While there is copious high-velocity \cojtwo\ emission detected in the vicinity of these objects \citep[see, e.g., Figure~3 from][]{stephens2018:masses}, indicating the likely presence of one or more outflows, we are unable to definitively identify individual outflows or assign driving sources.  

\vspace{0.1in}
\noindent \textbf{Per-emb~59: }
Per-emb~59 was identified as a protostar by \citet{enoch2009:protostars} based on the association of a mid-infrared \spitzer\ source with a 1.1~mm Bolocam continuum detection. However, as first noted by \citet{stephens2018:masses}, the protostellar status of this object is questionable due to its non-detection in our MASSES 230~GHz continuum observations \citep{pokhrel2018:masses}, our MASSES 230~GHz \cooojtwo\ observations \citep{stephens2018:masses}, and in the 8~mm VANDAM observations \citep{tobin2016:vandam}.  We see no evidence for an outflow in our MASSES \cojtwo\ observations \citep[see also Fig.~3 of ][for a similar result]{stephens2018:masses}. 

\vspace{0.1in}
\noindent \textbf{Per-emb~60: }
Per-emb~60 is not detected in the 1.3~mm continuum or \cooojtwo\ MASSES observations \citep{stephens2018:masses}, nor is it detected in the VANDAM 8~mm continuum \citep{tobin2016:vandam}.  It also has an extremely low and highly uncertain bolometric luminosity \citep{enoch2009:protostars,stephens2018:masses}.  For these reasons, \citet{stephens2018:masses} questioned the protostellar status of this object, although they did note the presence of relatively compact emission in single-dish continuum maps \citep[e.g.,][]{suresh2016:sharcii}.  We do not detect any evidence for an outflow driven by this object.

\vspace{0.1in}
\noindent \textbf{Per-emb~63: }
While we detect high-velocity \cojtwo\ emission in the vicinity of Per-emb~63, we do not detect any conclusive morphological evidence for an outflow driven by this object.

\vspace{0.1in}
\noindent \textbf{Per-emb~64: }
While we detect high-velocity \cojtwo\ emission in the vicinity of Per-emb~64, we do not detect any conclusive morphological evidence for an outflow driven by this object.

\vspace{0.1in}
\noindent \textbf{Per-emb~66: }
Per-emb~66, identified as a protostar by \citet{enoch2009:protostars}, is not associated with continuum detections by MASSES at 1.3~mm \citep{pokhrel2018:masses,stephens2018:masses} or by VANDAM at 8~mm \citep{tobin2016:vandam}.  We detect a compact structure in \cojtwo\ that could be tracing either a compact outflow or a large, rotating structure \citep[see also Figure~3 of][]{stephens2018:masses}, but we are unable to distinguish between these possibilities.  Even if we assumed it is tracing an outflow, all of the emission is within the 4$\arcsec$ radius from the driving source over which we mask emission to avoid excessive beam-smearing effects.

\vspace{0.1in}
\noindent \textbf{B1-bN: }
B1-bN, first identified by \citet{pezzuto2012:fhsc}, is either a very young Class 0 protostar or a first hydrostatic core.  \citet{gerin2015:b1b} detected a compact, east-west outflow driven by B1-bN in their methanol and formaldehyde observation, and while we see signs of this outflow in our \cojtwo\ observations, we are unable to clearly identify and separate this outflow in either position or velocity.  

\vspace{0.1in}
\noindent \textbf{L1448~IRS2E: }
L1448~IRS2E was first identified as a candidate first hydrostatic core by \citet{chen2010:fhsc}, who detected a compact 1.3~mm continuum source and a redshifted, jet-like feature in \cojtwo\ emission.  This redshifted feature is located within the redshifted lobe of the nearby Per-emb~22 outflow (see Figure \ref{fig_mosaic_l1448}) and is detected up to velocities of 25~\kms\ \citep[much too fast for outflows driven by first hydrostatic cores; e.g.,][]{machida2008:fhscoutflows,price2012:fhscoutflows,dunham2014:ppvi}, leading to substantial uncertainty in the nature of this object. While we detect this redshifted emission in our MASSES \cojtwo\ observations, we do not detect the compact source in our 1.3~mm continuum observations even though our sensitivity should be sufficient given the flux density reported by \citet{chen2010:fhsc}.  We thus agree with \citet{maureira2020:fhsc} that there is no compact source driving an outflow at this position.

\vspace{0.1in}
\noindent \textbf{L1451-mm: }
L1451-mm was first identified by \citet{pineda2011:fhsc} as a candidate first hydrostatic core, and current evidence confirms it is one of the best known candidates for a hydrostatic core \citep{maureira2017:l1451mm,maureira2020:fhsc}.  We detect a very slow, very compact outflow consistent with \citet[][see their Figure~7]{pineda2011:fhsc}.  However, as this outflow is essentially unresolved and contained entirely within 4$\arcsec$ of its driving source, we are unable to reliably measure opening and position angles.

\vspace{0.1in}
\noindent \textbf{SVS~13B: }
SVS~13B drives a collimated, jet-like outflow detected in SiO emission \citep{bachiller1998:svs13b}, with the blueshifted lobe extending to the south and the redshifted lobe extending to the north.  While we do see evidence for this same outflow in our MASSES \cojtwo\ data, the fact that it is located on the edge of the bright, blueshifted lobe of the outflow driven by Per-emb~44 (SVS~13A) makes it impossible to accurately separate and fit this outflow.

\color{black}
\section{Corrections for Source Inclination and Measurement Technique}\label{sec_app_corrections}

\begin{figure*}
    \resizebox{\hsize}{!}{\includegraphics{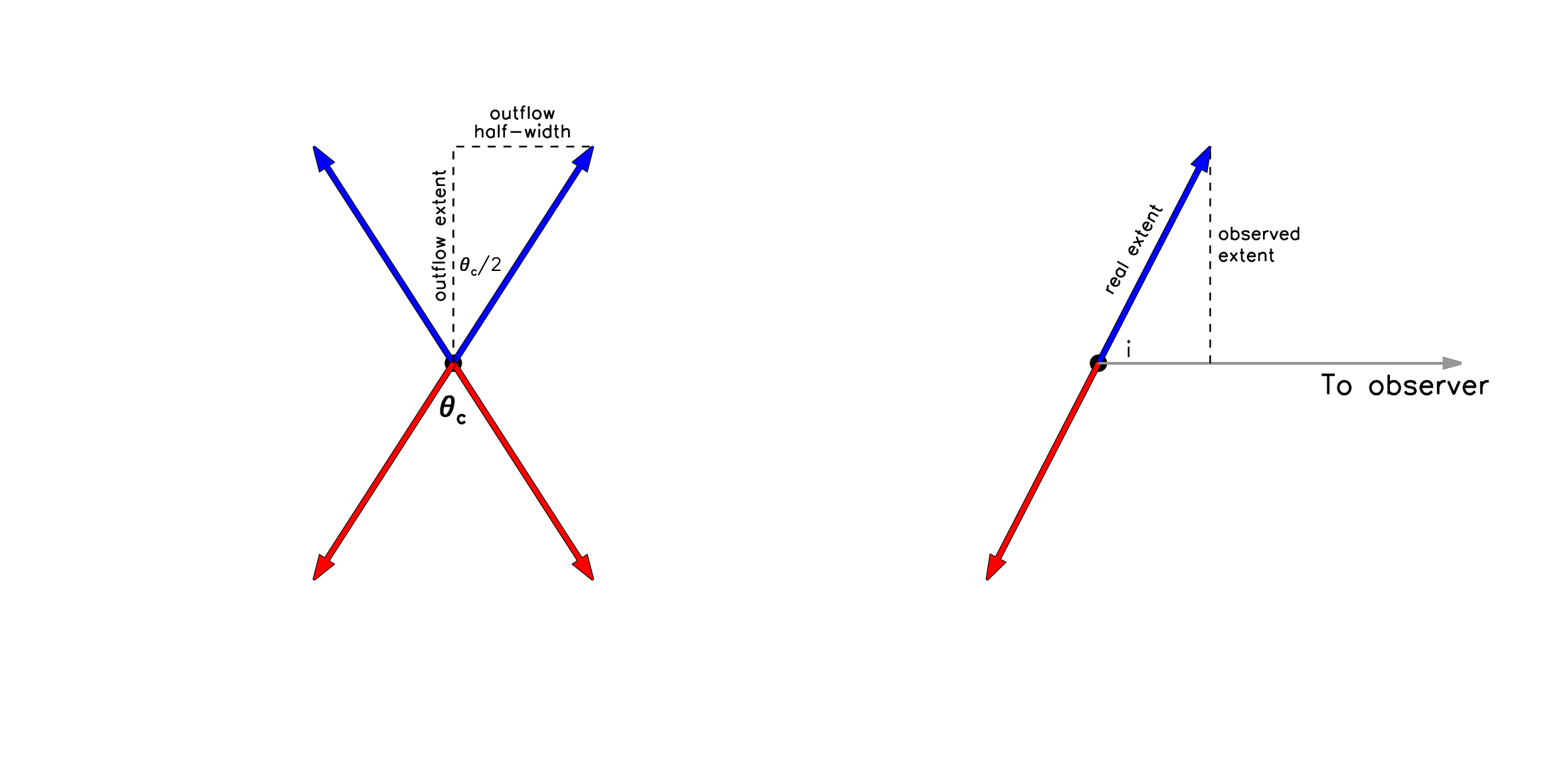}}
    \vspace{-0.8in}
    \caption{\color{black}{\it Left:}Geometry of an outflow viewed edge-on.  {\it Right:} Geometry of the real and observed extents of an outflow when viewed at an inclination angle $i$.\color{black}}
    \label{fig_corrections_inc}
\end{figure*}

\subsection{Corrections for Source Inclination}

Here we describe our method for correcting an outflow opening angles, $\theta_c$, for source inclination (or, alternatively, for projecting a corrected opening angle back onto the plane of the sky).  This method assumes conical shaped outflow cavities.

As seen in the left panel of Figure \ref{fig_corrections_inc}, for a conical outflow cavity,

\begin{equation}\label{eq_inc1}
    {\rm tan} \left(\frac{\theta_c}{2}\right) = \frac{\left(\rm outflow \, half{\rm -}width\right)}{\left(\rm outflow \, extent\right)} \, .
\end{equation}
Additionally, as seen in the right panel of Figure \ref{fig_corrections_inc}, for an observer viewing the outflow at an inclination $i$,

\begin{equation}\label{eq_inc2}
    \left({\rm observed \, extent}\right) = \left({\rm real \, extent}\right) {\rm sin}\left(i\right) \, ,
\end{equation}
Since the outflow extent is reduced by a factor of ${\rm sin}\left(i\right)$ but the outflow \color{black}half-width \color{black}is unchanged, the observed opening angle, $\theta_c^{\rm obs}$, is given by

\begin{equation}\label{eq_inc3}
    {\rm tan} \left(\frac{\theta_c^{\rm obs}}{2}\right) = \frac{\left(\rm real \, half{\rm -}width\right)}{\left(\rm observed \, extent\right)} = \frac{\left(\rm real \, half{\rm -}width\right)}{\left(\rm real \, extent\right)} \frac{1}{{\rm sin}\left(i\right)} \, .
\end{equation}
Comparing Equations \ref{eq_inc1} and \ref{eq_inc3}, the observed and intrinsic outflow opening angles are related by

\begin{equation}\label{eq_inclination}
    {\rm tan} \left(\frac{\theta_c}{2}\right) = {\rm tan} \left(\frac{\theta_c^{\rm obs}}{2}\right) {\rm sin}\left(i\right) \, .
\end{equation}
Equation \ref{eq_inclination} can be used to either correct the observed $\theta_c^{\rm obs}$ for inclination, or to project an inclination-corrected opening angle $\theta_c$ back onto the plane of the sky.

\subsection{Corrections For Measurement Technique}

In two of the previous studies to which we compare, namely \citet{hsieh2017:outflows} and \citet{habel2021:outflows}, curved outflow cavities given by power-law functions are fit to the cavity morphologies detected in infrared light.  Since the opening angle of such an outflow cavity depends on the height above the midplane where it is measured (wider near the midplane and narrower farther from the midplane), both studies tabulate the opening angles of cones that intersect the power-law cavity walls at fixed vertical heights above the midplane.  The specific vertical heights chosen in each study are 5000~au for \citet{hsieh2017:outflows} and 8000~au for \citet{habel2021:outflows}.

\begin{figure*}
    \resizebox{\hsize}{!}{\includegraphics{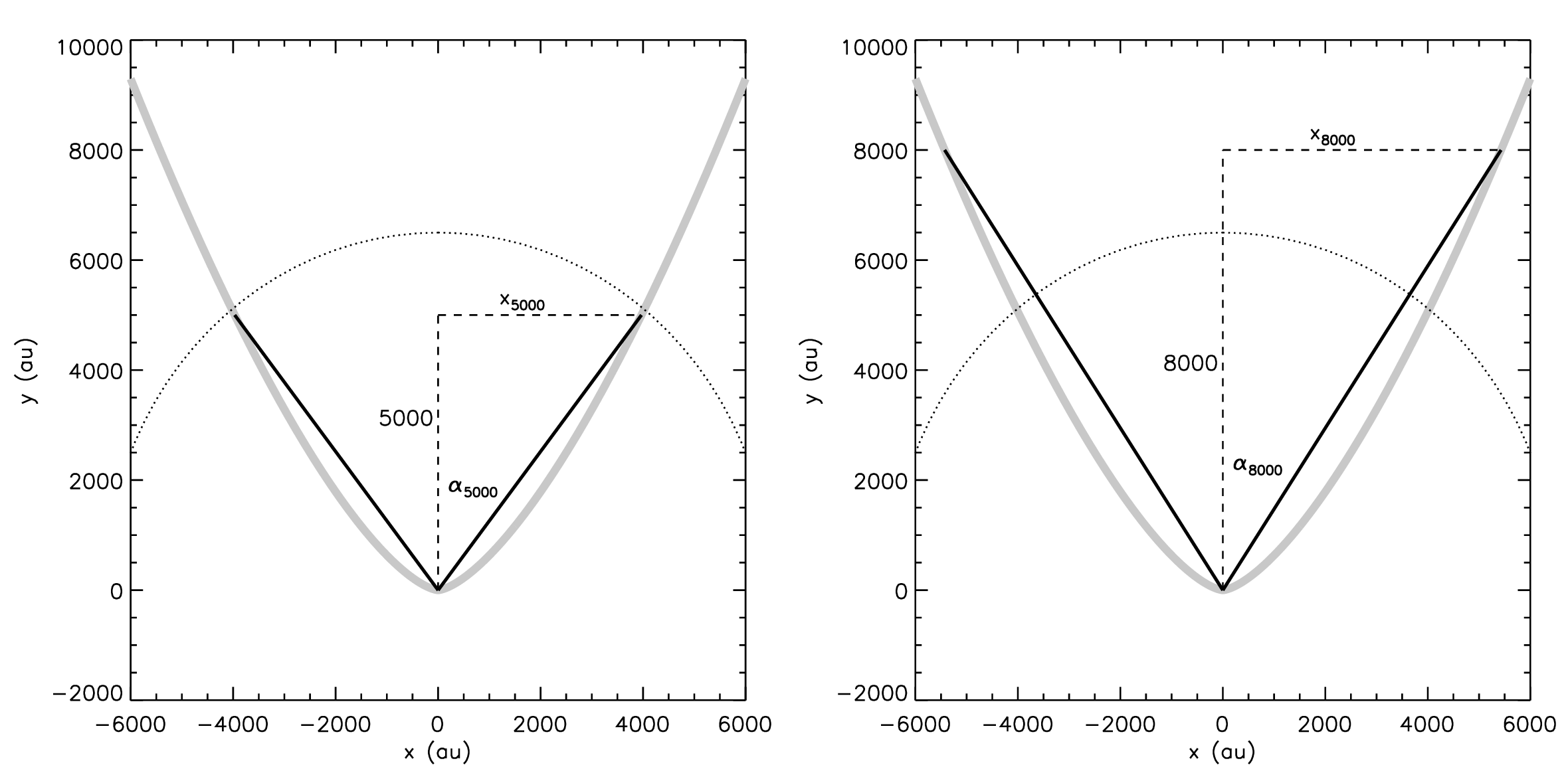}}
    \caption{\color{black}Comparison of curved (power-law) and conical outflow cavity shapes.  In both panels the dotted line shows an assumed core boundary of 6500~au (see text for details) and the thick gray line shows a power-law cavity given by $y=Ax^p$, where $A=0.02$ and $p=1.5$ in this specific example.  The solid black lines show cones that intersect the power-law cavity walls at fixed vertical heights above the midplane (5000~au in the left panel and 8000~au in the right panel).\color{black}}
    \label{fig_corrections_shape}
\end{figure*}

The left panel of Figure \ref{fig_corrections_shape} shows an example of this process for a power-law cavity given by:

\begin{equation}\label{eq_powerlaw}
    y=Ax^p \, ,
\end{equation}
where $A=0.02$ and $p=1.5$ in this specific example.  This example is meant to illustrate the process as it was followed by \citet{hsieh2017:outflows}, who also adopted $p=1.5$ in their fits.  For the specific example show in this figure, the opening half-angle is given by:

\begin{equation}
\alpha_{5000} = \arctan \left( \frac{x_{5000}}{5000} \right) = \arctan \left( \frac{3968.5}{5000} \right) = 38.4\degree \, ,
\end{equation}
giving a full opening angle of 76.8\degree\ \color{black}(we note here that we use $\theta$ to refer to full opening angles and $\alpha$ to refer to half opening angles).\color{black}

This method for measuring outflow opening angles is very different than the method used by us \citep[and by][]{hsieh2023:outflows}.  In an appendix to their paper, \citet{myers2023:outflows} compared the two methods and found that they show excellent statistical agreement, with the differences between the two methods smaller than the typical 5--15\degree\ uncertainties in the method used in this paper.  \citet{myers2023:outflows} performed this comparison using cones that intersect the power-law cavity walls at a fixed {\it radius} of 6500~au rather than a fixed {\it vertical height} of 5000~au. As seen in Figure \ref{fig_corrections_shape}, 
differences between these two cones are minimal (a cone drawn to the intersection of the 6500~au core boundary and the power-law cavity shape would be almost identical to the cone shown in the figure).  Thus, we do not perform any corrections for measurement technique before comparing our measuremments with opening angles measured by \citet{hsieh2017:outflows}.

However, as seen in the right panel of Figure \ref{fig_corrections_shape}, the vertical height of 8000~au used by \citet{habel2021:outflows} results in narrower cones.  To ensure an apples-to-apples comparison between their results and ours, We thus correct their reported opening half-angles ($\alpha_{8000}$) to those measured at a vertical height of 5000~au ($\alpha_{5000}$).  To perform this correction, we first solve Equation \ref{eq_powerlaw} for $A$ to calculate the specific value of $A$ for each outflow in the \citet{habel2021:outflows} sample as follows:

\begin{equation}\label{eq_shape1}
A = \frac{y}{x^p} = \frac{y}{\left[y\, {\rm tan}\left(\alpha\right)\right]^p} = \frac{8000}{\left[8000\, {\rm tan}\left(\alpha_{8000}\right)\right]^p} \, .
\end{equation}
Once the value of $A$ is known, we then solve Equation \ref{eq_powerlaw} for $x$ to calculate the horizontal coordinate where the power-law cavity reaches a vertical height of 5000~au:

\begin{equation}\label{eq_shape2}
    x_{5000} = \left(\frac{y}{A}\right)^\frac{1}{p} = \left(\frac{5000}{A}\right)^\frac{1}{p} \, .
\end{equation}
Finally, we calculate $\alpha_{5000}$ as:

\begin{equation}\label{eq_shape3}
    \alpha_{5000} = {\rm arctan} \left(\frac{x_{5000}}{y}\right) = {\rm arctan} \left(\frac{x_{5000}}{5000}\right) \, .
\end{equation}

\color{black}


\bsp	
\label{lastpage}
\end{document}